\documentclass[a4paper,11pt]{article}
\usepackage{jheppub} 
\usepackage{lineno}
\usepackage{float}
\usepackage{booktabs}
\usepackage{placeins}

\title{\boldmath A fluid dual to large \texorpdfstring{$D$}{D} membrane paradigm at first subleading order }

\author[a]{Supratim Halder,}
\author[a]{Manu Kurian,}
\author[a]{and Mangesh Mandlik}

\affiliation[a]{Department of Physics, Indian Institute of Technology
(Indian School of Mines) Dhanbad,\\ Jharkhand 826004, India}

\emailAdd{23dr0184@iitism.ac.in} 
\emailAdd{manukurian@iitism.ac.in}
\emailAdd{mandlik@iitism.ac.in}

\abstract{The large $D$ membrane paradigm establishes a duality between the dynamics of black holes and the evolution of a codimension-one timelike membrane in a non-gravitational background. In this work, we formulate the relativistic fluid dynamics dual to an uncharged black hole in asymptotically flat spacetime at the first subleading order in the $1/D$ expansion. Due to the absence of a timelike asymptotic boundary, unlike in AdS/CFT fluid-gravity duality, we construct an effective fluid intrinsically on the dynamical membrane, whose equations of motion are exactly equivalent to the subleading order membrane equations. By performing this fluid-dynamical analysis in the Landau frame, we show that the system behaves as a fluid influenced by an effective background force. Out-of-equilibrium viscous effects emerge naturally, allowing us to extract the fluid transport coefficients directly from the bulk viscous pressure and shear stress tensor. Furthermore, the fluid exhibits a negative pressure, capturing the intrinsic surface tension of the $(D-1)$-dimensional membrane worldvolume.
}

\begin{document}
\maketitle
\section{Introduction}
The large $D$ limit of General Relativity provides a perturbative framework for disentangling the nontrivial dynamics of black holes~\cite{Emparan:2013moa,Emparan:2014aba,Emparan:2015rva}. As the spacetime dimension approaches infinity ($D \to \infty$), the black hole's gravitational field strongly localises within a small boundary layer of thickness $\mathcal{O}(r_0/D)$ just outside the event horizon. This extreme geometric confinement cleanly factorises the quasinormal mode spectrum, completely decoupling the highly damped heavy modes from the long-lived light modes that drive the system's residual evolution. Because the frequencies of the heavy modes scale with $D$, their proportionately vast imaginary components ensure near-instantaneous dissipation. The surviving light degrees of freedom are comparatively few, whose effective non-linear behaviour is substantially more tractable than the full gravitational field equations. Leveraging this strict separation of temporal scales, the large $D$ membrane paradigm projects the complex bulk gravitational dynamics onto a continuous, codimension-one timelike hypersurface (the membrane hypersurface) embedded in a background spacetime~\cite{Bhattacharyya:2015dva,Bhattacharyya:2015fdk}. Consequently, this membrane operates as a purely non-gravitational effective field theory, dictating the non-linear, long-wavelength deformations of the black hole geometry~\cite{Bhattacharyya:2017hpj,Kundu:2018dvx,Dandekar:2016fvw}.

The translation of this effective field theory into a hydrodynamic framework is crucial for generalising the fluid-gravity correspondence. For black holes in Anti-de Sitter (AdS) spacetime, the fluid gravity duality is robustly defined~\cite{Bhattacharyya:2007vjd,Dandekar:2017aiv},  with subsequent literature proving the equivalence between the fluid-gravity derivative expansion and the large $D$ membrane-gravity correspondence at the asymptotic conformal boundary~\cite{Bhattacharyya:2018iwt,Bhattacharyya:2019mbz,Patra:2019hlq}. However, extending this correspondence to asymptotically flat spacetimes introduces a fundamental problem: the absence of a timelike asymptotic boundary prohibits mapping the fluid to spatial infinity. To resolve this, the effective relativistic fluid must be constructed intrinsically on the dynamical membrane's worldvolume. Recent work successfully implemented this fluid-dynamical construction at the leading order for a charged membrane~\cite{Halder:2026sdh}. The uncharged leading-order dynamics can be recovered trivially by taking the zero-charge limit of its Landau-frame formulation. Extending the analysis to subleading orders provides a more complete picture of the fluid's dissipative structure.

Building upon this foundation, this work extends the hydrodynamic duality for an uncharged large $D$ membrane to the first subleading order in the $1/D$ expansion. The earlier leading-order formalism successfully establishes the fundamental fluid kinematics and primary transport properties~\cite{Halder:2026sdh}, but the broader spectrum of dissipative relativistic hydrodynamics, including bulk viscous pressure and higher-derivative shear stresses, which naturally emerge from the subleading order formalism, remains unexplored.

In this work, we systematically utilise the higher-order (in $1/D$) corrections of the membrane stress-energy tensor to formulate relativistic fluid dynamics consistent with the subleading order membrane equations\footnote{The membrane equations are the geometric constraint equations which dictate the dynamics of the membrane in the dual membrane picture. They were initially introduced in \cite{Bhattacharyya:2015dva,Bhattacharyya:2015fdk}, building upon the foundational large $D$ framework established in \cite{Emparan:2013moa,Emparan:2013xia,Emparan:2013oza,Emparan:2014cia,Emparan:2014jca,Emparan:2014aba,Emparan:2015rva}. For the specific case of stationary solutions, alternative derivations of the membrane dynamics can be found in Refs.~\cite{Emparan:2015hwa,Suzuki:2015iha,Tanabe:2015isb,Tanabe:2016opw}. The subleading order membrane equation was first introduced in \cite{Dandekar:2016fvw}.}. By incorporating the subleading geometric corrections and evaluating the system in the Landau frame~\cite{Landau1959}, which rigidly aligns the fluid velocity with the local energy flow, we enable an unambiguous algebraic decomposition of the effective stress-energy tensor into its ideal thermodynamic sector and its transverse viscous corrections~\cite{Jaiswal:2016hex, Panda:2020zhr}. Similar to~\cite{Halder:2026sdh}, the fluid is driven by an effective background force. This analysis renders that the subleading geometric evolution of the large-$D$ membrane is dynamically equivalent to the forced non-equilibrium evolution of this relativistic fluid. After defining the global thermodynamic equilibrium, we redefine the local temperature to capture out-of-equilibrium dynamics and systematically extract the higher-order transport coefficients from the expressions for the bulk pressure and shear stress tensor.

This paper is organised as follows. Section~\ref{Large D membrane paradigm at subleading order} revisits the large $D$ membrane paradigm formalism to introduce the first subleading-order membrane equation. Section~\ref{Landau frame velocity and fluid's stress-energy tensor in the large $D$ Limit} defines the Landau velocity and the fluid's stress-energy tensor. In section~\ref{Thermodynamic State and the Macroscopic First Law}, we define the equilibrium thermodynamic variables (energy density, pressure, and temperature); section~\ref{Local thermodynamic equilibrium} then redefines the local temperature using a matching condition to establish the out-of-equilibrium energy density and pressure. In section~\ref{Out of equilibrium viscous effects}, we express the bulk viscous pressure and shear stress tensor in terms of the fluid's velocity and out-of-equilibrium temperature derived in the earlier sections to extract the transport coefficients. Section~\ref{Equation of motion} presents the fluid's energy and momentum conservation equations. Section~\ref{Entropy current} defines the entropy current. Section~\ref{Results and discussions} describes the transport coefficients for shear stress tensor and bulk pressure, followed by a discussion of our central results. Finally, in section~\ref{Summary and outlook}, we present the summary of the analysis with an outlook. 

\section{Large $D$ membrane paradigm at subleading order}\label{Large D membrane paradigm at subleading order}
\subsection{The large $D$ limit and the schwarzschild geometry}

In the limit $D \to \infty$, the gravitational dynamics of black holes separate into two distinct length scales: a larger scale of the order of the horizon radius $r_0$ that governs dynamics along the horizon-parallel directions, and a shorter radial scale of thickness $r_0/D$ to which all non-trivial gravitational effects are confined \cite{Emparan:2013moa}. This curvature confinement is explicit in the $D$-dimensional Schwarzschild metric
\begin{equation}
ds^2 = -f(r)dt^2 + \frac{dr^2}{f(r)} + r^2 d\Omega_{D-2}^2, \quad f(r) = 1 - \left(\frac{r_0}{r}\right)^{D-3}.
\end{equation}
Introducing the rescaled radial coordinate $R$ defined as $r = r_0\left(1 + \frac{R}{D-3}\right)$, the metric function limits to $f(R) = 1 - e^{-R}$ as $D \to \infty$.  Consequently, any deviation from the flat Minkowski background ($f(r)=1$) is confined to a very small region called the membrane region of order $r_0/D$ immediately outside the event horizon at $r=r_0$. Outside this highly localised region, the spacetime is flat.
The universal geometry of this membrane region is defined by the $D$ dimensional background metric ansatz \cite{Bhattacharyya:2015fdk}
\begin{align}\label{Leading}
G_{MN} = \eta_{MN} + \frac{(n_M - u_M)(n_N - u_N)}{\psi^{D-3}}.
\end{align}
Here, $\psi$ is a smooth real scalar profile defining the membrane shape. The quantity $u_M$ is a unit-normalised, time-like velocity field tangent to the membrane surfaces. The field $n_M$ is the space-like normal vector to the surfaces of constant $\psi$. For the Schwarzschild solution, setting the velocity to $u = -dt$ and the shape profile to $\psi = r/r_0$ exactly recovers the black hole metric in Kerr-Schild coordinates.

\subsection{Membrane embedding and kinematics}
Here in the background flat $D$-dimensional Minkowski spacetime ($\eta_{MN}$), the membrane is a dynamical, codimension-one timelike hypersurface, which is defined by the scalar shape function $\psi(X^M)$, as $\psi(X^M) = 1$. 
The outward-directed unit normal to this hypersurface is formulated as
  $$n_M = \frac{\partial_M \psi}{\sqrt{\eta^{AB}\partial_A \psi \partial_B \psi}}.$$Consequently, the induced metric on the membrane worldvolume is $$g_{MN} = \eta_{MN} - n_M n_N.$$
The embedding of the membrane is captured by the extrinsic curvature $K_{MN} = \nabla^{\text{(bg)}}_M n_N$ (where $\nabla^{\text{(bg)}}_M$ is the covariant derivative in background spacetime), and its corresponding trace $K$. Furthermore, the membrane velocity always lies in the membrane worldvolume, satisfying $g_{\mu\nu}u^\mu u^\nu = -1$. 
\subsection{Perturbation in order \texorpdfstring{$1/D$}{1/D}}

At leading order, an arbitrary membrane configuration is constructed by patching together local Schwarzschild black brane geometries (visible from Eq.~\eqref{Leading}). Because this initial ansatz only partially satisfies the vacuum Einstein equations, deriving exact, non-singular solutions requires a systematic perturbative expansion in $1/D$. Consequently, the full spacetime metric is expanded as
$$G_{MN} = \eta_{MN} + \sum_{k=0}^{\infty} \frac{h_{MN}^{(k)}}{(D-3)^k}.$$
Now, this perturbative metric ansatz should satisfy the Einstein equation at the desired order in $\mathcal{O}(1/D)$.

\subsection{Membrane equations at subleading order}
The membrane equations governing the membrane's dynamics are actually the Einstein constraint equations evaluated on the event horizon. To guarantee that the metric corrections remain regular in the $(n+1)^{\text{th}}$ perturbative order, these $n^{\text{th}}$ order constraint equations must be identically satisfied. At the first subleading order, these equations completely determine both the dynamical evolution of the membrane velocity field $u_M$ and its divergence given as Eq.~\eqref{eq:membrane:vector}~and~Eq.~\eqref{eq:membrane_scalar} respectively.
The geometric surface constraints governing the membrane equations can be written as the conservation equations of an effective stress-energy tensor. The framework dictates that the spacetime curvature, specifically the metric's deviation from the flat Minkowski background, is sourced entirely by this stress tensor living on the membrane worldvolume. In Ref.~\cite{Biswas:2019xip}, the explicit calculation of the membrane stress-energy tensor whose conservation reproduces the first subleading order membrane equations in a flat background yields the following form:
\begin{align}\label{Membrane:EM}
 T_{\mu\nu} &=\frac{1}{8\pi}( s_{1} u_{\mu} u_{\nu} + s_{2} g_{\mu\nu} + \mathcal{V}_{\mu} u_{\nu} + \mathcal{V}_{\nu} u_{\mu} + \mathcal{W}_{\mu\nu}),
\end{align}
with the quantities defined as follows:
\begin{align}
s_{1} &= \frac{K}{2} + \frac{1}{2}\left(\frac{\nabla^{2}K}{K^{2}} - \frac{1}{K}K_{\alpha\beta}K^{\alpha\beta}\right) \nonumber \\
&\quad + \frac{1}{K} \bigg[ -u\cdot K\cdot K\cdot u - 13\left(\frac{u\cdot\nabla K}{K}\right)^{2} + 2u^{\alpha}K_{\alpha\beta}\left(\frac{\nabla^{\beta}K}{K}\right) + 14\left(\frac{u\cdot\nabla K}{K}\right)(u\cdot K\cdot u) \nonumber \\
&\quad - \frac{K}{D}\left(\frac{u\cdot\nabla K}{K}\right) + \frac{K}{D}(u\cdot K\cdot u) + \frac{1}{K^{3}}\nabla^{2}(\nabla^{2}K) - 4(u\cdot K\cdot u)^{2}- 2\left(\frac{\nabla_{\alpha}K}{K}\right)\left(\frac{\nabla^{\alpha}K}{K}\right) \bigg] \nonumber \\
&\quad + \frac{1}{K}(2\zeta(3)-1) \bigg[ -\frac{K}{D}\left(\frac{u\cdot\nabla K}{K} - u\cdot K\cdot u\right) - u\cdot K\cdot K\cdot u + 2\left(\frac{\nabla_{\alpha}K}{K}\right)u^{\beta}K_{\beta}^{\alpha} \nonumber \\
&\quad - \left(\frac{u\cdot\nabla K}{K}\right)^{2} + 2\left(\frac{u\cdot\nabla K}{K}\right)(u\cdot K\cdot u) - \left(\frac{\nabla^{\alpha}K}{K}\right)\left(\frac{\nabla_{\alpha}K}{K}\right) - (u\cdot K\cdot u)^{2} \bigg], \label{eq:S1}
\end{align}
\begin{align}
s_{2} &= -\frac{1}{2}(u\cdot K\cdot u) - \frac{1}{2K}K^{\alpha\beta}K_{\alpha\beta} - \frac{1}{K}\left(\frac{u\cdot\nabla K}{K} - \frac{1}{2}(u\cdot K\cdot u) - \frac{K}{2D}\right)(u\cdot K\cdot u) \nonumber \\
&\quad + \frac{1}{K}K^{\alpha\beta}(\nabla_{\alpha}u_{\beta}) - \frac{2}{K}u_{\alpha}K^{\alpha\beta}\left(\frac{1}{2}\frac{\nabla_{\beta}K}{K} - \frac{\nabla^{2}u_{\beta}}{K}\right), \label{eq:S2}
\end{align}
\begin{align}
\mathcal{V}_{\mu} &= \frac{1}{2}\left(\frac{\nabla_{\mu}K}{K}\right) - \left(\frac{\nabla^{2}u_{\mu}}{K}\right) + \frac{1}{K}K_{\mu}^{\alpha}K_{\alpha\beta}u^{\beta} - \frac{1}{K^{3}}\nabla^{2}(\nabla^{2}u_{\mu}) + \frac{1}{K}\nabla_{\mu}\left(\frac{u\cdot\nabla K}{K}\right) \nonumber \\
&\quad + \frac{1}{K}\left(\frac{\nabla^{2}u_{\mu}}{K}\right)\left(-2(u\cdot K\cdot u) + 4\frac{u\cdot\nabla K}{K}  - \frac{K}{D}\right) + \frac{1}{2K}\left(\frac{\nabla_{\mu}K}{K}\right)(u\cdot K\cdot u) ,\label{eq:Vmu}
\end{align}
\begin{align}
\mathcal{W}_{\mu\nu} &= \frac{1}{2}K_{\mu\nu} - \frac{1}{2}(\nabla_{\mu}u_{\nu} + \nabla_{\nu}u_{\mu}) - \frac{1}{K}K_{\mu\nu}(u\cdot K\cdot u) + \frac{1}{2K}(\nabla_{\mu}u_{\nu} + \nabla_{\nu}u_{\mu})(u\cdot K\cdot u) \nonumber \\
&\quad + \frac{1}{2K} \bigg[ \nabla_{\mu}\left(\frac{\nabla^{2}u_{\nu}}{K}\right) + \nabla_{\nu}\left(\frac{\nabla^{2}u_{\mu}}{K}\right) + \nabla_{\mu}(u^{\alpha}K_{\alpha\nu}) + \nabla_{\nu}(u^{\alpha}K_{\alpha\mu}) - 2\nabla_{\mu}\left(\frac{\nabla_{\nu}K}{K}\right) \bigg] \nonumber \\
&\quad - \frac{1}{K}(\nabla^{\alpha}u_{\mu})(\nabla_{\alpha}u_{\nu}) - \frac{1}{K}\left(\frac{\nabla^{2}u_{\mu}}{K}\right)\left(\frac{\nabla^{2}u_{\nu}}{K}\right).\label{eq:Wmunu}
\end{align}
Imposing the conservation of the aforementioned stress-energy tensor (Eq.~\eqref{Membrane:EM}) yields the membrane equations at the first subleading order. The two membrane equations derived by the membrane velocity component and perpendicular to the membrane velocity component of the conservation equation. They can be expressed as follows\footnote{Where the first bracket at the suffix represents the symmetrisation of the index, and the projector $P_{\mu \nu}$ is defined as $P_{\mu \nu}=g_{\mu \nu}+u_{\mu} u_{\nu}$. }:
\begin{align}
&\nabla\cdot u = \frac{1}{2K} \left( \nabla_{(\alpha}u_{\beta)} \nabla_{(\gamma} u_{\delta)} P^{\beta\gamma} P^{\alpha\delta} \right)+ \mathcal{O}\left(\frac{1}{D}\right)^{2}, \label{eq:membrane_scalar}
\end{align}
\begin{align}\label{eq:membrane:vector}
&\Big[ \frac{\nabla^2 u_\alpha}{K} - \frac{\nabla_\alpha K}{K} + u^\beta K_{\beta\alpha} - u \cdot \nabla u_\alpha \Big] P^\alpha_\gamma \notag\\
&+ \bigg[ \Big( -\frac{u^\gamma K_{\gamma\beta} K^\beta_\alpha}{K} \Big) + \Big( \frac{\nabla^2 \nabla^2 u_\alpha}{K^3} - \frac{u \cdot \nabla K \nabla_\alpha K}{K^3} - \frac{\nabla^\beta K \nabla_\beta u_\alpha}{K^2} - 2 \frac{K^{\gamma\delta} \nabla_\gamma \nabla_\delta u_\alpha}{K^2} \Big) \notag \\
& + \Big( -\frac{\nabla_\alpha \nabla^2 K}{K^3} + \frac{\nabla_\alpha ( K_{\beta\gamma} K^{\beta\gamma} K )}{K^3} \Big) + 3 \frac{(u \cdot K \cdot u)(u \cdot \nabla u_\alpha)}{K} - 3 \frac{(u \cdot K \cdot u)(u^\beta K_{\beta\alpha})}{K} \notag \\
& - 6 \frac{(u \cdot \nabla K)(u \cdot \nabla u_\alpha)}{K^2} + 6 \frac{(u \cdot \nabla K)(u^\beta K_{\beta\alpha})}{K^2} + \frac{3}{(D-3)} u \cdot \nabla u_\alpha - \frac{3}{(D-3)} u^\beta K_{\beta\alpha} \bigg] P^\alpha_\gamma \notag\\
& =\mathcal{O}\left(\frac{1}{D}\right)^{2}.
\end{align}
 The divergence operator in the large $D$ limit can elevate the dimensional scaling of tensorial structures by an implicit factor of $D$; therefore, constructing a hydrodynamic framework consistent with the subleading-order membrane equation requires retaining terms in the stress-energy tensor strictly up to the second subleading order, $\mathcal{O}(1/D)$. This allows us to safely suppress the terms $\mathcal{O}(1/D^2)$. Subject to this truncation scheme, we proceed to formulate the macroscopic fluid analysis by defining the corresponding thermodynamic and kinematic variables derived from the large-D membrane paradigm. In this work, we adopt the Landau frame to analyse the effective fluid dynamics.

\section{Landau velocity and fluid's stress-energy tensor at large $D$ limit}\label{Landau frame velocity and fluid's stress-energy tensor in the large $D$ Limit}
In the Landau frame, the fluid four-velocity is defined to align with the flow of energy. Consequently, the velocity field is a timelike eigenvector of the energy-momentum tensor, with the local energy density as its corresponding eigenvalue. To determine the Landau velocity, we first project the different components of the membrane energy-momentum tensor from Eq.~\eqref{Membrane:EM} parallel and orthogonal to the membrane velocity $u^\mu$ as follows:
\begin{align}\label{eq:stress_tensor}
8\pi T_{\mu\nu} &= s_1 u_\mu u_\nu + s_2 g_{\mu\nu} + \mathcal{V}_\mu u_\nu + \mathcal{V}_\nu u_\mu + \mathcal{W}_{\mu\nu},\notag \\
8\pi T_{\mu\nu} &= (s_1 - 2\mathcal{V}^\alpha u_\alpha + \mathcal{W}^{\alpha\beta} u_\alpha u_\beta) u_\mu u_\nu\notag \\
&\quad + s_2 g_{\mu\nu} + (\mathcal{V}^\alpha P_{\mu\alpha} - \mathcal{W}^{\alpha\beta} P_{\mu\alpha} u_\beta) u_\nu\notag \\
&\quad + (\mathcal{V}^\beta P_{\nu\beta} - \mathcal{W}^{\alpha\beta} P_{\nu\beta} u_\alpha) u_\mu + \mathcal{W}^{\alpha\beta} P_{\mu\alpha} P_{\nu\beta} + \mathcal{O}(1/D^2)\notag \\
&= e u_\mu u_\nu + s_2 g_{\mu\nu} + l_\mu u_\nu + l_\nu u_\mu + r_{\mu\nu} + \mathcal{O}(1/D^2).
\end{align}
The variables $e$, $l_\mu$, $r_{\mu\nu}$ are defined in Table~\ref{tab:variables}.
\begin{table}[htbp]
    \centering
    \renewcommand{\arraystretch}{1.25}
    \begin{tabular}{|l|l|l|}
        \hline
        Scalar sector & $e$ & $ s_1 - 2\,\mathcal{V}^\alpha u_\alpha + \mathcal{W}^{\alpha\beta}u_\alpha u_\beta$ \\
        \hline
        Vector sector & $l_\mu$ & $ \mathcal{V}^\alpha P_{\mu\alpha} - \mathcal{W}^{\alpha\beta}P_{\mu\alpha}u_\beta$ \\
        \hline
        Tensor sector & $r_{\mu\nu}$ & $ \mathcal{W}^{\alpha\beta}P_{\mu\alpha}P_{\nu\beta}$ \\
        \hline
    \end{tabular}
    \caption{Decomposition of variables.}
    \label{tab:variables}
\end{table}

Now we define the fluid stress-energy tensor $T^{\text{(fluid)}}_{\mu\nu}$ as follows:
\begin{align}\label{T}
    8\pi T^{\text{(fluid)}}_{\mu\nu}= e u_\mu u_\nu +s_2 g_{\mu\nu} + l_\mu u_\nu + l_\nu u_\mu + \tau_{\mu\nu}\,,
\end{align}
where 
\begin{align}
    \tau_{\mu\nu}= r_{\mu\nu}-\bigg( \frac{1}{2} K_{\alpha\beta} P^{\alpha}_{\mu} P^{\beta}_{\nu}- \frac{u \cdot K \cdot u}{K} K_{\alpha\beta} P^{\alpha}_{\mu} P^{\beta}_{\nu}\bigg)
\end{align}
The Landau frame velocity $U^\mu$ is defined by enforcing the exact eigenvalue condition for the fluid stress-energy tensor. In Eq.~\eqref{T}, $u_\mu$ is the unit-normalised membrane velocity ($u^\mu u_\mu = -1$), and the transversality conditions $u^\mu l_\mu = 0$ and $u^\mu \tau_{\mu\nu} = 0$ are satisfied. 
To maintain exact normalisation $U^\mu U_\mu = -1$ non-perturbatively, we decompose the Landau velocity along and  orthogonal to membrane velocity ($u_\mu$) using a shift vector $v^\mu$ (with $u_\mu v^\mu = 0$) and a normalisation factor $Y = (1 - v^2)^{-1/2}$ as follows:
\begin{equation}\label{U}
    U^\mu = Y (u^\mu + v^\mu)\,,
\end{equation}
where $v^2 \equiv v_\mu v^\mu$. The Landau frame condition leads to
\begin{equation} \label{eq:landau_cond}
    8\pi T^{\text{(fluid)}}_{\mu\nu} U^\nu = -8\pi \mathcal{E} U_\mu\,,
\end{equation}
with $\mathcal{E}$ as the proper energy density. Substituting Eq.~\eqref{eq:stress_tensor} into Eq.~\eqref{eq:landau_cond}, we obtain the following algebraic relation:
\begin{equation} \label{eq:exact_algebraic}
    -e u^\mu + s_2 u^\mu - l^\mu + s_2 v^\mu + (l_\nu v^\nu)u^\mu + \tau^\mu_{~~\nu} v^\nu = -8\pi\mathcal{E} (u^\mu + v^\mu)\,.
\end{equation}
By projecting Eq.~\eqref{eq:exact_algebraic} parallel to the membrane velocity (contracting with $-u_\mu$) to extract the energy density as follows:
\begin{equation} \label{eq:energy_density}
    8\pi\mathcal{E} = e - s_2 - l_\nu v^\nu\,.
\end{equation}
Similarly, by projecting Eq.~\eqref{eq:exact_algebraic} orthogonal to $u^\mu$ (contracting with $P^\alpha_\mu = \delta^\alpha_\mu + u^\alpha u_\mu$) and substituting Eq.~\eqref{eq:energy_density} we obtain
\begin{equation} \label{eq:shift_constraint}
    e v^\alpha - (l_\nu v^\nu) v^\alpha + \tau^\alpha_{~~\nu} v^\nu = l^\alpha\,.
\end{equation}
We now systematically solve Eq.~\eqref{eq:shift_constraint} using the Large $D$ scaling rules. The scaling of the tensor components dictates that $e \sim \mathcal{O}(D)$, while $l^\alpha \sim \mathcal{O}(1)$ and $\tau^\alpha_{~~\nu} \sim \mathcal{O}(1)$. For Eq.~\eqref{eq:shift_constraint} to balance, the shift must scale as $v^\alpha \sim \mathcal{O}(1/D)$. Expanding $v^\alpha$ perturbatively in $1/D$, we obtain the leading order contribution as
\begin{equation}
    v^\alpha_{(1)} = \frac{l^\alpha}{e}\,.
\end{equation}
At the next subleading order, we evaluate the relative scalings of the remaining terms in Eq.~\eqref{eq:shift_constraint}. The contraction $(l_\nu v^\nu) $ scales as $\mathcal{O}(1/D)$. Thus, it vanishes at this truncation order. Solving for $v^\alpha$ perturbatively yields
\begin{equation} \label{eq:spatial_shift_subleading}
    v^\alpha = \frac{l^\alpha}{e} - \frac{\tau^\alpha_{~~\nu} l^\nu}{e^2}+\mathcal{O}(1/D^3)\,.
\end{equation}
To compute the corresponding normalisation factor up to $\mathcal{O}(1/D^2)$, we construct $v^2$ using the leading order contribution
\begin{equation}
    v^2 = \frac{l_\nu l^\nu}{e^2} + \mathcal{O}(1/D^3)\,.
\end{equation}
The Taylor expansion of the normalisation factor gives
\begin{align} \label{eq:lorentz_factor}
Y& \approx 1 + \frac{1}{2}v^2 \notag\\
    & = 1 + \frac{l_\nu l^\nu}{2e^2}+\mathcal{O}(1/D^3)\,.
\end{align}
Finally, by substituting Eq.~\eqref{eq:spatial_shift_subleading} and Eq.~\eqref{eq:lorentz_factor} in Eq.~\eqref{U}  we obtain the Landau frame velocity up to $\mathcal{O}(1/D^2)$ as
\begin{equation}\label{LandauVelocity}
    U^\mu = u^\mu + \frac{l_\nu l^\nu}{2e^2} u^\mu + \frac{l^\mu}{e} - \frac{\tau^\mu_{~~\nu} l^\nu}{e^2}\,.
\end{equation}
We further define the Landau projection operator as follows:
\begin{align}
    \Delta_{\mu\nu}=g_{\mu\nu}+U_{\mu}U_{\nu}.
\end{align}
Substituting the Landau velocity from Eq.~\eqref{LandauVelocity} into Eq.~\eqref{eq:stress_tensor} yields the following expression for the membrane stress-energy tensor up to $\mathcal{O}(1/D)$ can be expressed as
\begin{align}\label{EMTensor}
&8\pi T_{\mu\nu}= e \left( 1 - \frac{l^2}{e^2} \right) U_{\mu} U_{\nu} + s_{2} g_{\mu\nu} - \frac{1}{2} (\nabla_{\alpha} U_{\beta} + \nabla_{\beta} U_{\alpha}) \Delta^{\alpha}_{\mu} \Delta^{\beta}_{\nu}\notag \\
&\quad + \left[ - \frac{1}{K} (\nabla^{\sigma} U_{\alpha}) (\nabla_{\sigma} U_{\beta}) - \frac{3}{2K^3} \nabla^2 U_{\alpha} \nabla^2 U_{\beta} - \frac{1}{K^2} (U_{\sigma} \nabla^2 U^{\sigma}) ( \nabla_{\alpha} U_{\beta} + \nabla_{\beta} U_{\alpha} ) \right.\notag \\
&\quad \left. + \frac{1}{2K^2} (U \cdot \nabla K) (\nabla_{\alpha} U_{\beta} + \nabla_{\beta} U_{\alpha}) + \frac{1}{2K} ( \nabla_{\alpha} (U \cdot \nabla U_{\beta}) + \nabla_{\beta} (U \cdot \nabla U_{\alpha}) ) \right.\notag \\
&\quad \left. - \frac{1}{2K^2} ( \nabla_{\alpha} (\nabla^2 U_{\beta}) + \nabla_{\beta} (\nabla^2 U_{\alpha}) ) + \frac{1}{K^3} ( (\nabla^2 U_{\beta})(\nabla_{\alpha} K) + (\nabla^2 U_{\alpha})(\nabla_{\beta} K) ) \right.\notag \\
&\quad \left. -\frac{1}{2K^2}((\nabla^2 U_\alpha) U\cdot\nabla U_\beta+(\nabla^2 U_\beta) U\cdot\nabla U_\alpha) \right] \Delta^{\alpha}_{\mu} \Delta^{\beta}_{\nu}+ \frac{1}{2} K_{\alpha\beta} P^{\alpha}_{\mu} P^{\beta}_{\nu}  \notag \\
&\quad - \frac{u \cdot K \cdot u}{K} K_{\alpha\beta} P^{\alpha}_{\mu} P^{\beta}_{\nu}+ \mathcal{O}(1/D^2).
\end{align}
We refer to Appendix~\ref{Landau_substitution} for details. Now, the fluid stress-energy tensor $T_{\mu\nu}^{\text{(fluid)}}$ can be written as 
\begin{align}\label{rawEMtensor}
    T_{\mu\nu}^{\text{(fluid)}}&=\frac{e}{8\pi} \left( 1 - \frac{l^2}{e^2} \right) U_{\mu} U_{\nu} + \frac{s_2}{8\pi} g_{\mu\nu}+\frac{\tau_{\mu\nu}}{8\pi},\notag\\ 
 &= \mathcal{E} U_{\mu} U_{\nu} + \frac{s_2}{8\pi} \Delta_{\mu\nu} + \tilde{\tau}_{\mu\nu} ,
\end{align}
where $\tilde{\tau}_{\mu\nu}=\frac{\tau_{\mu\nu}}{8\pi}$ and the energy density of the fluid $\mathcal{E}$ takes the following form:
\begin{align}
    \mathcal{E}&=\frac{1}{8\pi}\left( e - \frac{l^2}{e} -s_2\right).\notag 
\end{align}
The explicit geometric expression of $\mathcal{E}$ is listed in Appendix~\ref{Explicit geometric expression of thermodynamic quantities}.
The expression of $\tilde{\tau}_{\mu\nu}$ becomes
\begin{align}\label{Exp:tau}
    \tilde{\tau}_{\mu\nu}&=- \frac{1}{16\pi} (\nabla_{\alpha} U_{\beta} + \nabla_{\beta} U_{\alpha}) \Delta^{\alpha}_{\mu} \Delta^{\beta}_{\nu}\notag \\
&\quad + \left[ - \frac{1}{8\pi K} (\nabla^{\sigma} U_{\alpha}) (\nabla_{\sigma} U_{\beta}) - \frac{3}{16\pi K^3} \nabla^2 U_{\alpha} \nabla^2 U_{\beta} - \frac{1}{8\pi K^2} (U_{\sigma} \nabla^2 U^{\sigma}) ( \nabla_{\alpha} U_{\beta} + \nabla_{\beta} U_{\alpha} ) \right.\notag \\
&\quad \left. + \frac{1}{16\pi K^2} (U \cdot \nabla K) (\nabla_{\alpha} U_{\beta} + \nabla_{\beta} U_{\alpha}) + \frac{1}{16\pi K} ( \nabla_{\alpha} (U \cdot \nabla U_{\beta}) + \nabla_{\beta} (U \cdot \nabla U_{\alpha}) ) \right.\notag \\
&\quad \left. - \frac{1}{16\pi K^2} ( \nabla_{\alpha} (\nabla^2 U_{\beta}) + \nabla_{\beta} (\nabla^2 U_{\alpha}) ) + \frac{1}{8\pi K^3} ( (\nabla^2 U_{\beta})(\nabla_{\alpha} K) + (\nabla^2 U_{\alpha})(\nabla_{\beta} K) ) \right.\notag \\
&\quad \left. -\frac{1}{16\pi K^2}((\nabla^2 U_\alpha) U\cdot\nabla U_\beta+(\nabla^2 U_\beta) U\cdot\nabla U_\alpha) \right] \Delta^{\alpha}_{\mu} \Delta^{\beta}_{\nu}+ \mathcal{O}(1/D^2).
\end{align}
To separate the fluid stress tensor into its pure trace and traceless symmetric sectors, we need to define the standard kinematic variables and projection operators. The fluid expansion scalar $\theta$ and the acceleration vector $A_{\mu}$ are defined as
\begin{align}
    \theta &\equiv \nabla_{\alpha} U^{\alpha}, \\
    A_{\mu} &\equiv U \cdot \nabla U_{\mu},
\end{align}
respectively. The traceless projection of a rank-2 tensor  $X_{\alpha\beta}$ when it is transverse to $U_\mu$ can be defined as
\begin{equation}
    X_{\langle \mu\nu \rangle} \equiv \left[ \frac{1}{2}\left(\Delta_\mu^\alpha \Delta_\nu^\beta + \Delta_\nu^\alpha \Delta_\mu^\beta\right) - \frac{1}{D-2}\Delta_{\mu\nu}\Delta^{\alpha\beta} \right] X_{\alpha\beta}.
\end{equation}
Following this definition, the first-order traceless shear tensor takes the following form:
\begin{equation}
    \sigma_{\mu\nu} = \nabla_{\langle \mu} U_{\nu \rangle} = \frac{1}{2}(\nabla_{\alpha} U_{\beta} + \nabla_{\beta} U_{\alpha}) \Delta^{\alpha}_{\mu} \Delta^{\beta}_{\nu} - \frac{\theta}{D-2}\Delta_{\mu\nu}.
\end{equation}
Using these definitions, we decompose $\tilde{\tau}_{\mu\nu}$ into a scalar trace part $\tilde{\pi}$ and a traceless symmetric shear part $\pi_{\mu\nu}$ as
\begin{equation}\label{tau}
    \tilde{\tau}_{\mu\nu} = \pi_{\mu\nu} + \tilde{\pi} \Delta_{\mu\nu} + \mathcal{O}(1/D^2).
\end{equation}
Extracting the trace by contracting $\tilde{\tau}_{\mu\nu}$ with $\Delta^{\mu\nu}$ and dividing by the spatial dimension $(D-2)$ yields the complete scalar contribution as
\begin{align*}
    \tilde{\pi} &= - \frac{\theta}{8\pi(D-2)} + \frac{\nabla_{\alpha} A^{\alpha}}{8\pi K (D-2)} - \frac{\nabla_\alpha (\nabla^2 U^\alpha)}{8\pi K^2 (D-2)}\notag\\& - \frac{(\nabla^\sigma U^\alpha)(\nabla_\sigma U_\alpha)}{8\pi K (D-2)} - \frac{3 (\nabla^2 U^\alpha)(\nabla^2 U_\alpha)}{16\pi K^3 (D-2)} - \frac{(U_\sigma \nabla^2 U^\sigma)\theta}{4\pi K^2 (D-2)}\notag\\
    &\quad + \frac{(U \cdot \nabla K)\theta}{8\pi K^2 (D-2)} + \frac{(\nabla^2 U^\alpha)(\nabla_\alpha K)}{4\pi K^3 (D-2)}-\frac{(\nabla^2 U^\alpha)(U\cdot\nabla U_\alpha)}{8\pi K^2 (D-2)}. \end{align*}
The first three terms of the above expression is $\mathcal{O}(1/D)$, where the remaining terms are $\mathcal{O}(1/D^2)$. Hence, the leading order terms of the trace part can be written as 
    \begin{align}\label{Exp:pi:tilde}
    \tilde{\pi}=- \frac{\theta}{8\pi(D-2)}+ \frac{\nabla_{\alpha} A^{\alpha}}{8\pi K (D-2)} - \frac{\nabla_\alpha (\nabla^2 U^\alpha)}{8\pi K^2 (D-2)}+\mathcal{O}(1/D^2).
\end{align}
The geometric expression of the traceless shear tensor $\pi_{\mu\nu}$ becomes
\begin{align}\label{PimunuRaw}
    \pi_{\mu\nu} &= - \frac{1}{8\pi} \sigma_{\mu\nu} - \frac{1}{8\pi K} (\nabla^\sigma U_{\langle \mu})(\nabla_{|\sigma|} U_{\nu \rangle}) - \frac{3}{16\pi K^3} \nabla^2 U_{\langle \mu} \nabla^2 U_{\nu \rangle} \notag \\
    &\quad - \frac{1}{4\pi K^2} (U_\sigma \nabla^2 U^\sigma) \sigma_{\mu\nu} + \frac{1}{8\pi K^2} (U \cdot \nabla K) \sigma_{\mu\nu} + \frac{1}{8\pi K} \nabla_{\langle \mu} A_{\nu \rangle} \notag \\
    &\quad - \frac{1}{8\pi K^2} \nabla_{\langle \mu} (\nabla^2 U_{\nu \rangle}) + \frac{1}{4\pi K^3} \nabla_{\langle \mu} K \nabla^2 U_{\nu \rangle} -\frac{1}{8\pi K^2} A_{\langle \mu} \nabla^2 U_{\nu \rangle}.
\end{align}
To systematically extract the transport coefficients, we must express the bare higher-derivative geometric structures found in the subleading large $D$ stress tensor entirely in terms of the fundamental first-order hydrodynamic variables and their independent gradients. 
 The antisymmetric vorticity tensor can be defined as \begin{align}
     \omega_{\mu\nu}=\frac{1}{2}(\nabla_{\alpha} U_{\beta} - \nabla_{\beta} U_{\alpha}) \Delta^{\alpha}_{\mu} \Delta^{\beta}_{\nu}.
 \end{align} 
We further define the purely spatial gradient of a scalar $\Phi$ and the spatial divergence of a tensor $X_{\mu\nu}$ as
\begin{align}
    \nabla^\perp_\mu \Phi &\equiv \Delta_\mu^\rho \nabla_\rho \Phi,
\end{align}
\begin{equation}
    (\nabla \cdot X)^\perp_\mu \equiv \Delta_\mu^\rho \nabla^\lambda X_{\lambda\rho}.
\end{equation}
After explicit calculation (we refer to Appendix~\ref{Calculation for Shear part} for details), up to order in $1/D$, the expression of $\pi_{\mu\nu}$ can be written as
\begin{align}\label{Exp:pi:mu:nu}
\pi_{\mu\nu}=&- \frac{1}{8\pi} \sigma_{\mu\nu} - \frac{1}{8\pi K} \Big( \sigma^{ \alpha}_{\langle \mu} \sigma_{|\alpha|\nu\rangle} + \omega^{ \alpha}{}_{\langle \mu} \omega_{|\alpha|\nu\rangle} +2 \sigma_{\ \langle\mu}^{\alpha} \omega_{|\alpha|\nu\rangle}  - A_{\langle \mu} A_{\nu \rangle} \Big) \notag \\
    &\quad - \frac{3}{16\pi K^3} \Big( (\nabla \cdot \sigma)^\perp_{\langle \mu} (\nabla \cdot \sigma)^\perp_{\nu \rangle} + (\nabla \cdot \omega)^\perp_{\langle \mu} (\nabla \cdot \omega)^\perp_{\nu \rangle} + 2 (\nabla \cdot \sigma)^\perp_{\langle \mu} (\nabla \cdot \omega)^\perp_{\nu \rangle} \Big) \notag \\
    &\quad + \frac{1}{8\pi K^2} (U \cdot \nabla K) \sigma_{\mu\nu}+ \frac{1}{8\pi K} \nabla^{\perp}_{\langle \mu} A_{\nu \rangle} - \frac{1}{8\pi K^2} \Big( \nabla^{\perp}_{\langle \mu} (\nabla \cdot \sigma)^\perp_{\nu \rangle} + \nabla^{\perp}_{\langle \mu} (\nabla \cdot \omega)^\perp_{\nu \rangle} \Big) \notag \\
    &\quad+ \frac{1}{4\pi K^3} \Big( \nabla^\perp_{\langle \mu} K (\nabla \cdot \sigma)^\perp_{\nu \rangle} + \nabla^\perp_{\langle \mu} K (\nabla \cdot \omega)^\perp_{\nu \rangle} \Big)\notag\\ &\quad -\frac{1}{8\pi K^2} \bigg(A_{\langle \mu}  (\nabla \cdot \sigma)^\perp_{\nu \rangle} + A_{\langle \mu}  (\nabla \cdot \omega)^\perp_{\nu \rangle}\bigg) + \mathcal{O}(1/D^2).
\end{align}
Finally, using Eq.~\eqref{rawEMtensor}~and~\eqref{tau} the stress-energy tensor $T_{\mu\nu}^{\text{(fluid)}}$ becomes
\begin{align}\label{Fluid:EM}
    T_{\mu\nu}^{\text{(fluid)}} &= \mathcal{E} U_{\mu} U_{\nu} + \bigg(\frac{s_2}{8\pi}+\tilde{\pi}\bigg) \Delta_{\mu\nu} + \pi_{\mu\nu}.
\end{align}
Note that the construction of the fluid is such that the conservation equation of $T_{\mu\nu}$ from Eq.~\eqref{EMTensor} is mathematically equivalent to the equation of motion of this fluid in a force field with force density 
\begin{align}\label{Force}
    f_\nu=\frac{1}{8\pi}\nabla_\mu(\frac{u \cdot K \cdot u}{K} K_{\alpha\beta} P^{\alpha\mu} P^{\beta}_{\nu}-\frac{1}{2} K_{\alpha\beta} P^{\alpha\mu} P^{\beta}_{\nu}).
\end{align} 
So the dynamics of this forced fluid form a mathematically isomorphic dual picture to the subleading-order dynamics of the large-D membrane paradigm. The equation of motion of the fluid is given as
\begin{align}
    \nabla_\mu  (T^{\text{(fluid)}})^{\mu\nu}=f^\nu.
\end{align}
This is a background geometric force $f_\nu$ arises due to the extrinsic curvature of the membrane hypersurface as a direct consequence of fluid's embedding. The leading order expression of the force density in Eq.~\eqref{Force} exactly matches the uncharged version of the previous leading order analysis done in Ref.~\cite{Halder:2026sdh}.
\section{Thermodynamic state and the macroscopic first law}\label{Thermodynamic State and the Macroscopic First Law}
At the first subleading order in $1/D$, the effective stress-energy tensor of the uncharged large $D$ membrane fluid is endowed with energy density $\mathcal{E}$ and pressure $p$. To establish the thermodynamic consistency of this dual fluid, we evaluate its equilibrium state variables. The global thermodynamic equilibrium of the membrane fluid is mapped with the membrane corresponding to the Schwarzschild black hole.  For a static membrane configuration (at the Schwarzschild limit)  parameterised by the horizon length scale $r_0$, the energy density and pressure are explicitly extracted from the stress-energy tensor component. At the limit of the Schwarzschild background, the expression of the stress-energy tensor becomes \begin{align}
    T_{\mu\nu}^{\text{(fluid)}}=\frac{K}{16\pi}U_\mu U_\nu-\frac{
K}{16\pi (D-2)}\Delta_{\mu\nu}.
\end{align}
Here, the expression of the Landau velocity $U^\mu$ at the equilibrium limit becomes identical with the membrane velocity $u^\mu$ ($U^\mu \to u^\mu$). Using the above expression of stress-energy tensor, the values of equilibrium energy density and pressure can be written as
\begin{align}\mathcal{E} &=\frac{K}{16\pi} =\frac{D-2}{16\pi r_0}, \label{eq:eps_val}\\ p &=-\frac{
K}{16\pi (D-2)}= -\frac{1}{16\pi r_0}. \label{eq:p_val}
\end{align}
The above expressions can be verified from  Appendix~\ref{Explicit geometric expression of thermodynamic quantities}.
Following the Bekenstein-Hawking area theorem, the equilibrium entropy density ($s$) of the dual membrane fluid is $s = 1/4$. Because the equilibrium fluid is constructed from a stationary state, it behaves as a perfectly homogeneous thermodynamic system. For any such system in absolute thermal equilibrium, the macroscopic state variables are intrinsically tied together by the Euler relation, $\mathcal{E} + p = T s$. We utilise this exact algebraic constraint to predict the local temperature $T$ of the effective fluid strictly from the geometric stress tensor data as follows:
\begin{equation}
\label{eq:euler_temp}T = \frac{\mathcal{E} + p}{s} = \frac{\frac{D-2}{16\pi r_0} - \frac{1}{16\pi r_0}}{1/4} = \frac{D-3}{4\pi r_0}=C_D\frac{K}{4\pi},
\end{equation}
with $C_D=\frac{D-3}{D-2}.$ This predicted fluid temperature exactly matches the known Hawking temperature of an uncharged large $D$ static black hole. The immediate recovery of the correct Hawking temperature directly from the subleading pressure and energy density validates the precise thermodynamic structure of this fluid mapping. We now turn to the global differential thermodynamics. The fluid is spatially localised on the $(D-1)$-dimensional membrane hypersurface; therefore, its total thermodynamic volume corresponds to the area of the geometric horizon. So the volume where the fluid lives is $V = \Omega_{D-2} r_0^{D-2}$. The total energy and total entropy variables are given by $E = \mathcal{E} V$ and $S = sV$. Multiplying the local Euler relation by $V$ yields
\begin{equation} \label{eq:global_euler}
E + p V = T S.
\end{equation}
To determine the First Law governing a quasi-static transition between neighboring equilibrium states parameterized by $r_0$ and $r_0+dr_0$, we take the exact exterior derivative of Eq. (\ref{eq:global_euler}):\begin{equation} \label{eq:diff_euler}dE + d(p V) = T dS + S dT.\end{equation}For a conventional fluid, the local Gibbs-Duhem relation ($V dp = S dT$) dictates that $d(pV) - SdT = p dV$, immediately reducing Eq. (\ref{eq:diff_euler}) to the standard macroscopic first law, $dE + p dV = T dS$.

However, in the large $D$ membrane formalism, all intensive and extensive variables are parameterised by the single length scale $r_0$. The large-$D$ scaling strictly mandates $V \propto r_0^{D-2}$ alongside $P \propto r_0^{-1}$ and $T \propto r_0^{-1}$. Evaluating the total differentials explicitly yields \begin{align}d(p V) &= d \left( - \frac{\Omega_{D-2}}{16\pi} r_0^{D-3} \right) = - \frac{D-3}{16\pi} \Omega_{D-2} r_0^{D-4} dr_0, \\
S dT &= \left( \frac{\Omega_{D-2} r_0^{D-2}}{4} \right) \left( - \frac{D-3}{4\pi r_0^2} dr_0 \right) = - \frac{D-3}{16\pi} \Omega_{D-2} r_0^{D-4} dr_0.\end{align}This reveals an absolute geometric identity $d(pV) \equiv S dT$. The Gibbs-Duhem relation is structurally violated, which is also observed in the leading-order charged membrane fluid~\cite{Halder:2026sdh}. Substituting this identity back into Eq. (\ref{eq:diff_euler}), the $d(pV)$ and $SdT$ terms identically annihilate one another, yielding the exact global first law for the membrane fluid:\begin{equation}\label{1stlaw}
dE = T dS.\end{equation}
It is important to note that our fluid is characterised by an effective negative pressure (Eq.~\eqref{eq:p_val}). Negative pressure in a bulk fluid implies an inward pull from all spatial directions. However, because the membrane fluid is confined to a $(D-1)$-dimensional hypersurface, its negative thermodynamic pressure functions purely as an intrinsic surface tension that resists the geometric expansion of the membrane.

To formally distinguish the membrane fluid's first law from that of a conventional fluid, we must examine the specific functional dependence. In classical thermodynamics, the identity $dE = T dS - p dV$ rests on the assumption that entropy and volume are independent degrees of freedom. Writing $E = E(S, V)$ yields the exact differential
$$dE = \left(\frac{\partial E}{\partial S}\right)_V dS + \left(\frac{\partial E}{\partial V}\right)_S dV$$
Defining the conjugate variables $T \equiv (\partial E / \partial S)_V$ and $p \equiv -(\partial E / \partial V)_S$ separates the thermal energy ($T dS$) and internal mechanical work ($-p dV$). This derivation explicitly requires the ability to mathematically vary volume while holding entropy constant.

For the effective membrane fluid, this independence collapses. Because $S = V/4$, the internal energy is a function of only one extensive variable, $E = E(V)$ (or $E = E(S)$). In this degenerate state space, partial derivatives such as $(\partial E / \partial V)_S$ are mathematically undefined, rendering the classical separation of $dE$ into distinct $T dS$ and $-p dV$ terms invalid, leading to the expression of first law in Eq.~\eqref{1stlaw}.

In the above equilibrium analysis, the following expressions hold true
\begin{align}
    \mathcal{E}(T)&=\frac{T}{4C_D},\label{Exp:mathcal{E}(T)}\\p(T)&=-\frac{\mathcal{E}(T)}{(D-2)}.\label{Exp:p(T)}
\end{align}
\section{Local thermodynamic equilibrium}\label{Local thermodynamic equilibrium}
Perturbing a fluid from global equilibrium induces gradients in thermodynamic parameters such as temperature and pressure; these gradients subsequently drive non-trivial fluid dynamics. To formulate the hydrodynamic description, we parameterise the non-equilibrium state using local thermodynamic fields. However, outside of global equilibrium, the exact definitions of local parameters such as temperature, pressure, and energy density are inherently ambiguous. Resolving this ambiguity requires the choice of a specific hydrodynamic frame.\footnote{Although local thermodynamic variables are intrinsically frame-dependent, the full stress-energy tensor remains unambiguously defined across all frames.} Nevertheless, all valid frame choices must seamlessly reduce to the exact thermodynamic expressions in the global equilibrium limit, consistent with the approach in Ref.~\cite{Kovtun:2019hdm}. An analogous temperature redefinition was utilized in Ref.~\cite{Halder:2026sdh} to extract the leading-order transport coefficients of the charged membrane fluid. Accordingly, we define the modified local temperature $T'$ as
\begin{align}\label{Temp:Redefinition}
    T'(x)=T(x)+\delta T=C_D \frac{K(x)}{4\pi}+\delta T.
\end{align}
The redefinition is such that the out-of-equilibrium energy density $\mathcal{E}$ can be expressed as a function of temperature $T'$, in the following way:
\begin{align}\label{E1}
    \mathcal{E}(T')&=\frac{T'(x)}{4C_D}.
\end{align}
The geometric expression of out-of-equilibrium energy density is given as (we refer to Appendix~\ref{Explicit geometric expression of thermodynamic quantities} for details):
\begin{align}\label{E2}
    \mathcal{E}(T') &= \frac{K}{16\pi} + \frac{1}{8\pi}\left( \frac{\nabla^2 K}{2K^2} - \frac{u \cdot \nabla K}{K} + 2 \frac{u_\alpha \nabla^2 u^\alpha}{K} + u \cdot K \cdot u \right)\nonumber
    \\&\quad + \frac{1}{8\pi K} \bigg[ -3u\cdot K\cdot K\cdot u - 13\left(\frac{u\cdot\nabla K}{K}\right)^{2} + 3u^{\alpha}K_{\alpha\beta}\left(\frac{\nabla^{\beta}K}{K}\right) + 14\left(\frac{u\cdot\nabla K}{K}\right)(u\cdot K\cdot u) \nonumber \\
&\quad - \frac{K}{D}\left(\frac{u\cdot\nabla K}{K}\right) + \frac{K}{2D}(u\cdot K\cdot u) + \frac{1}{K^{3}}\nabla^{2}(\nabla^{2}K) - \frac{7}{2}(u\cdot K\cdot u)^{2}- 2\left(\frac{\nabla_{\alpha}K}{K}\right)\left(\frac{\nabla^{\alpha}K}{K}\right)  \nonumber \\
&\quad + (2\zeta(3)-1) \bigg( -\frac{K}{D}\left(\frac{u\cdot\nabla K}{K} - u\cdot K\cdot u\right) - u\cdot K\cdot K\cdot u + 2\left(\frac{\nabla_{\alpha}K}{K}\right)u^{\beta}K_{\beta}^{\alpha} - \left(\frac{u\cdot\nabla K}{K}\right)^{2} \nonumber \\
&\quad + 2\left(\frac{u\cdot\nabla K}{K}\right)(u\cdot K\cdot u) - \left(\frac{\nabla^{\alpha}K}{K}\right)\left(\frac{\nabla_{\alpha}K}{K}\right) - (u\cdot K\cdot u)^{2}\bigg)+  \frac{2u^\mu}{K^{2}}\nabla^{2}(\nabla^{2}u_{\mu}) \nonumber \\
&\quad - 2u^\mu\nabla_{\mu}\left(\frac{u\cdot\nabla K}{K}\right) - 2u^\mu\left(\frac{\nabla^{2}u_{\mu}}{K}\right)\left(-2(u\cdot K\cdot u) + 4\frac{u\cdot\nabla K}{K}  - \frac{K}{D}\right)+u^\mu u^\nu \nabla_{\mu}\left(\frac{\nabla^{2}u_{\nu}}{K}\right)\nonumber\\&\quad   +u^\mu u^\nu  \nabla_{\mu}(u^{\alpha}K_{\alpha\nu})  - u^\mu u^\nu \nabla_{\mu}\left(\frac{\nabla_{\nu}K}{K}\right)-u^\mu u^\nu \left(\frac{\nabla^{2}u_{\mu}}{K}\right)\left(\frac{\nabla^{2}u_{\nu}}{K}\right)- K^{\alpha\beta}(\nabla_{\alpha}u_{\beta})\notag\\
&\quad - \frac{2}{K}u_{\alpha}K^{\alpha\beta}\nabla^{2}u_{\beta}+ \left( \frac{\nabla_\mu K \nabla^2 u_\nu}{K^2} - \frac{\nabla_\mu K \nabla_\nu K}{2K^2} - \frac{\nabla^2 u_\mu \nabla^2 u_\nu}{2K^2} \right) P^{\mu\nu}\bigg]+ \mathcal{O}(1/D^2).
\end{align}
Using Eqs.~\eqref{Temp:Redefinition},~\eqref{E1},~and~\eqref{E2} the value of $\delta T$ can be evaluated as follows:
\begin{align}\label{delta:T}
    \delta T &=\frac{4C_D}{8\pi}\left( \frac{\nabla^2 K}{2K^2} - \frac{u \cdot \nabla K}{K} + 2 \frac{u_\alpha \nabla^2 u^\alpha}{K} + u \cdot K \cdot u \right)\nonumber
    \\&\quad + \frac{C_D}{8\pi K} \bigg[ -3u\cdot K\cdot K\cdot u - 13\left(\frac{u\cdot\nabla K}{K}\right)^{2} + 3u^{\alpha}K_{\alpha\beta}\left(\frac{\nabla^{\beta}K}{K}\right) + 14\left(\frac{u\cdot\nabla K}{K}\right)(u\cdot K\cdot u) \nonumber \\
&\quad - \frac{K}{D}\left(\frac{u\cdot\nabla K}{K}\right) + \frac{K}{2D}(u\cdot K\cdot u) + \frac{1}{K^{3}}\nabla^{2}(\nabla^{2}K) - \frac{7}{2}(u\cdot K\cdot u)^{2}- 2\left(\frac{\nabla_{\alpha}K}{K}\right)\left(\frac{\nabla^{\alpha}K}{K}\right)  \nonumber \\
&\quad + (2\zeta(3)-1) \bigg( -\frac{K}{D}\left(\frac{u\cdot\nabla K}{K} - u\cdot K\cdot u\right) - u\cdot K\cdot K\cdot u + 2\left(\frac{\nabla_{\alpha}K}{K}\right)u^{\beta}K_{\beta}^{\alpha} - \left(\frac{u\cdot\nabla K}{K}\right)^{2} \nonumber \\
&\quad + 2\left(\frac{u\cdot\nabla K}{K}\right)(u\cdot K\cdot u) - \left(\frac{\nabla^{\alpha}K}{K}\right)\left(\frac{\nabla_{\alpha}K}{K}\right) - (u\cdot K\cdot u)^{2}\bigg)+  \frac{2u^\mu}{K^{2}}\nabla^{2}(\nabla^{2}u_{\mu}) \nonumber \\
&\quad - 2u^\mu\nabla_{\mu}\left(\frac{u\cdot\nabla K}{K}\right) - 2u^\mu\left(\frac{\nabla^{2}u_{\mu}}{K}\right)\left(-2(u\cdot K\cdot u) + 4\frac{u\cdot\nabla K}{K}  - \frac{K}{D}\right)+u^\mu u^\nu \nabla_{\mu}\left(\frac{\nabla^{2}u_{\nu}}{K}\right)\nonumber\\&\quad   +u^\mu u^\nu  \nabla_{\mu}(u^{\alpha}K_{\alpha\nu})  - u^\mu u^\nu \nabla_{\mu}\left(\frac{\nabla_{\nu}K}{K}\right)-u^\mu u^\nu \left(\frac{\nabla^{2}u_{\mu}}{K}\right)\left(\frac{\nabla^{2}u_{\nu}}{K}\right)- K^{\alpha\beta}(\nabla_{\alpha}u_{\beta})\notag\\
&\quad - \frac{2}{K}u_{\alpha}K^{\alpha\beta}\nabla^{2}u_{\beta}+ \left( \frac{\nabla_\mu K \nabla^2 u_\nu}{K^2} - \frac{\nabla_\mu K \nabla_\nu K}{2K^2} - \frac{\nabla^2 u_\mu \nabla^2 u_\nu}{2K^2} \right) P^{\mu\nu}\bigg]+ \mathcal{O}(1/D^2).
\end{align}
By enforcing the Landau matching condition, we get the subleading order out-of-equilibrium correction of temperature such that the energy density $\mathcal{E}(T')$ mimics its exact equilibrium functional form. Consequently, the out-of-equilibrium pressure is constrained to preserve the equilibrium equation of state, yielding
\begin{align}
    p(T')&=-\frac{\mathcal{E}(T')}{(D-2)},\notag\\
    &=-\frac{K}{16\pi(D-2)} + \frac{1}{8\pi(D-2)}\left( \frac{\nabla^2 K}{2K^2} - \frac{u \cdot \nabla K}{K} + 2 \frac{u_\alpha \nabla^2 u^\alpha}{K} + u \cdot K \cdot u \right)\notag\\&\qquad+\mathcal{O}(1/D^2).
\end{align}
By absorbing the gradient corrections into the redefined temperature $T'$, the Landau matching condition preserves the expression of the equation of state. This isolates the ideal fluid background, ensuring that all remaining terms in the stress-energy tensor correspond to purely dissipative fluxes. This unambiguous separation is required to extract the dissipative transport coefficients in the subsequent sections.
\section{Out of equilibrium viscous effects}\label{Out of equilibrium viscous effects}
In this section, we use out-of-equilibrium temperature ($T'$), energy density ($\mathcal{E}(T')$), and pressure ($p(T')$) to extract the expression of dissipative quantities.
\subsection{Bulk viscous pressure}\label{Bulk viscosity}
In Eq.~\eqref{Fluid:EM}, the scalar coefficient of the Landau projection operator $\Delta_{\mu\nu}$ yields the additive contribution of the thermodynamic pressure and the bulk pressure of the fluid, implying
\begin{align*}
   p+ \Pi&=\frac{s_2}{8\pi}+\tilde{\pi},\\ \Pi&=\frac{s_2}{8\pi}+\tilde{\pi}+\frac{\mathcal{E}(T')}{(D-2)}.
\end{align*}
By substituting the geometric expression in the above equation, the expression of bulk pressure can be written in terms of geometric quantities up to $\mathcal{O}(1/D)$ as
\begin{align}\label{pigeo}
\Pi &= -\frac{1}{16\pi} (u \cdot k \cdot u) + \frac{1}{8\pi K} \left[ K^{\alpha \beta} \nabla_{\alpha} u_{\beta} - u_{\alpha} K^{\alpha \beta} \left( \frac{\nabla_{\beta} K}{K} - \frac{2 \nabla^2 u_{\beta}}{K} \right) \right.\notag \\
&\quad \left. - \left( \frac{u \cdot \nabla K}{K} - \frac{u \cdot k \cdot u}{2} - \frac{K}{2D} \right) u \cdot k \cdot u \right] \notag\\
&\quad - \frac{1}{8\pi (D-2)} \left( \frac{\nabla^2 K}{2 K^2} - \frac{u \cdot \nabla K}{K} + 2 \frac{u_{\alpha} \nabla^2 u^{\alpha}}{K} + u \cdot k \cdot u \right) + \widetilde{\pi} + \mathcal{O}(1/D^2).
\end{align}
After explicit algebraic substitution (we refer to Appendix~\ref{Bulk_Part} for detailed calculation),  the expression of $\Pi$ can be written in terms of the Landau velocity derivatives (Eq.~\eqref{pi:final}), as follows:
\begin{align}\label{Exp:pi}
\Pi &= - \frac{1}{16\pi K} \nabla \cdot A - \frac{\dot{K}}{8\pi K} + \frac{1}{8\pi K^2} \nabla_{\alpha} \nabla_{\gamma} \sigma^{\alpha \gamma} \notag\\
&\quad + \frac{1}{16\pi K} (\sigma_{\mu\nu}\sigma^{\mu\nu} - \omega_{\mu\nu}\omega^{\mu\nu}) + \frac{1}{8\pi (D-2)} \frac{\dot{K}}{K} + \frac{(\nabla \cdot A)^2}{8\pi K^3} \nonumber \\
&\quad + \frac{\dot{K}}{8\pi K^4} \nabla_{\alpha} \nabla_{\lambda} \sigma^{\lambda\alpha} - \frac{\theta}{8\pi K^2} \nabla \cdot A - \frac{1}{8\pi K^2} U^{\alpha} \nabla_{\alpha} (\nabla \cdot A) \nonumber \\
&\quad - \frac{1}{16\pi(D-2)} \frac{\nabla^2 K}{K^2} - \frac{\nabla^2 K}{4\pi K^5} \nabla_{\alpha} \nabla_{\lambda} \sigma^{\lambda\alpha} - \frac{\theta}{8\pi(D-2)} - \frac{1}{8\pi K^2 (D-2)} \nabla_{\alpha} \nabla_{\lambda} \sigma^{\lambda\alpha} \nonumber \\
&\quad + \frac{1}{16\pi K^3} \Bigl[ 3(\nabla^{\lambda} \sigma_{\lambda}^{\;\;\alpha})(\nabla^{\gamma} \sigma_{\gamma\alpha}) + 2 (\nabla^{\lambda} \sigma_{\lambda}^{\;\;\alpha})(\nabla^{\gamma} \omega_{\gamma\alpha}) - 5 (\nabla^{\lambda} \omega_{\lambda}^{\;\;\alpha})(\nabla^{\gamma} \omega_{\gamma\alpha}) \Bigr] \nonumber \\
&\quad - \frac{1}{8\pi K^3} (\nabla^{\lambda} \sigma_{\lambda}^{\;\;\alpha}-\nabla^{\lambda} \omega_{\lambda}^{\;\;\alpha}) \nabla_{\alpha} K - \frac{1}{8\pi K^2} A^{\gamma} (\nabla^{\lambda} \sigma_{\lambda\gamma} - \nabla^{\lambda} \omega_{\lambda\gamma}) \nonumber \\
&\quad - \frac{U^\beta}{16\pi K^3} \nabla_\beta (\nabla_{\alpha} \nabla_{\lambda} \sigma^{\lambda\alpha}) + \frac{1}{8\pi K^4} \nabla^2 (\nabla_{\alpha} \nabla_{\lambda} \sigma^{\lambda\alpha}) + \mathcal{O}(1/D^2).
\end{align}
Now, we will replace the geometric quantity $K$ with the fluid's variable. The first three terms in the above expression are of leading order, $\mathcal{O}(1)$, and the remaining terms are $\mathcal{O}(1/D)$. The quantity $K$ can be substituted with its leading-order equivalent, $K = \frac{4\pi T'}{C_D}$ (prominent from Eq.~\eqref{Temp:Redefinition}), strictly within the subleading $\mathcal{O}(1/D)$ terms. Any corrections to this substitution within the subleading sector are trivially relegated to $\mathcal{O}(1/D^2)$. Conversely, substituting $K$ into the leading $\mathcal{O}(1)$ terms demands exact algebraic tracking to accurately capture all residual $\mathcal{O}(1/D)$ contributions. To evaluate these leading-order terms, $K$ must be replaced by the exact relation $K = \frac{4\pi}{C_D}(T' - \delta T)$, incorporating the $\mathcal{O}(1)$ components of $\delta T$ defined in Eq.~\eqref{delta:T}. We address the leading-order expansion as follows:
\begin{align}
    &-\frac{1}{16\pi K} \nabla \cdot A - \frac{1}{8\pi K} \dot{K} + \frac{1}{8\pi K^2 } \nabla_\alpha \nabla_\lambda \sigma^{\lambda\alpha} \nonumber \\ 
    &= -\frac{C_D}{64\pi^2 T'}\left(1 + \frac{\delta T}{T'}\right) \nabla \cdot A - \frac{C_D}{32\pi^2 T'}\left(1 + \frac{\delta T}{T'}\right) U^\alpha \nabla_\alpha \left( \frac{4\pi}{C_D} (T' - \delta T) \right) \nonumber \\
    &\quad + \frac{C_D^2}{128\pi^3 T'^2} \left(1 + \frac{2\delta T}{T'} \right) \nabla_\alpha \nabla_\lambda \sigma^{\lambda\alpha} + \mathcal{O}(1/D^2) \nonumber \\
    &= -\frac{C_D}{64\pi^2 T'} \nabla \cdot A - \frac{\dot{T}'}{8\pi T'} + \frac{C_D^2}{128\pi^3 T'^2} \nabla_\alpha \nabla_\lambda \sigma^{\lambda\alpha}  - \frac{C_D}{64\pi^2 T'^2} \delta T \nabla \cdot A \nonumber \\
    &\quad- \frac{\delta T}{8\pi T'^2} \dot{T}' + \frac{1}{8\pi T'} U^\alpha \nabla_\alpha \delta T  + \frac{C_D^2}{64\pi^3 T'^3} \delta T \nabla_\alpha \nabla_\lambda \sigma^{\lambda\alpha} + \mathcal{O}(1/D^2). \label{eq:expanded_K_full} 
    \end{align}
From the exact geometric definition, the subleading order temperature perturbation have the following form:
\begin{align}
    \delta T &= \frac{C_D}{2\pi} \left( \frac{\nabla^2 K}{2K^2} - \frac{u \cdot \nabla K}{K} + 2 \frac{u_\alpha \nabla^2 u^\alpha}{K} + u \cdot K \cdot u \right). \label{eq:delta_T_geometric_full}
    \end{align}
Substituting the leading-order expressions for the fluid velocity ($u_\mu \to U_\mu$) and temperature ($K \to 4\pi T'/C_D$) in the expression of $\delta T$, and explicitly dropping the dynamically suppressed $\mathcal{O}(1/D)$ contribution of $\frac{u_\alpha \nabla^2 u^\alpha}{K}$, yields the exact corrected fluid-frame perturbation
\begin{align}
    \delta T &= \frac{C_D}{2\pi} \left( \frac{C_D}{8\pi} \frac{\nabla^2 T'}{T'^2} - \frac{\dot{T}'}{T'} + \frac{C_D}{4\pi} \frac{\nabla \cdot A}{T'} \right) + \mathcal{O}(1/D). \label{eq:delta_T_fluid_full}
\end{align}
To evaluate Eq.~\eqref{eq:expanded_K_full}, we must explicitly compute the comoving derivative of the perturbation, $U^\alpha \nabla_\alpha \delta T$, as
\begin{align}
    U^\alpha \nabla_\alpha \delta T &= \frac{C_D^2}{16\pi^2} \left[ \frac{U^\alpha \nabla_\alpha (\nabla^2 T')}{T'^2} - \frac{2 \dot{T}' \nabla^2 T'}{T'^3} \right] - \frac{C_D}{2\pi} \left[ \frac{\ddot{T}'}{T'} - \frac{\dot{T}'^2}{T'^2} \right] \nonumber \\
    &\quad + \frac{C_D^2}{8\pi^2} \left[ \frac{U^\alpha \nabla_\alpha (\nabla \cdot A)}{T'} - \frac{\dot{T}' \nabla \cdot A}{T'^2} \right] + \mathcal{O}(1/D). \label{eq:delta_T_derivative}
    \end{align}
Substituting both $\delta T$ and $U^\alpha \nabla_\alpha \delta T$ into Eq.~\eqref{eq:expanded_K_full} and expanding all products into independent tensorial structures produces the following form:
\begin{align}
    &-\frac{1}{16\pi K} \nabla \cdot A - \frac{1}{8\pi K} \dot{K} + \frac{1}{8\pi K^2} \nabla_\alpha \nabla_\lambda \sigma^{\lambda\alpha} \nonumber \\
    &= -\frac{C_D}{64\pi^2 T'} \nabla \cdot A - \frac{\dot{T}'}{8\pi T'} + \frac{C_D^2}{128\pi^3 T'^2} \nabla_\alpha \nabla_\lambda \sigma^{\lambda\alpha} \nonumber \\
    &\quad - \frac{C_D}{16\pi^2 T'^2} \ddot{T}' + \frac{C_D}{8\pi^2 T'^3} \dot{T}'^2 + \frac{C_D^2}{128\pi^3 T'^3} U^\alpha \nabla_\alpha (\nabla^2 T') + \frac{C_D^2}{64\pi^3 T'^2} U^\alpha \nabla_\alpha (\nabla \cdot A) \nonumber \\
    &\quad - \frac{C_D^3}{512\pi^4 T'^3} (\nabla \cdot A)^2  - \frac{3 C_D^2}{128\pi^3 T'^4} \dot{T}' (\nabla^2 T') - \frac{3 C_D^2}{128\pi^3 T'^3} \dot{T}' (\nabla \cdot A) \nonumber \\
    &\quad - \frac{C_D^3}{1024\pi^4 T'^4} (\nabla^2 T')(\nabla \cdot A)  + \frac{C_D^4}{1024\pi^5 T'^5} (\nabla^2 T') \nabla_\alpha \nabla_\lambda \sigma^{\lambda\alpha} \nonumber \\
    &\quad  - \frac{C_D^3}{128\pi^4 T'^4} \dot{T}' \nabla_\alpha \nabla_\lambda \sigma^{\lambda\alpha}+ \frac{C_D^4}{512\pi^5 T'^4} (\nabla \cdot A) \nabla_\alpha \nabla_\lambda \sigma^{\lambda\alpha} + \mathcal{O}(1/D^2). \label{eq:final_evaluated_complete}
\end{align}
Substituting all the $K$ in Eq.~\eqref{Exp:pi}, in terms of the fluid's thermodynamic and kinematic variables, yields the following form:
\begin{align}
\Pi &= -\frac{C_D}{64\pi^2 T'} \nabla \cdot A - \frac{\dot{T}'}{8\pi T'} + \frac{C_D^2}{128\pi^3 T'^2} \nabla_\alpha \nabla_\lambda \sigma^{\lambda\alpha} \nonumber \\
&\quad - \frac{C_D}{16\pi^2 T'^2} \ddot{T}' + \frac{C_D}{8\pi^2 T'^3} \dot{T}'^2 + \frac{C_D^2}{128\pi^3 T'^3} U^\alpha \nabla_\alpha (\nabla^2 T')  + \frac{C_D^2}{128\pi^3 T'^2} U^\alpha \nabla_\alpha (\nabla \cdot A)\nonumber \\
&\quad - \frac{3 C_D^2}{128\pi^3 T'^4} \dot{T}' (\nabla^2 T') - \frac{3 C_D^2}{128\pi^3 T'^3} \dot{T}' (\nabla \cdot A) - \frac{C_D^3}{1024\pi^4 T'^4} (\nabla^2 T')(\nabla \cdot A) \nonumber \\
&\quad - \frac{3 C_D^3}{512\pi^4 T'^4} \dot{T}' \nabla_\alpha \nabla_\lambda \sigma^{\lambda\alpha} + \frac{C_D^4}{512\pi^5 T'^4} (\nabla \cdot A) \nabla_\alpha \nabla_\lambda \sigma^{\lambda\alpha} \nonumber \\
&\quad + \frac{C_D}{64\pi^2 T'} (\sigma_{\mu\nu}\sigma^{\mu\nu} - \omega_{\mu\nu}\omega^{\mu\nu}) + \frac{1}{8\pi (D-2)} \frac{\dot{T}'}{T'}- \frac{C_D \nabla^2 T'}{64\pi^2 (D-2) T'^2} \nonumber \\
&\quad  - \frac{\theta}{8\pi(D-2)} - \frac{C_D^2}{128\pi^3 T'^2 (D-2)} \nabla_\alpha \nabla_\lambda \sigma^{\lambda\alpha}- \frac{C_D^2}{128\pi^3 T'^2} \theta (\nabla \cdot A) \nonumber \\
&\quad + \frac{C_D^3}{1024\pi^4 T'^3} \Bigl[ 3(\nabla^{\lambda} \sigma_{\lambda}^{\;\;\alpha})(\nabla^{\gamma} \sigma_{\gamma\alpha}) + 2 (\nabla^{\lambda} \sigma_{\lambda}^{\;\;\alpha})(\nabla^{\gamma} \omega_{\gamma\alpha}) - 5 (\nabla^{\lambda} \omega_{\lambda}^{\;\;\alpha})(\nabla^{\gamma} \omega_{\gamma\alpha}) \Bigr] \nonumber \\
&\quad - \frac{C_D^2}{128\pi^3 T'^3} (\nabla^{\lambda} \sigma_{\lambda}^{\;\;\alpha}-\nabla^{\lambda} \omega_{\lambda}^{\;\;\alpha}) \nabla_{\alpha} T' - \frac{C_D^2}{128\pi^3 T'^2} A^{\gamma} (\nabla^{\lambda} \sigma_{\lambda\gamma} - \nabla^{\lambda} \omega_{\lambda\gamma}) \nonumber \\
&\quad - \frac{C_D^3}{1024\pi^4 T'^3} U^\beta \nabla_\beta (\nabla_{\alpha} \nabla_{\lambda} \sigma^{\lambda\alpha}) + \frac{C_D^4}{2048\pi^5 T'^4} \nabla^2 (\nabla_{\alpha} \nabla_{\lambda} \sigma^{\lambda\alpha}) + \mathcal{O}(1/D^2).
\end{align}
Now, in the large $D$ limit, a rigorous basis reduction is required within the bulk pressure expression. In the leading-order expression for the fluid expansion scalar ($\theta = \nabla_\mu U^\mu$),  the leading-order term of $U^\mu$ (which is $u^\mu$, as, $U^\mu = u^\mu + l^\mu/e + \mathcal{O}(1/D^2)$) does not contribute, because the scalar membrane equation enforces $\nabla_\mu u^\mu = \mathcal{O}(1/D)$. The leading $\theta$ contribution purely comes from the sub-leading order term of $U^\mu$ (which is $l^\mu/e$), which causes a basis degeneracy. To eliminate basis degeneracy, $\theta$ must be systematically replaced by its equivalent geometric structure. Utilising $e = K/2 + \mathcal{O}(1)$ and standard large $D$ scaling, the exact contribution of $\theta$ is derived as follows:
\begin{align*}     \theta = \nabla_\mu U^\mu      &= \nabla_\mu u^\mu + \nabla_\mu \left( \frac{l^\mu}{e} \right) + \mathcal{O}(1/D) \\     &= \nabla_\mu \left( \frac{l^\mu}{e} \right) + \mathcal{O}(1/D) \\     &= \frac{2}{K} \nabla_\mu l^\mu + \mathcal{O}(1/D) \\    &= \frac{2}{K} \nabla_\mu \left( - \frac{\nabla^2 u^\sigma}{2K} P^\mu_\sigma \right) + \mathcal{O}(1/D) \\     &= - \frac{1}{K} \nabla_\sigma \left( \frac{\nabla^2 U^\sigma}{K} \right) + \mathcal{O}(1/D) \\     &= -\frac{\nabla_\sigma \nabla^2 U^\sigma}{K^2} + \mathcal{O}(1/D) \\     &= -\frac{\nabla_a \nabla_b \sigma^{ab}}{K^2} + \mathcal{O}(1/D). 
\end{align*}
In the above calculation, we have substituted the leading order expression of $l^\mu$ from Appendix~\ref{Explicit geometric expression of thermodynamic quantities} and used the leading order vector membrane equation (Eq.~\eqref{eq:membrane:vector}). The required leading-order substitution to eliminate the basis redundancy is $\theta \to -\frac{\nabla_a \nabla_b \sigma^{ab}}{K^2}$.
Implementing this the modified expression of bulk pressure becomes
\begin{align}\label{PI}
\Pi &= -\frac{C_D}{64\pi^2 T'} \nabla \cdot A - \frac{\dot{T}'}{8\pi T'} + \frac{C_D^2}{128\pi^3 T'^2} \nabla_\alpha \nabla_\lambda \sigma^{\lambda\alpha} \nonumber \\
&\quad - \frac{C_D}{16\pi^2 T'^2} \ddot{T}' + \frac{C_D}{8\pi^2 T'^3} \dot{T}'^2 + \frac{C_D^2}{128\pi^3 T'^3} U^\alpha \nabla_\alpha (\nabla^2 T') + \frac{C_D^2}{128\pi^3 T'^2} U^\alpha \nabla_\alpha (\nabla \cdot A) \nonumber \\
&\quad - \frac{3 C_D^2}{128\pi^3 T'^4} \dot{T}' (\nabla^2 T') - \frac{3 C_D^2}{128\pi^3 T'^3} \dot{T}' (\nabla \cdot A) - \frac{C_D^3}{1024\pi^4 T'^4} (\nabla^2 T')(\nabla \cdot A) \nonumber \\
&\quad - \frac{3 C_D^3}{512\pi^4 T'^4} \dot{T}' \nabla_\alpha \nabla_\lambda \sigma^{\lambda\alpha} + \frac{5 C_D^4}{2048\pi^5 T'^4} (\nabla \cdot A) \nabla_\alpha \nabla_\lambda \sigma^{\lambda\alpha} \nonumber \\
&\quad + \frac{C_D}{64\pi^2 T'} (\sigma_{\mu\nu}\sigma^{\mu\nu} - \omega_{\mu\nu}\omega^{\mu\nu}) + \frac{1}{8\pi (D-2)} \frac{\dot{T}'}{T'} \nonumber - \frac{C_D \nabla^2 T'}{64\pi^2 (D-2) T'^2} \nonumber \\
&\quad + \frac{C_D^3}{1024\pi^4 T'^3} \Bigl[ 3(\nabla^{\lambda} \sigma_{\lambda}^{\;\;\alpha})(\nabla^{\gamma} \sigma_{\gamma\alpha}) + 2 (\nabla^{\lambda} \sigma_{\lambda}^{\;\;\alpha})(\nabla^{\gamma} \omega_{\gamma\alpha}) - 5 (\nabla^{\lambda} \omega_{\lambda}^{\;\;\alpha})(\nabla^{\gamma} \omega_{\gamma\alpha}) \Bigr] \nonumber \\
&\quad - \frac{C_D^2}{128\pi^3 T'^3} (\nabla^{\lambda} \sigma_{\lambda}^{\;\;\alpha}-\nabla^{\lambda} \omega_{\lambda}^{\;\;\alpha}) \nabla_{\alpha} T' - \frac{C_D^2}{128\pi^3 T'^2} A^{\gamma} (\nabla^{\lambda} \sigma_{\lambda\gamma} - \nabla^{\lambda} \omega_{\lambda\gamma}) \nonumber \\
&\quad - \frac{C_D^3}{1024\pi^4 T'^3} U^\beta \nabla_\beta (\nabla_{\alpha} \nabla_{\lambda} \sigma^{\lambda\alpha}) + \frac{C_D^4}{2048\pi^5 T'^4} \nabla^2 (\nabla_{\alpha} \nabla_{\lambda} \sigma^{\lambda\alpha}) + \mathcal{O}(1/D^2).
\end{align}
This is the full expression for the bulk pressure up to order $\mathcal{O}(1/D)$; however, to obtain the first subleading-order membrane equation, not all terms contribute. As $\Pi$ is multiplied by the Landau projector $\Delta_{\mu \nu}$ in the expression of fluid stress-energy tensor $T_{\mu\nu}^{\text{fluid}}$, when the divergence operator acts on it, only the leading order $\mathcal{O}(1)$ term of the bulk pressure $\Pi$ will contribute to the first subleading order membrane equation (since $\nabla_\mu \Delta^{\mu\nu}\sim \mathcal{O}(1)$). For this reason, only the first three terms of Eq.~\eqref{PI} will effectively contribute to the bulk pressure for the first subleading order fluid analysis. The rest of the terms will dictate the expression of bulk pressure in the second subleading order analysis. 
\subsection{Shear stress tensor}\label{Shear viscosity}
The shear stress tensor $\pi_{\mu\nu}$ evaluated in Eq.~\eqref{Exp:pi:mu:nu} can now be written entirely in terms of the effective fluid variables. The leading-order contribution of $\pi_{\mu\nu}$ (which is $-\frac{1}{8\pi}\sigma_{\mu\nu}$) is independent of $K$. In the subleading $\mathcal{O}(1/D)$ terms $K = \frac{4\pi T'}{C_D}$, can be substituted directly, as the remaining term in the expression of $K$ (which contains the contribution of $\delta T$ ) is suppressed by $\mathcal{O}(1/D^2)$ in the expression of $\pi_{\mu\nu}$. By substituting this the complete fluid-frame shear stress tensor in terms of independent tensorial structures can be written as
\begin{align}\label{eq:pi_substituted}
\pi_{\mu\nu} &= - \frac{1}{8\pi} \sigma_{\mu\nu}  - \frac{C_D}{32\pi^2 T'} \Big( \sigma^{ \alpha}_{\langle \mu} \sigma_{|\alpha|\nu\rangle} + \omega^{ \alpha}{}_{\langle \mu} \omega_{|\alpha|\nu\rangle} + 2 \sigma_{\ \langle\mu}^{\alpha} \omega_{|\alpha|\nu\rangle} - A_{\langle \mu} A_{\nu \rangle} \Big) \notag \\
    &\quad - \frac{3 C_D^3}{1024\pi^4 T'^3} \Big( (\nabla \cdot \sigma)^\perp_{\langle \mu} (\nabla \cdot \sigma)^\perp_{\nu \rangle} + (\nabla \cdot \omega)^\perp_{\langle \mu} (\nabla \cdot \omega)^\perp_{\nu \rangle} \Big) \notag \\
    &\quad - \frac{3 C_D^3}{512\pi^4 T'^3} (\nabla \cdot \sigma)^\perp_{\langle \mu} (\nabla \cdot \omega)^\perp_{\nu \rangle} + \frac{C_D}{32\pi^2 T'^2} \dot{T}' \sigma_{\mu\nu} + \frac{C_D}{32\pi^2 T'} \nabla^{\perp}_{\langle \mu} A_{\nu \rangle} \notag \\
    &\quad - \frac{C_D^2}{128\pi^3 T'^2} \Big( \nabla^{\perp}_{\langle \mu} (\nabla \cdot \sigma)^\perp_{\nu \rangle} + \nabla^{\perp}_{\langle \mu} (\nabla \cdot \omega)^\perp_{\nu \rangle} \Big) \notag \\
    &\quad + \frac{C_D^2}{64\pi^3 T'^3} \Big( \nabla^\perp_{\langle \mu} T' (\nabla \cdot \sigma)^\perp_{\nu \rangle} + \nabla^\perp_{\langle \mu} T' (\nabla \cdot \omega)^\perp_{\nu \rangle} \Big) \notag\\ &\quad -\frac{C_D^2}{128\pi^3 T'^2} \bigg(A_{\langle \mu}  (\nabla \cdot \sigma)^\perp_{\nu \rangle} + A_{\langle \mu}  (\nabla \cdot \omega)^\perp_{\nu \rangle}\bigg) + \mathcal{O}(1/D^2).
\end{align}
\section{Equation of motion}\label{Equation of motion}
The stress-energy tensor of the fluid described in Eq.~\eqref{Fluid:EM} can now be written in terms of $p$ and $\Pi$ as
\begin{align}\label{EM:Fluid:final}
     T_{\mu\nu}^{\text{(fluid)}} &= \mathcal{E} U_{\mu} U_{\nu} + \bigg(p+\Pi\bigg) \Delta_{\mu\nu} + \pi_{\mu\nu}. 
\end{align}
The conservation equation in the presence of an external force density $f^\mu$ is given by
\begin{equation}
    \nabla_\mu T^{\mu\nu}_{\text{(fluid)}} = f^\nu.
\end{equation}
Projecting the conservation equation parallel to the fluid flow by contracting with the Landau velocity vector $U_\nu$ yields the scalar energy conservation equation as
\begin{equation}
    U^\mu \nabla_\mu \mathcal{E} + (\mathcal{E} + p + \Pi)\nabla_\mu U^\mu + \pi^{\mu\nu}\nabla_\mu U_\nu = -U_\mu f^\mu \, . \label{eq:energy_conservation}
\end{equation}
Similarly projecting the conservation equation transverse to the fluid flow using the spatial projector $\Delta^\alpha_\nu \equiv \delta^\alpha_\nu + U^\alpha U_\nu$ yields the momentum conservation equation as
\begin{equation}
    (\mathcal{E} + p + \Pi) U^\mu \nabla_\mu U^\alpha + \Delta^{\alpha\mu}\nabla_\mu(p + \Pi) + \Delta^\alpha_\nu \nabla_\mu \pi^{\mu\nu} = \Delta^\alpha_\mu f^\mu \, . \label{eq:momentum_conservation}
\end{equation}
\section{Entropy current}\label{Entropy current}
The large $D$ membrane paradigm guarantees the existence of a local entropy current $J_S^\mu = \frac{u^\mu}{4}$, the divergence of which satisfies $\nabla_\mu J_S^\mu \sim \mathcal{O}(1/D) \geq 0$. To formulate the macroscopic fluid entropy current in the Landau frame, the membrane velocity $u^\mu$ must be algebraically inverted in terms of the Landau velocity $U^\mu$. Using Eq.~\eqref{LandauVelocity} we obtain
\begin{equation}
    u^\mu = U^\mu - \frac{l^2}{2e^2} U^\mu - \frac{l^\mu}{e} + \frac{\tau^\mu_{~\nu} l^\nu}{e^2} + \mathcal{O}(1/D^3).
\end{equation}
Substituting this transformation into the geometric entropy current yields the effective fluid entropy current in the Landau frame as
\begin{align*}
    J_S^\mu &= \frac{1}{4} \left( 1 - \frac{l^2}{2e^2} \right)U^\mu - \frac{1}{4e} \left( l^\mu - \frac{{\tau^\mu}_\nu l^\nu}{e} \right) + \mathcal{O}(1/D^3).
\end{align*}
To isolate the purely transverse and longitudinal components with respect to $U^\mu$, we will project the term $l^\mu/e$ along and perpendicular to the Landau velocity using the Landau transverse projector $\Delta^{\alpha\mu} = g^{\alpha\mu} + U^\alpha U^\mu$. 
Now the following quantity can be written as
\begin{equation}
    \frac{l^\mu}{e} = \frac{l_\alpha}{e} P^{\alpha\mu} = \frac{l_\alpha}{e} (g^{\alpha\mu} + u^\alpha u^\mu).
\end{equation}
Decomposing $l_\alpha$ into its $\mathcal{O}(1)$ and $\mathcal{O}(1/D)$ components, and substituting $u^\alpha = U^\alpha - l^\alpha/e + \mathcal{O}(1/D^2)$, we obtain
\begin{equation}
    \frac{l^\mu}{e} = \frac{l_\alpha^{(1)}}{e} \left( \Delta^{\alpha\mu} - \frac{U^\alpha l^\mu}{e} - \frac{l^\alpha U^\mu}{e} \right) + \frac{l_\alpha^{(1/D)}}{e} \Delta^{\alpha\mu} + \mathcal{O}(1/D^3).
\end{equation}
Since $l_\alpha u^\alpha = 0$, we have $l_\alpha^{(1)} U^\alpha \sim \mathcal{O}(1/D)$, which suppresses the $l_{\alpha}U^\alpha l^\mu / e^2$ term to $\mathcal{O}(1/D^3)$. The expansion strictly reduces to
\begin{equation}
    \frac{l^\mu}{e} = \frac{l_\alpha}{e} \Delta^{\alpha\mu} - \frac{l^2}{e^2} U^\mu + \mathcal{O}(1/D^3).
\end{equation}
Similarly, projecting the tensor correction yields $\frac{\tau^\mu_{~\nu} l^\nu}{e^2} = \frac{\tau_{\alpha\nu} l^\nu}{e^2} \Delta^{\alpha\mu} + \mathcal{O}(1/D^3)$. Substituting these projections back into the inverted velocity equation exactly determines the fluid entropy current $J_S^\mu$ as
\begin{align}
    J_S^\mu &= \frac{1}{4} \left( 1 + \frac{l^2}{2e^2} \right) U^\mu - \frac{1}{4} \left( \frac{l_\alpha}{e} - \frac{\tau_{\alpha\nu} l^\nu}{e^2} \right) \Delta^{\alpha\mu} + \mathcal{O}(1/D^3),\notag\\&=s' U^\mu +s^\mu+ \mathcal{O}(1/D^3),
\end{align}
where $s^\mu=- \frac{1}{4} \left( \frac{l_\alpha}{e} - \frac{\tau_{\alpha\nu} l^\nu}{e^2} \right) \Delta^{\alpha\mu}.$
Because the divergence operator $\nabla_\mu$ on $J_S^\mu$ may scale as $\mathcal{O}(D)$ in the large $D$ limit, retaining terms up to $\mathcal{O}(1/D^2)$ guarantees that the local entropy production $\nabla_\mu J_S^\mu$ is strictly positive up to $\mathcal{O}(1/D)$. The physical out-of-equilibrium entropy density $s'$ is defined as the longitudinal projection of the entropy current along the macroscopic fluid velocity, $s' \equiv -U_\mu J_S^\mu$. The expression of $s'$ takes the following form:
\begin{equation}
    s' = \frac{1}{4} \left( 1 + \frac{l^2}{2e^2} \right).
\end{equation}
The transverse to velocity component of entropy current is defined as $s^\mu$.
The out-of-equilibrium physical entropy density deviates from the static constant value of $1/4$. At equilibrium (at Schwarschild limit), the extra parts ($s^\mu$ and the correction in entropy density) vanish, resulting in a static entropy density ($s=1/4$), consistent with the global first law and thermodynamic Euler relation.

Note that the membrane velocity $u_\mu$ has a direct proportionality relation with the entropy current $J_S^\mu$ at the leading order, in that sense $u^\mu$ can be defined as the fluid's velocity in a different frame called entropy frame at this order, which traces the entropy current\footnote{This is conceptually analogous to the Eckart frame in classical hydrodynamics where the velocity vector traces the particle flow~\cite{Eckart1940}.}. Because the dual gravitational calculation mathematically enforces this proportionality, the out-of-equilibrium entropy density of the fluid is geometrically constrained to remain fixed at its equilibrium value of $s = 1/4$ in that frame at this order.

\section{Results and discussions}\label{Results and discussions}
In this section, we classify the transport coefficients of the shear stress tensor and the bulk pressure according to their derivative order, followed by a brief discussion. The coefficients corresponding to the shear stress tensor and the bulk pressure are compiled in Table~\ref{tab:shear_stress_coefficients} and Table~\ref{tab:transport_coeffs}, respectively.

\begin{table}[H]
    \centering
    \renewcommand{\arraystretch}{1.4}
    \begin{tabular}{cll}
        \toprule
        Hydrodynamic Derivative Order & Independent Tensorial Structure & Transport Coefficient \\
        \midrule
        
        $\mathcal{O}(\partial^1)$ & $\sigma_{\mu\nu}$ & $-\frac{1}{8\pi}$ \\
        \midrule
        
        $\mathcal{O}(\partial^2)$ & $\sigma^{ \alpha}_{\langle \mu} \sigma_{|\alpha|\nu\rangle}$ & $-\frac{C_D}{32\pi^2 T'}$ \\
        $\mathcal{O}(\partial^2)$ & $\omega^{ \alpha}{}_{\langle \mu} \omega_{|\alpha|\nu\rangle}$ & $-\frac{C_D}{32\pi^2 T'}$ \\
        $\mathcal{O}(\partial^2)$ & $\sigma_{\ \langle\mu}^{\alpha} \omega_{|\alpha|\nu\rangle}$ & $-\frac{C_D}{16\pi^2 T'}$ \\
        $\mathcal{O}(\partial^2)$ & $A_{\langle \mu} A_{\nu \rangle}$ & $+\frac{C_D}{32\pi^2 T'}$ \\
        $\mathcal{O}(\partial^2)$ & $\dot{T}' \sigma_{\mu\nu}$ & $+\frac{C_D}{32\pi^2 T'^2}$ \\
        $\mathcal{O}(\partial^2)$ & $\nabla^{\perp}_{\langle \mu} A_{\nu \rangle}$ & $+\frac{C_D}{32\pi^2 T'}$ \\
        \midrule
        
        $\mathcal{O}(\partial^3)$ & $\nabla^{\perp}_{\langle \mu} (\nabla \cdot \sigma)^\perp_{\nu \rangle}$ & $-\frac{C_D^2}{128\pi^3 T'^2}$ \\
        $\mathcal{O}(\partial^3)$ & $\nabla^{\perp}_{\langle \mu} (\nabla \cdot \omega)^\perp_{\nu \rangle}$ & $-\frac{C_D^2}{128\pi^3 T'^2}$ \\
        $\mathcal{O}(\partial^3)$ & $\nabla^\perp_{\langle \mu} T' (\nabla \cdot \sigma)^\perp_{\nu \rangle}$ & $+\frac{C_D^2}{64\pi^3 T'^3}$ \\
        $\mathcal{O}(\partial^3)$ & $\nabla^\perp_{\langle \mu} T' (\nabla \cdot \omega)^\perp_{\nu \rangle}$ & $+\frac{C_D^2}{64\pi^3 T'^3}$ \\
        $\mathcal{O}(\partial^3)$ & $A_{\langle \mu} (\nabla \cdot \sigma)^\perp_{\nu \rangle}$ & $-\frac{C_D^2}{128\pi^3 T'^2}$ \\
        $\mathcal{O}(\partial^3)$ & $A_{\langle \mu} (\nabla \cdot \omega)^\perp_{\nu \rangle}$ & $-\frac{C_D^2}{128\pi^3 T'^2}$ \\
        \midrule
        
        $\mathcal{O}(\partial^4)$ & $(\nabla \cdot \sigma)^\perp_{\langle \mu} (\nabla \cdot \sigma)^\perp_{\nu \rangle}$ & $-\frac{3 C_D^3}{1024\pi^4 T'^3}$ \\
        $\mathcal{O}(\partial^4)$ & $(\nabla \cdot \omega)^\perp_{\langle \mu} (\nabla \cdot \omega)^\perp_{\nu \rangle}$ & $-\frac{3 C_D^3}{1024\pi^4 T'^3}$ \\
        $\mathcal{O}(\partial^4)$ & $(\nabla \cdot \sigma)^\perp_{\langle \mu} (\nabla \cdot \omega)^\perp_{\nu \rangle}$ & $-\frac{3 C_D^3}{512\pi^4 T'^3}$ \\
        \bottomrule
    \end{tabular}
    \caption{Corresponding tensorial structures and their corresponding transport coefficients extracted from the shear stress tensor $\pi_{\mu\nu}$.}
    \label{tab:shear_stress_coefficients}
\end{table}
The transport coefficients, along with the associated tensorial structures in Table~\ref{tab:shear_stress_coefficients}, determine the structure of the shear stress tensor for the membrane fluid at first subleading order. Because the large-$D$ expansion and the hydrodynamic gradient expansion are formally independent, evaluating the subleading order theory systematically generates fluid dynamical corrections that span multiple derivative orders. At first order in derivatives, the coefficient associated with the shear tensor $\sigma_{\mu\nu}$ is $-\frac{1}{8\pi}$. By matching this against the relativistic first derivative order constitutive relation for shear stress, $-2\eta \sigma_{\mu\nu}$, the shear viscosity of the membrane fluid is determined to be $\eta = \frac{1}{16\pi}$, which is consistent with the leading order analysis done in Refs.~\cite{Halder:2026sdh, Bhattacharyya:2016nhn}.

\begin{table}[H]
\centering
\renewcommand{\arraystretch}{1.2}
\begin{tabular}{ccc}
\toprule
Hydrodynamic Derivative Order & Independent Tensorial Structure & Transport Coefficient \\
\midrule
$\mathcal{O}(\partial^1)$ & $\dot{T}'$ & $-\frac{D-3}{8\pi (D-2) T'}$ \\
\midrule
$\mathcal{O}(\partial^2)$ & $\nabla \cdot A$ & $-\frac{C_D}{64\pi^2 T'}$ \\
\midrule
$\mathcal{O}(\partial^3)$ & $\nabla_\alpha \nabla_\lambda \sigma^{\lambda\alpha}$ & $+\frac{C_D^2}{128\pi^3 T'^2}$ \\
\bottomrule
\end{tabular}
\caption{Corresponding tensorial structures and their corresponding transport coefficients extracted from the bulk pressure $\Pi$.}
\label{tab:transport_coeffs}
\end{table}

The transport coefficients in Table~\ref{tab:transport_coeffs}, associated with the corresponding tensorial structures, describe the bulk pressure of the membrane fluid at first subleading order in $1/D$. Similar to the shear part, the large-$D$ expansion (which is independent of derivative expansion in usual hydrodynamics) produces higher derivative terms here.

In standard first-order hydrodynamics~\cite{Eckart1940,Landau1959,Romatschke:2017ejr}, the bulk viscous pressure is governed exclusively by the expansion scalar $\theta$. As is evident from the energy conservation equation Eq.~\eqref{eq:energy_conservation} at first derivative order, the comoving derivative of the energy density, $\dot{\mathcal{E}}$, is strictly proportional to the divergence of the Landau velocity $\theta$ in the absence of external forces. Utilising the thermodynamic equation of state, this kinematic constraint allows the comoving temperature derivative, $\dot{T}'$, to be entirely absorbed into $\theta$. However, for the forced fluid system and $1/D$ expansion scheme considered here, the full energy conservation equation (Eq.~\eqref{eq:energy_conservation}) invalidates this algebraic substitution, necessitating its explicit inclusion as an independent contribution to the bulk pressure at the first derivative order.

It can be seen that the membrane fluid has no contribution from the expansion scalar $\theta=\nabla \cdot U$, resulting in no bulk viscosity in this order. Extraction of the subleading order contribution of bulk pressure $\Pi$ in  Eq.~\eqref{PI} enables us to verify that even for the second subleading order analysis of membrane fluid, $\theta$ does not contribute to the expression of $\Pi$. This happens due to the direct consequence of the scalar membrane equation ($\nabla \cdot u=\mathcal{O}(1/D)$), which forces the leading-order component of the Landau velocity ($U^\mu$), which is $u^\mu$, to contribute to the subleading order expression of $\theta$. It leads this first-derivative contribution to be suppressed by the $1/D$ expansion, relating it to a tensorial structure of higher-derivative contributions ($\nabla_\mu \nabla_\nu \sigma^{\mu\nu}$) at that order (in $1/D$) as discussed in section~\ref{Bulk viscosity}. That is the reason for getting no bulk viscosity even in the second subleading order analysis. As bulk viscosity is frame independent, the same thing is visible more clearly using the notion of the entropy frame defined in section~\ref{Entropy current}.  In that frame, the fluid velocity is $u^\mu$, so the expansion scalar $\tilde{\theta}=\nabla \cdot u$ is solely $\mathcal{O}(1/D)$. This scaling suppresses the contribution of the explicit factor of $\tilde{\theta}$ into order $\mathcal{O}(1/D^2)$ term at the stress-energy tensor. This implies that no bulk viscosity contributes at that order, which directly revalidates our result here.

\section{Summary and outlook}\label{Summary and outlook}
In summary, we have extracted the transport coefficients for the shear stress tensor and the bulk pressure of the uncharged membrane fluid at the first subleading order in the Landau fluid frame, which are listed in Table~\ref{tab:shear_stress_coefficients} and Table~\ref{tab:transport_coeffs}, respectively. Similar to the leading-order charged membrane fluid \cite{Halder:2026sdh}, this fluid is influenced by an effective background force and does not obey the Gibbs-Duhem relation. The pressure of the fluid has a negative sign, representing the surface tension of the codimension-one membrane hypersurface on which the fluid is localised. At this order in $1/D$, the expansion scalar does not contribute to the bulk pressure, indicating that the fluid lacks a bulk viscosity. Further investigation shows that even at the second subleading order, the bulk viscosity remains absent due to large $D$ perturbation suppression.

Looking ahead, it would be interesting to extend the fluid dynamical analysis to the charged large $D$ membrane paradigm at subleading order. Furthermore, constructing the boundary dual fluid for this system, which is inherently Carrollian, would enable a direct comparison between the bulk and boundary transport coefficients.
\FloatBarrier

\acknowledgments
The authors express their gratitude to the citizens of India for their continued support of scientific research. S.H. is supported by an Institute Fellowship from the Indian Institute of Technology (ISM) Dhanbad. M.K. acknowledges financial support from the Department of Science and Technology (DST), Government of India, through the INSPIRE-Faculty grant (DST/INSPIRE/04/2024/001794), and the Faculty Research Scheme (Project No. MISC 0240) at IIT (ISM) Dhanbad. We also thank the Department of Physics at IIT (ISM) Dhanbad for providing essential research facilities.
\appendix
\section{Appendix}
\subsection{Different components of tensor sector}\label{Different_Components_of_r}
The tensor sector of Table~\ref{tab:variables}, consists of the tensor $r_{\mu\nu}$, can be further split it into two components, $\mathcal{O}(1)$ which is $ r_{\mu\nu(1)}$ and another is $\mathcal{O}(1/D)$ which is $ r_{\mu\nu(1/D)}$. We have,
\begin{align*}
r_{\mu\nu} &= \mathcal{W}^{\alpha\beta} P_{\alpha\mu} P_{\beta\nu}= r_{\mu\nu(1)} + r_{\mu\nu(1/D)},
\end{align*}
Using the expression of $\mathcal{W}^{\alpha \beta}$ from Eq.~\eqref{eq:Wmunu} we have
\begin{align*}
r_{\mu\nu(1)} &= \frac{1}{2} K_{\alpha\beta} P^{\alpha}_{\mu} P^{\beta}_{\nu} - \frac{1}{2} (\nabla_{\alpha}u_{\beta} + \nabla_{\beta}u_{\alpha}) P^{\alpha}_{\mu} P^{\beta}_{\nu},
\end{align*}
\begin{align}\label{exp:r}
r_{\mu\nu(1/D)} &= -\frac{1}{K} K_{\alpha\beta} P^{\alpha}_{\mu} P^{\beta}_{\nu} (u \cdot K \cdot u) + \frac{u \cdot K \cdot u}{2K} (\nabla_{\alpha}u_{\beta} + \nabla_{\beta}u_{\alpha}) P^{\alpha}_{\mu} P^{\beta}_{\nu} \notag\\
&\quad + \frac{P^{\alpha}_{\mu} P^{\beta}_{\nu}}{2K} \left[ \nabla_{\alpha} \left( \frac{\nabla^{2} u_{\beta}}{K} \right) + \nabla_{\beta} \left( \frac{\nabla^{2} u_{\alpha}}{K} \right) + \nabla_{\alpha} (u^{\sigma} K_{\sigma\beta}) \right. \notag\\
&\quad \left. + \nabla_{\beta} (u^{\sigma} K_{\sigma\alpha}) - 2 \nabla_{\alpha} \left( \frac{\nabla_{\beta}K}{K} \right) \right] - \frac{1}{K} \left[ (\nabla^{\sigma} u_{\alpha}) (\nabla_{\sigma} u_{\beta}) + \frac{\nabla^{2} u_{\alpha}}{K} \frac{\nabla^{2} u_{\beta}}{K} \right] P^{\alpha}_{\mu} P^{\beta}_{\nu}.
\end{align}
The term
\begin{align*}
\nabla_{\alpha} \left[ u^{\sigma} K_{\sigma\rho} \right] P^{\alpha}_{\mu} P^{\rho}_{\nu} 
&= \nabla_{\alpha} \left[ u^{\sigma} K_{\sigma}^{\rho} (g_{\rho\beta}) \right] P^{\alpha}_{\mu} P^{\beta}_{\nu}= \nabla_{\alpha} \left[ - (u \cdot K \cdot u) u_{\beta} + u^{\sigma} K_{\sigma\rho} P^{\rho}_{\beta} \right] P^{\alpha}_{\mu} P^{\beta}_{\nu}.
\end{align*}
Using the vector membrane equation (Eq.~\eqref{eq:membrane:vector}), we obtain
\begin{align*}
\nabla_{\alpha} \left[ u^{\sigma} K_{\sigma\rho} \right] P^{\alpha}_{\mu} P^{\rho}_{\nu}&= \nabla_{\alpha} \left[ - (u \cdot K \cdot u) u_{\beta} + \left( - \frac{\nabla^{2} u_{\rho}}{K} + \frac{\nabla_{\rho} K}{K} + u \cdot \nabla u_{\rho} \right) P^{\rho}_{\beta} \right] P^{\alpha}_{\mu} P^{\beta}_{\nu} + \mathcal{O}(1/D), \\
&= - P^{\alpha}_{\mu} P^{\beta}_{\nu} (u \cdot K \cdot u) (\nabla_{\alpha} u_{\beta}) \\
&\quad + P^{\alpha}_{\mu} P^{\beta}_{\nu} \left[ - \nabla_{\alpha} \left( P^{\rho}_{\beta} \left( \frac{\nabla^{2} u_{\rho}}{K} \right) \right) + \nabla_{\alpha} \left( P^{\rho}_{\beta} \left( \frac{\nabla_{\rho} K}{K} \right) \right) + \nabla_{\alpha} (u \cdot \nabla u_{\beta}) \right].
\end{align*}
Substituting the expression of $\nabla_{\alpha} \left[ u^{\sigma} K_{\sigma\rho} \right] P^{\alpha}_{\mu} P^{\rho}_{\nu}$ in Eq.~\eqref{exp:r} yields
\begin{align*}
r_{\mu\nu(1/D)} &= -\frac{1}{K} K_{\alpha\beta} P^{\alpha}_{\mu} P^{\beta}_{\nu} (u \cdot K \cdot u) - \frac{1}{K} \left[ (\nabla^{\sigma} u_{\alpha}) (\nabla_{\sigma} u_{\beta}) + \frac{\nabla^{2} u_{\alpha}}{K} \frac{\nabla^{2} u_{\beta}}{K} \right] P^{\alpha}_{\mu} P^{\beta}_{\nu} \\
&\quad + \frac{P^{\alpha}_{\mu} P^{\beta}_{\nu}}{2K} \bigg[ - \nabla_{\alpha} \left(\frac{u_{\beta} (u_{\sigma} \nabla^{2} u^{\sigma})}{K}\right) - \nabla_{\beta} \left(\frac{ u_{\alpha} (u_{\sigma} \nabla^{2} u^{\sigma}}{K}) \right) \\
&\quad + \nabla_{\alpha} \left( u_{\beta} \left( \frac{u \cdot \nabla K}{K} \right) \right) + \nabla_{\beta} \left( u_{\alpha} \left( \frac{u \cdot \nabla K}{K} \right) \right)+ \nabla_{\alpha} (u \cdot \nabla u_{\beta}) + \nabla_{\beta} (u \cdot \nabla u_{\alpha}) \bigg] + \mathcal{O}(1/D^2), \\
&= - \frac{u \cdot K \cdot u}{K} K_{\alpha\beta} P^{\alpha}_{\mu} P^{\beta}_{\nu} - \frac{1}{K} \left[ (\nabla^{\sigma} u_{\alpha}) (\nabla_{\sigma} u_{\beta}) + \frac{\nabla^{2} u_{\alpha}}{K} \frac{\nabla^{2} u_{\beta}}{K} \right] P^{\alpha}_{\mu} P^{\beta}_{\nu} \\
&\quad + \frac{P^{\alpha}_{\mu} P^{\beta}_{\nu}}{2K} \bigg[ \left( -\frac{u_{\sigma} \nabla^{2} u^{\sigma}}{K} + \frac{u \cdot \nabla K}{K} \right) (\nabla_{\alpha} u_{\beta} + \nabla_{\beta} u_{\alpha})+ \nabla_{\alpha} (u \cdot \nabla u_{\beta}) + \nabla_{\beta} (u \cdot \nabla u_{\alpha}) \bigg]+ \mathcal{O}(1/D^2).
\end{align*}
\subsection{Membrane stress-energy tensor in the Landau frame}\label{Landau_substitution}
We start with, $8\pi T_{\mu\nu} = e u_{\mu} u_{\nu} + s_{2} g_{\mu\nu} + l_{\mu} u_{\nu} + l_{\nu} u_{\mu} + r_{\mu\nu(1)} + r_{\mu\nu(1/D)}$.
Utilizing different components of $r_{\mu\nu}$ as in Appendix~\ref{Different_Components_of_r} we obtain, 
\begin{flalign*}
8\pi T_{\mu\nu}&= e \left( U_{\mu} - U_{\mu} \frac{l^2}{2 e^2} - \frac{l_{\mu}}{e} + \frac{\tau_{\mu}^{\alpha} l_{\alpha}}{e^2} \right) \left( U_{\nu} - U_{\nu} \frac{l^2}{2 e^2} - \frac{l_{\nu}}{e} + \frac{\tau_{\nu}^{\alpha} l_{\alpha}}{e^2} \right) \\
&\quad + s_{2} g_{\mu\nu} + l_{\mu} \left( U_{\nu} - \frac{l_{\nu}}{e} \right) + l_{\nu} \left( U_{\mu} - \frac{l_{\mu}}{e} \right) - \frac{1}{2} \left[ \nabla_{\alpha} U_{\beta} + \nabla_{\beta} U_{\alpha} \right] P^{\alpha}_{\mu} P^{\beta}_{\nu} \\
&\quad + \frac{1}{2} \left[ \nabla_{\alpha}\bigg( \frac{l_{\beta}}{e}\bigg) + \nabla_{\beta} \bigg(\frac{l_{\alpha}}{e}\bigg) \right] P^{\alpha}_{\mu} P^{\beta}_{\nu}+ \frac{1}{2} K_{\alpha\beta} P^{\alpha}_{\mu} P^{\beta}_{\nu} + r_{\mu\nu(1/D)} + \mathcal{O}(1/D^2), &&
\end{flalign*}
\begin{flalign*}
&= e \left( U_{\mu} - U_{\mu} \frac{l^2}{2 e^2} - \frac{l_{\mu}}{e} + \frac{\tau_{\mu}^{\alpha} l_{\alpha}}{e^2} \right) \left( U_{\nu} - U_{\nu} \frac{l^2}{2 e^2} - \frac{l_{\nu}}{e} + \frac{\tau_{\nu}^{\alpha} l_{\alpha}}{e^2} \right) + s_{2} g_{\mu\nu} \\
&\quad + l_{\mu} \left( U_{\nu} - \frac{l_{\nu}}{e} \right) + l_{\nu} \left( U_{\mu} - \frac{l_{\mu}}{e} \right) - \frac{1}{2} \left[ \nabla_{\alpha} U_{\beta} + \nabla_{\beta} U_{\alpha} \right] P^{\alpha}_{\mu} P^{\beta}_{\nu} + \frac{1}{2} K_{\alpha\beta} P^{\alpha}_{\mu} P^{\beta}_{\nu} \\
&\quad + \left[  \frac{1}{2} \left
(\nabla_{\alpha}\bigg( \frac{l_{\beta}}{e}\bigg) + \nabla_{\beta} \bigg(\frac{l_{\alpha}}{e}\bigg) \right) P^{\alpha}_{\mu} P^{\beta}_{\nu} + r_{\mu\nu(1/D)} \right] + \mathcal{O}(1/D^2), &&
\end{flalign*}
\begin{flalign*}
&= e \left( U_{\mu} - U_{\mu} \frac{l^2}{2 e^2} - \frac{l_{\mu}}{e} + \frac{\tau_{\mu}^{\alpha} l_{\alpha}}{e^2} \right) \left( U_{\nu} - U_{\nu} \frac{l^2}{2 e^2} - \frac{l_{\nu}}{e} + \frac{\tau_{\nu}^{\alpha} l_{\alpha}}{e^2} \right)  + s_{2} g_{\mu\nu}\\
&\quad + l_{\mu} \left( U_{\nu} - \frac{l_{\nu}}{e} \right) + l_{\nu} \left( U_{\mu} - \frac{l_{\mu}}{e} \right) - \frac{1}{2} \left[ \nabla_{\alpha} U_{\beta} + \nabla_{\beta} U_{\alpha} \right] \left( \delta^{\alpha}_{\mu} + \left(U^{\alpha} - \frac{l^{\alpha}}{e}\right)\left(U_{\mu} - \frac{l_{\mu}}{e}\right) \right) \\
&\qquad \left( \delta^{\beta}_{\nu} + \left(U^{\beta} - \frac{l^{\beta}}{e}\right)\left(U_{\nu} - \frac{l_{\nu}}{e}\right) \right) + \frac{1}{2} K_{\alpha\beta} P^{\alpha}_{\mu} P^{\beta}_{\nu} \\
&\quad + \left[  \frac{1}{2} \left
(\nabla_{\alpha}\bigg( \frac{l_{\beta}}{e}\bigg) + \nabla_{\beta} \bigg(\frac{l_{\alpha}}{e}\bigg) \right) P^{\alpha}_{\mu} P^{\beta}_{\nu} + r_{\mu\nu(1/D)} \right] + \mathcal{O}(1/D^2), &&
\end{flalign*}
\begin{flalign*}
&= e \left[ U_{\mu} U_{\nu} \left( 1 - \frac{l^2}{2e^2} \right) - {\frac{l_{\mu} U_{\nu}}{e}} + \frac{\tau_{\mu}^{\rho} l_{\rho}}{e^2} U_{\nu} - \frac{l^2}{2e^2} U_{\mu} U_{\nu} - {\frac{U_{\mu} l_{\nu}}{e}} + \frac{l_{\mu} l_{\nu}}{e^2} + \frac{\tau_{\nu}^{\alpha} l_{\alpha}}{e^2} U_{\mu} \right] + s_{2} g_{\mu\nu}\\
&\quad + l_{\nu} \left({U_{\mu}} - \frac{l_{\mu}}{e} \right) + l_{\mu} \left( {U_{\nu}} - \frac{l_{\nu}}{e} \right) - \frac{1}{2} \left[ \nabla_{\alpha} U_{\beta} + \nabla_{\beta} U_{\alpha} \right] \\
&\qquad\left( \delta^{\alpha}_{\mu} + \left(U^{\alpha} - \frac{l^{\alpha}}{e}\right)\left(U_{\mu} - \frac{l_{\mu}}{e}\right) \right) \left( \delta^{\beta}_{\nu} + \left(U^{\beta} - \frac{l^{\beta}}{e}\right)\left(U_{\nu} - \frac{l_{\nu}}{e}\right) \right) + \frac{1}{2} K_{\alpha\beta} P^{\alpha}_{\mu} P^{\beta}_{\nu} \\
&\quad + \left[  \frac{1}{2} \left
(\nabla_{\alpha}\bigg( \frac{l_{\beta}}{e}\bigg) + \nabla_{\beta} \bigg(\frac{l_{\alpha}}{e}\bigg) \right) P^{\alpha}_{\mu} P^{\beta}_{\nu} + r_{\mu\nu(1/D)} \right] + \mathcal{O}(1/D^2), &&
\end{flalign*}
\begin{flalign*}
&= e \left( 1 - \frac{l^2}{e^2} \right) U_{\mu} U_{\nu} + \frac{\tau_{\mu}^{\rho} l_{\rho}}{e} U_{\nu} + \frac{\tau_{\nu}^{\alpha} l_{\alpha}}{e} U_{\mu}  + s_{2} g_{\mu\nu} - \frac{1}{2} \left[ \nabla_{\alpha} U_{\beta} + \nabla_{\beta} U_{\alpha} \right] \\
&\qquad \left( \delta^{\alpha}_{\mu} + \left(U^{\alpha} - \frac{l^{\alpha}}{e}\right)\left(U_{\mu} - \frac{l_{\mu}}{e}\right) \right) \left( \delta^{\beta}_{\nu} + \left(U^{\beta} - \frac{l^{\beta}}{e}\right)\left(U_{\nu} - \frac{l_{\nu}}{e}\right) \right) \\
&\quad + \frac{1}{2} K_{\alpha\beta} P^{\alpha}_{\mu} P^{\beta}_{\nu} + \underbrace{\left[  \frac{1}{2} \left
(\nabla_{\alpha}\bigg( \frac{l_{\beta}}{e}\bigg) + \nabla_{\beta} \bigg(\frac{l_{\alpha}}{e}\bigg) \right) P^{\alpha}_{\mu} P^{\beta}_{\nu} - \frac{l_{\mu} l_{\nu}}{e} + r_{\mu\nu(1/D)} \right]}_{\Pi_{\mu\nu(1/D)}} + \mathcal{O}(1/D^2), &&
\end{flalign*}
\begin{flalign*}
&= e \left( 1 - \frac{l^2}{e^2} \right) U_{\mu} U_{\nu} + \frac{\tau_{\mu}^{\rho} l_{\rho}}{e} U_{\nu} + \frac{\tau_{\nu}^{\alpha} l_{\alpha}}{e} U_{\mu} + s_{2} g_{\mu\nu} - \frac{1}{2} \left[ \nabla_{\alpha} U_{\beta} + \nabla_{\beta} U_{\alpha} \right] \\
&\quad \left( \delta^{\alpha}_{\mu} + U^{\alpha} U_{\mu} - \frac{l^{\alpha} U_{\mu}}{e} - \frac{U^{\alpha} l_{\mu}}{e} \right) \left( \delta^{\beta}_{\nu} + U^{\beta} U_{\nu} - \frac{l^{\beta} U_{\nu}}{e} - \frac{U^{\beta} l_{\nu}}{e} \right) + \frac{1}{2} K_{\alpha\beta} P^{\alpha}_{\mu} P^{\beta}_{\nu} + \Pi_{\mu\nu(1/D)} + \mathcal{O}(1/D^2). &&
\end{flalign*}
By using the $\Delta_{\alpha\mu} = g_{\alpha\mu} + U_{\alpha} U_{\mu},$ and the following identity,
\begin{flalign*}    
\frac{\tau_{\mu}^{\rho} l_{\rho}}{e} U_{\nu}&= - \frac{1}{2e} (\nabla^{\delta} U_{\sigma} + \nabla_{\sigma} U^{\delta}) P_{\mu\delta} P^{\sigma}_{\rho} l^{\rho} U_{\nu}+ \mathcal{O}(1/D^2), \\
&= - \frac{1}{2e} (\nabla^{\delta} U_{\sigma} + \nabla_{\sigma} U^{\delta}) l^{\sigma} U_{\nu} (g_{\mu\delta} + U_{\mu} U_{\delta}) + \mathcal{O}(1/D^2), \\
&= - \frac{1}{2e} (\nabla_{\mu} U_{\sigma} + \nabla_{\sigma} U_{\mu}) l^{\sigma} U_{\nu} - \frac{1}{2e} U_{\mu} (U \cdot \nabla U_{\sigma}) l^{\sigma} U_{\nu} + \mathcal{O}(1/D^2), &&
\end{flalign*}
we obtain 
\begin{flalign}\label{EM_appendix}
8\pi T_{\mu\nu}&= e \left( 1 - \frac{l^2}{e^2} \right) U_{\mu} U_{\nu} + s_{2} g_{\mu\nu} - \frac{1}{2} (\nabla_{\alpha} U_{\beta} + \nabla_{\beta} U_{\alpha}) \Delta^{\alpha}_{\mu} \Delta^{\beta}_{\nu}\notag \\
&\quad+\frac{l_\mu U\cdot\nabla U_\nu+l_\nu U\cdot\nabla U_\mu}{2e} + \Pi_{\mu\nu(1/D)} + \frac{1}{2} K_{\alpha\beta} P^{\alpha}_{\mu} P^{\beta}_{\nu}+ \mathcal{O}(1/D^2).&&
\end{flalign}
Now the term
\begin{align*}
\Pi_{\mu\nu(1/D)} &= \frac{1}{2} \left
(\nabla_{\alpha}\bigg( \frac{l_{\beta}}{e}\bigg) + \nabla_{\beta} \bigg(\frac{l_{\alpha}}{e}\bigg) \right) P^{\alpha}_{\mu} P^{\beta}_{\nu} - \frac{l_{\mu} l_{\nu}}{e} + r_{\mu\nu(1/D)}.
\end{align*}
By substituting
$$l_{\mu} = \mathcal{V}^{\alpha} P_{\alpha\mu} - W^{\alpha\beta} P_{\mu\beta} u_{\alpha} 
= \left[ \frac{1}{2} \frac{\nabla^{\alpha} K}{K} P_{\alpha\mu} - P_{\alpha\mu} \frac{\nabla^2 u^{\alpha}}{K} \right]+ \frac{1}{2} u \cdot \nabla u_{\mu} - \frac{1}{2} K^{\alpha\beta} P_{\mu\beta} u_{\alpha} + \mathcal{O}(1/D),$$ and using leading order vector membrane Eq.~\eqref{eq:membrane:vector}, $l_{\mu}= - \frac{1}{2} \frac{\nabla^2 u^{\sigma}}{K} P_{\sigma\mu} + \mathcal{O}(1/D),$ we express
\begin{flalign*}
\Pi_{\mu\nu(1/D)} &= \frac{1}{2} \left( \nabla_{\alpha}\left(\frac{l_{\beta}}{e}\right) + \nabla_{\beta}\left(\frac{l_{\alpha}}{e}\right) \right) P^{\alpha}_{\mu} P^{\beta}_{\nu} - \frac{l_{\mu} l_{\nu}}{e} + r_{\mu\nu(1/D)} + \mathcal{O}(1/D^2), \\
&= \left[ \nabla_{\alpha}\left(\frac{l_{\beta}}{K}\right) + \nabla_{\beta}\left(\frac{l_{\alpha}}{K}\right) \right] P^{\alpha}_{\mu} P^{\beta}_{\nu} - 2 \frac{l_{\mu} l_{\nu}}{K} + r_{\mu\nu(1/D)} + \mathcal{O}(1/D^2), \\
&= -\frac{P^{\alpha}_{\mu} P^{\beta}_{\nu}}{2} \left[ \nabla_{\alpha} \left( \frac{\nabla^2 u^{\sigma}}{K^2} P_{\sigma\beta} \right) + \nabla_{\beta} \left( \frac{\nabla^2 u^{\sigma}}{K^2} P_{\sigma\alpha} \right) \right] \\
&\quad - \frac{1}{2 K^3} (\nabla^2 u^{\sigma})(\nabla^2 u^{\alpha}) P_{\sigma\mu} P_{\alpha\nu} + r_{\mu\nu(1/D)} + \mathcal{O}(1/D^2), &&
\end{flalign*}
\begin{flalign*}
&\hspace{1.5 cm}= - \frac{P^{\alpha}_{\mu} P^{\beta}_{\nu}}{2} \left( \frac{\nabla^2 u^{\sigma}}{K^2} \right) u_{\sigma} (\nabla_{\alpha} u_{\beta} + \nabla_{\beta} u_{\alpha})- \frac{P^{\alpha}_{\mu} P_{\sigma\nu}}{2} \nabla_{\alpha} \left( \frac{\nabla^2 u^{\sigma}}{K^2} \right) \\
&\hspace{1.5 cm}\quad  - \frac{P^{\beta}_{\nu} P_{\sigma\mu}}{2} \nabla_{\beta} \left( \frac{\nabla^2 u^{\sigma}}{K^2} \right)  - \frac{1}{2K^3} (\nabla^2 u^{\sigma})(\nabla^2 u^{\alpha}) P_{\sigma\mu} P_{\alpha\nu} + r_{\mu\nu(1/D)} + \mathcal{O}(1/D^2), &&
\end{flalign*}
\begin{flalign*}
&\hspace{1.5 cm}= - \frac{P^{\alpha}_{\mu} P^{\beta}_{\nu}}{2} \left( \frac{\nabla^2 u^{\sigma}}{K^2} \right) u_{\sigma} (\nabla_{\alpha} u_{\beta} + \nabla_{\beta} u_{\alpha})  - \frac{1}{2K^3} (\nabla^2 u^{\sigma})(\nabla^2 u^{\alpha}) P_{\sigma\mu} P_{\alpha\nu} \\
&\hspace{1.5 cm}\quad - \frac{1}{2K^2} P^{\alpha}_{\mu} P^{\beta}_{\nu} \left( \nabla_{\alpha} (\nabla^2 u_{\beta}) + \nabla_{\beta} (\nabla^2 u_{\alpha}) \right) \\
&\hspace{1.5 cm}\quad + \frac{2}{K^3} P^{\alpha}_{\mu} P^{\beta}_{\nu} \left( (\nabla^2 u_{\beta})(\nabla_{\alpha} K) + (\nabla^2 u_{\alpha})(\nabla_{\beta} K) \right) + r_{\mu\nu(1/D)} + \mathcal{O}(1/D^2). &&
\end{flalign*}
Substituting the value of $r_{\mu\nu(1/D)}$  in the above expression from Appendix~\eqref{Different_Components_of_r}, we obtain:
\begin{flalign*}
\Pi_{\mu\nu(1/D)}&= - \frac{u \cdot K \cdot u}{K} K_{\alpha\beta} P^{\alpha}_{\mu} P^{\beta}_{\nu}+ \left[ - \frac{1}{K} (\nabla^{\sigma} u_{\alpha}) (\nabla_{\sigma} u_{\beta}) - \frac{3}{2K^3} \nabla^2 u_{\alpha} \nabla^2 u_{\beta} \right. \\
&\quad \left. - \frac{1}{K^2} (u_{\sigma} \nabla^2 u^{\sigma}) ( \nabla_{\alpha} u_{\beta} + \nabla_{\beta} u_{\alpha} )  + \frac{1}{2K^2} (u \cdot \nabla K) (\nabla_{\alpha} u_{\beta} + \nabla_{\beta} u_{\alpha}) \right. \\
&\quad \left. + \frac{1}{2K} ( \nabla_{\alpha} (u \cdot \nabla u_{\beta}) + \nabla_{\beta} (u \cdot \nabla u_{\alpha}) )  - \frac{1}{2K^2} ( \nabla_{\alpha} (\nabla^2 u_{\beta}) + \nabla_{\beta} (\nabla^2 u_{\alpha}) ) \right. \\
&\quad \left. + \frac{1}{K^3} ( (\nabla^2 u_{\beta})(\nabla_{\alpha} K) + (\nabla^2 u_{\alpha})(\nabla_{\beta} K) ) \right] P^{\alpha}_{\mu} P^{\beta}_{\nu} + \mathcal{O}(1/D^2). &&
\end{flalign*}
Further, substituting the above expression in Eq.~\eqref{EM_appendix} yields
\begin{flalign*}
8\pi T_{\mu\nu}&= e \left( 1 - \frac{l^2}{e^2} \right) U_{\mu} U_{\nu} + s_{2} g_{\mu\nu} - \frac{1}{2} (\nabla_{\alpha} U_{\beta} + \nabla_{\beta} U_{\alpha}) \Delta^{\alpha}_{\mu} \Delta^{\beta}_{\nu}\notag \\
&\quad + \left[ - \frac{1}{K} (\nabla^{\sigma} U_{\alpha}) (\nabla_{\sigma} U_{\beta}) - \frac{3}{2K^3} \nabla^2 U_{\alpha} \nabla^2 U_{\beta} - \frac{1}{K^2} (U_{\sigma} \nabla^2 U^{\sigma}) ( \nabla_{\alpha} U_{\beta} + \nabla_{\beta} U_{\alpha} ) \right.\notag \\
&\quad \left. + \frac{1}{2K^2} (U \cdot \nabla K) (\nabla_{\alpha} U_{\beta} + \nabla_{\beta} U_{\alpha}) + \frac{1}{2K} ( \nabla_{\alpha} (U \cdot \nabla U_{\beta}) + \nabla_{\beta} (U \cdot \nabla U_{\alpha}) ) \right.\notag \\
&\quad \left. - \frac{1}{2K^2} ( \nabla_{\alpha} (\nabla^2 U_{\beta}) + \nabla_{\beta} (\nabla^2 U_{\alpha}) ) + \frac{1}{K^3} ( (\nabla^2 U_{\beta})(\nabla_{\alpha} K) + (\nabla^2 U_{\alpha})(\nabla_{\beta} K) ) \right.\notag \\
&\quad \left. -\frac{(\nabla^2 U_\alpha) U\cdot\nabla U_\beta+(\nabla^2 U_\beta) U\cdot\nabla U_\alpha}{2K}\right] \Delta^{\alpha}_{\mu} \Delta^{\beta}_{\nu} + \frac{1}{2} K_{\alpha\beta} P^{\alpha}_{\mu} P^{\beta}_{\nu}- \frac{u \cdot \kappa \cdot u}{K} K_{\alpha\beta} P^{\alpha}_{\mu} P^{\beta}_{\nu}+ \mathcal{O}(1/D^2).&&
\end{flalign*}
\subsection{Explicit geometric expression of thermodynamic quantities}\label{Explicit geometric expression of thermodynamic quantities}

\begin{align}
e &= s_1 - 2 \nu_\mu u^\mu + W_{\mu\nu} u^\mu u^\nu\nonumber \\
&= \frac{K}{2} + \left( \frac{\nabla^2 K}{2K^2} - \frac{K}{2(D-2)} - \frac{u \cdot \nabla K}{K} + 2 \frac{u_\alpha \nabla^2 u^\alpha}{K} + \frac{u \cdot K \cdot u}{2} \right)\nonumber\\&\quad + \frac{1}{K} \bigg[ -3u\cdot K\cdot K\cdot u - 13\left(\frac{u\cdot\nabla K}{K}\right)^{2} + 2u^{\alpha}K_{\alpha\beta}\left(\frac{\nabla^{\beta}K}{K}\right) + 13\left(\frac{u\cdot\nabla K}{K}\right)(u\cdot K\cdot u) \nonumber \\
&\quad - \frac{K}{D}\left(\frac{u\cdot\nabla K}{K}\right) + \frac{K}{D}(u\cdot K\cdot u) + \frac{1}{K^{3}}\nabla^{2}(\nabla^{2}K) - 5(u\cdot K\cdot u)^{2}- 2\left(\frac{\nabla_{\alpha}K}{K}\right)\left(\frac{\nabla^{\alpha}K}{K}\right) \bigg] \nonumber \\
&\quad + \frac{1}{K}(2\zeta(3)-1) \bigg[ -\frac{K}{D}\left(\frac{u\cdot\nabla K}{K} - u\cdot K\cdot u\right) - u\cdot K\cdot K\cdot u + 2\left(\frac{\nabla_{\alpha}K}{K}\right)u^{\beta}K_{\beta}^{\alpha} \nonumber \\
&\quad - \left(\frac{u\cdot\nabla K}{K}\right)^{2} + 2\left(\frac{u\cdot\nabla K}{K}\right)(u\cdot K\cdot u) - \left(\frac{\nabla^{\alpha}K}{K}\right)\left(\frac{\nabla_{\alpha}K}{K}\right) - (u\cdot K\cdot u)^{2} \nonumber \\
&\quad+  \frac{2u^\mu}{K^{3}}\nabla^{2}(\nabla^{2}u_{\mu}) - \frac{2u^\mu}{K}\nabla_{\mu}\left(\frac{u\cdot\nabla K}{K}\right) - \frac{2u^\mu}{K}\left(\frac{\nabla^{2}u_{\mu}}{K}\right)\left(-2(u\cdot K\cdot u) + 4\frac{u\cdot\nabla K}{K}  - \frac{K}{D}\right)\nonumber\\&\quad + \frac{1}{K} \bigg[ u^\mu u^\nu \nabla_{\mu}\left(\frac{\nabla^{2}u_{\nu}}{K}\right) +u^\mu u^\nu  \nabla_{\mu}(u^{\alpha}K_{\alpha\nu})  - u^\mu u^\nu \nabla_{\mu}\left(\frac{\nabla_{\nu}K}{K}\right)-u^\mu u^\nu \left(\frac{\nabla^{2}u_{\mu}}{K}\right)\left(\frac{\nabla^{2}u_{\nu}}{K}\right) \bigg].
\end{align}
\begin{align}
\frac{s_{2}}{8\pi} &=  - \frac{K}{16\pi(D-2)} -\frac{1}{16\pi}(u\cdot K\cdot u)- \frac{1}{8\pi K}\left(\frac{u\cdot\nabla K}{K} - \frac{1}{2}(u\cdot K\cdot u) - \frac{K}{2D}\right)(u\cdot K\cdot u) \nonumber \\
&\quad + \frac{1}{8\pi K}K^{\alpha\beta}(\nabla_{\alpha}u_{\beta}) - \frac{2}{8\pi K}u_{\alpha}K^{\alpha\beta}\left(\frac{1}{2}\frac{\nabla_{\beta}K}{K} - \frac{\nabla^{2}u_{\beta}}{K}\right). \label{eq:S2}
\end{align}
In the above substitution, we used the identity $K_{\alpha\beta}K^{\alpha\beta}=\frac{K^2}{D-2}$. Further, we have
\begin{align}
l_\alpha &= P_\alpha^\mu (\mathcal{V}_\mu - \mathcal{W}_{\mu\nu} u^\nu), \nonumber \\
&= P_\alpha^\mu \Bigg[ \frac{1}{2}\left(\frac{\nabla_\mu K}{K}\right) - \left(\frac{\nabla^2 u_\mu}{K}\right) + \frac{1}{K} K_\mu^\sigma K_{\sigma\beta} u^\beta - \frac{1}{K^3} \nabla^2(\nabla^2 u_\mu) + \frac{1}{K} \nabla_\mu \left(\frac{u \cdot \nabla K}{K}\right) \nonumber \\
&\quad + \frac{1}{K} \left(\frac{\nabla^2 u_\mu}{K}\right) \left( -2(u \cdot K \cdot u) + 4\frac{u \cdot \nabla K}{K} - \frac{K}{D} +\left( \frac{u^\nu \nabla^2 u_\nu}{K} \right)\right) + \frac{1}{2K} \left(\frac{\nabla_\mu K}{K}\right)(u \cdot K \cdot u) \nonumber \\
&\quad - \frac{1}{2}K_{\mu\nu}u^\nu + \frac{1}{2}(u \cdot \nabla u_\mu) + \frac{1}{K} K_{\mu\nu} u^\nu (u \cdot K \cdot u) - \frac{1}{2K} (u \cdot \nabla u_\mu) (u \cdot K \cdot u) \nonumber \\
&\quad - \frac{1}{2K} \left( u^\nu \nabla_\mu \left(\frac{\nabla^2 u_\nu}{K}\right) + u \cdot \nabla \left(\frac{\nabla^2 u_\mu}{K}\right) + u^\nu \nabla_\mu (u^\sigma K_{\sigma\nu}) + u \cdot \nabla (u^\sigma K_{\sigma\mu}) - 2 u^\nu \nabla_\mu \left(\frac{\nabla_\nu K}{K}\right) \right) \Bigg],\nonumber \\
&=\Bigg( \frac{1}{2}\left(\frac{\nabla_\mu K}{K}\right) - \left(\frac{\nabla^2 u_\mu}{K}\right) - \frac{1}{2}K_{\mu\nu}u^\nu + \frac{1}{2}(u \cdot \nabla u_\mu)\Bigg) P_\alpha^\mu \notag \\&\quad + \Bigg[\frac{1}{K} K_\mu^\sigma K_{\sigma\beta} u^\beta - \frac{1}{K^3} \nabla^2(\nabla^2 u_\mu) + \frac{1}{K} \nabla_\mu \left(\frac{u \cdot \nabla K}{K}\right) \nonumber \\
&\quad + \frac{1}{K} \left(\frac{\nabla^2 u_\mu}{K}\right) \left( -2(u \cdot K \cdot u) + 4\frac{u \cdot \nabla K}{K} - \frac{K}{D} +\left( \frac{u^\nu \nabla^2 u_\nu}{K} \right)\right) + \frac{1}{2K} \left(\frac{\nabla_\mu K}{K}\right)(u \cdot K \cdot u) \nonumber \\
&\quad+ \frac{1}{K} K_{\mu\nu} u^\nu (u \cdot K \cdot u) - \frac{1}{2K} (u \cdot \nabla u_\mu) (u \cdot K \cdot u) - \frac{1}{2K} \Bigg( u^\nu \nabla_\mu \left(\frac{\nabla^2 u_\nu}{K}\right) \nonumber \\
&\quad + u \cdot \nabla \left(\frac{\nabla^2 u_\mu}{K}\right) + u^\nu \nabla_\mu (u^\sigma K_{\sigma\nu}) + u \cdot \nabla (u^\sigma K_{\sigma\mu}) - 2 u^\nu \nabla_\mu \left(\frac{\nabla_\nu K}{K}\right) \Bigg) \Bigg] P_\alpha^\mu, 
\end{align}
\begin{align}
    \mathcal{E}&=\frac{1}{8\pi}\left( e - \frac{l^2}{e} -s_2\right),\notag\\ \quad &= \frac{K}{16\pi} + \frac{1}{8\pi}\left( \frac{\nabla^2 K}{2K^2} - \frac{u \cdot \nabla K}{K} + 2 \frac{u_\alpha \nabla^2 u^\alpha}{K} + u \cdot K \cdot u \right)\nonumber
    \\&\quad + \frac{1}{8\pi K} \bigg[ -3u\cdot K\cdot K\cdot u - 13\left(\frac{u\cdot\nabla K}{K}\right)^{2} + 3u^{\alpha}K_{\alpha\beta}\left(\frac{\nabla^{\beta}K}{K}\right) + 14\left(\frac{u\cdot\nabla K}{K}\right)(u\cdot K\cdot u) \nonumber \\
&\quad - \frac{K}{D}\left(\frac{u\cdot\nabla K}{K}\right) + \frac{K}{2D}(u\cdot K\cdot u) + \frac{1}{K^{3}}\nabla^{2}(\nabla^{2}K) - \frac{7}{2}(u\cdot K\cdot u)^{2}- 2\left(\frac{\nabla_{\alpha}K}{K}\right)\left(\frac{\nabla^{\alpha}K}{K}\right)  \nonumber \\
&\quad + (2\zeta(3)-1) \bigg( -\frac{K}{D}\left(\frac{u\cdot\nabla K}{K} - u\cdot K\cdot u\right) - u\cdot K\cdot K\cdot u + 2\left(\frac{\nabla_{\alpha}K}{K}\right)u^{\beta}K_{\beta}^{\alpha} - \left(\frac{u\cdot\nabla K}{K}\right)^{2} \nonumber \\
&\quad + 2\left(\frac{u\cdot\nabla K}{K}\right)(u\cdot K\cdot u) - \left(\frac{\nabla^{\alpha}K}{K}\right)\left(\frac{\nabla_{\alpha}K}{K}\right) - (u\cdot K\cdot u)^{2}\bigg)+  \frac{2u^\mu}{K^{2}}\nabla^{2}(\nabla^{2}u_{\mu}) \nonumber \\
&\quad - 2u^\mu\nabla_{\mu}\left(\frac{u\cdot\nabla K}{K}\right) - 2u^\mu\left(\frac{\nabla^{2}u_{\mu}}{K}\right)\left(-2(u\cdot K\cdot u) + 4\frac{u\cdot\nabla K}{K}  - \frac{K}{D}\right)+u^\mu u^\nu \nabla_{\mu}\left(\frac{\nabla^{2}u_{\nu}}{K}\right)\nonumber\\&\quad   +u^\mu u^\nu  \nabla_{\mu}(u^{\alpha}K_{\alpha\nu})  - u^\mu u^\nu \nabla_{\mu}\left(\frac{\nabla_{\nu}K}{K}\right)-u^\mu u^\nu \left(\frac{\nabla^{2}u_{\mu}}{K}\right)\left(\frac{\nabla^{2}u_{\nu}}{K}\right)- K^{\alpha\beta}(\nabla_{\alpha}u_{\beta})\notag\\
&\quad - \frac{2}{K}u_{\alpha}K^{\alpha\beta}\nabla^{2}u_{\beta} - \frac{\nabla^2 u_\mu \nabla^2 u_\nu}{2K^2}  P^{\mu\nu}\bigg]+ \mathcal{O}(1/D^2).
\end{align}
\subsection{Calculation for shear part}\label{Calculation for Shear part}
The expression of traceless shear stress tension is given in Eq.~\eqref{PimunuRaw}. Here, we perform the term-by-term analysis of it, which is a necessary prerequisite for extracting the transport coefficients.
\subsubsection*{1. Derivation of $(\nabla^\sigma U_{\langle \mu})(\nabla_{|\sigma|} U_{\nu \rangle})$}
\begin{align}
    (\nabla^\sigma U_{\alpha})(\nabla_{\sigma} U_{\beta})\Delta^\alpha_\mu\Delta^\beta_\nu&=(\nabla^\sigma U_\mu)(\nabla_\sigma U_\nu),\notag\\ &= g^{\sigma\rho} (\nabla_\rho U_\mu)(\nabla_\sigma U_\nu), \notag \\
    &= \left(\Delta^{\sigma\rho} - U^\sigma U^\rho\right) (\nabla_\rho U_\mu)(\nabla_\sigma U_\nu), \notag \\
    &= \Delta^{\sigma\rho} (\nabla_\rho U_\mu)(\nabla_\sigma U_\nu) - (U^\rho \nabla_\rho U_\mu)(U^\sigma \nabla_\sigma U_\nu), \notag \\
    &= \Delta^{\sigma\rho} (\nabla_\rho U_\mu)(\nabla_\sigma U_\nu) - A_\mu A_\nu. 
\end{align}
The velocity gradient decomposes exactly as:
\begin{equation}
    \nabla_\sigma U_\alpha = -A_\alpha U_\sigma + \sigma_{\sigma\alpha} + \omega_{\sigma\alpha} + \frac{\theta}{D-2}\Delta_{\sigma\alpha}.
\end{equation}
Now,
\begin{align}
    \Delta^{\sigma\rho} (\nabla_\rho U_\mu)(\nabla_\sigma U_\nu) &= \left( \sigma^\sigma_{\ \mu} + \omega^\sigma_{\ \mu} + \frac{\theta}{D-2}\Delta^\sigma_\mu \right) \left( \sigma_{\sigma\nu} + \omega_{\sigma\nu} + \frac{\theta}{D-2}\Delta_{\sigma\nu} \right), \notag \\
    &= \sigma^\sigma_{\ \mu}\sigma_{\sigma\nu} + \sigma^\sigma_{\ \mu}\omega_{\sigma\nu} + \frac{\theta}{D-2}\sigma_{\mu\nu} + \omega^\sigma_{\ \mu}\sigma_{\sigma\nu} + \omega^\sigma_{\ \mu}\omega_{\sigma\nu} \notag \\
    &\quad + \frac{\theta}{D-2}\omega_{\mu\nu} + \frac{\theta}{D-2}\sigma_{\mu\nu} + \frac{\theta}{D-2}\omega_{\mu\nu} + \frac{\theta^2}{(D-2)^2}\Delta_{\mu\nu}.
\end{align}
\begin{align}
   (\nabla^\sigma U_{\langle \mu})(\nabla_{|\sigma|} U_{\nu \rangle}) &= \left[ \Delta^{\sigma\rho} (\nabla_\rho U_\mu)(\nabla_\sigma U_\nu) - A_\mu A_\nu \right]_{\langle \mu\nu \rangle}, \notag \\
    &= \sigma_{\ \langle \mu}^{\sigma} \sigma_{|\sigma|\nu\rangle} + \omega^{ \sigma}_{\ \langle \mu} \omega_{|\sigma|\nu\rangle} + (\sigma_{\ \mu}^{ \sigma} \omega_{\sigma\nu} + \omega_{\ \mu}^{ \sigma} \sigma_{\sigma\nu}) \notag \\
    &\quad + \frac{2\theta}{D-2}\sigma_{\mu\nu} - A_{\langle \mu} A_{\nu \rangle},\notag\\
    &= \sigma_{\ \langle \mu}^{ \sigma} \sigma_{|\sigma|\nu\rangle} + \omega_{\ \langle \mu}^{ \sigma} \omega_{|\sigma|\nu\rangle} + 2 \sigma_{\ \langle\mu}^{\sigma} \omega_{|\sigma|\nu\rangle} \notag \\
    &\quad - A_{\langle \mu} A_{\nu \rangle} +\mathcal{O}(1/D).
\end{align}
\subsubsection*{2. Derivation of the spatially projected Laplacian $(\nabla^2 U)^\perp_\nu$}
\begin{align}
    \nabla^2 U_\nu &= \nabla^\lambda (\nabla_\lambda U_\nu), \notag \\
    &= \nabla^\lambda \left( -A_\nu U_\lambda + \sigma_{\lambda\nu} + \omega_{\lambda\nu} + \frac{\theta}{D-2}\Delta_{\lambda\nu} \right), \notag \\
    &= - \nabla^\lambda (A_\nu U_\lambda) + \nabla^\lambda \sigma_{\lambda\nu} + \nabla^\lambda \omega_{\lambda\nu} + \frac{1}{D-2}\nabla^\lambda(\theta \Delta_{\lambda\nu}).\label{Exp:Lap:U} 
\end{align}
Now we separately evaluate these two specific components
\begin{align}
    - \nabla^\lambda (A_\nu U_\lambda) &= - (U^\lambda \nabla_\lambda A_\nu) - A_\nu (\nabla^\lambda U_\lambda) = - \dot{A}_\nu - \theta A_\nu, \\
    \nabla^\lambda(\theta \Delta_{\lambda\nu}) &= \nabla^\perp_\nu \theta + \theta (\nabla^\lambda \Delta_{\lambda\nu}) = \nabla^\perp_\nu \theta + \theta (U_\nu \theta + A_\nu).
\end{align}
Substitution of the above terms in Eq.~\eqref{Exp:Lap:U} yields
\begin{align}
    (\nabla^2 U)^\perp_\nu &\equiv \Delta_\nu^\rho \nabla^2 U_\rho, \notag \\
    &= \Delta_\nu^\rho \left( - \dot{A}_\rho - \theta A_\rho + \nabla^\lambda \sigma_{\lambda\rho} + \nabla^\lambda \omega_{\lambda\rho} + \frac{1}{D-2} \left[ \nabla^\perp_\rho \theta + \theta^2 U_\rho + \theta A_\rho \right] \right), \notag \\
    &= - \dot{A}^\perp_\nu - \theta A_\nu + (\nabla \cdot \sigma)^\perp_\nu + (\nabla \cdot \omega)^\perp_\nu + \frac{1}{D-2} \left( \nabla^\perp_\nu \theta + \theta A_\nu \right), \notag \\
    &= (\nabla \cdot \sigma)^\perp_\nu + (\nabla \cdot \omega)^\perp_\nu - \dot{A}^\perp_\nu - \left(\frac{D-3}{D-2}\right)\theta A_\nu + \frac{1}{D-2}\nabla^\perp_\nu \theta.\label{Exp:Lap:U:Perp}
\end{align}
\subsubsection*{3. Derivation of $U_\sigma \nabla^2 U^\sigma$}
As $U^\mu$ is a timelike vector ($U^\mu U_\mu=-1$) we have
\begin{align}
    \nabla_\lambda (U_\sigma U^\sigma) &= 2 U^\sigma \nabla_\lambda U_\sigma = 0. \notag
\end{align}
Due to this identity, the following differential also vanishes
\begin{align}
    \nabla^\lambda (U^\sigma \nabla_\lambda U_\sigma) &= (\nabla^\lambda U^\sigma)(\nabla_\lambda U_\sigma) + U^\sigma \nabla^\lambda \nabla_\lambda U_\sigma = 0 \notag \\
    &\implies U_\sigma \nabla^2 U^\sigma = - (\nabla^\lambda U^\sigma)(\nabla_\lambda U_\sigma).
\end{align}
Now,
\begin{align}
    (\nabla^\lambda U^\sigma)(\nabla_\lambda U_\sigma) &= g^{\lambda\rho}(\nabla_\lambda U^\sigma)(\nabla_\rho U_\sigma),\notag \\
    &= \left(\Delta^{\lambda\rho} - U^\lambda U^\rho\right)(\nabla_\lambda U^\sigma)(\nabla_\rho U_\sigma), \notag \\
    &= \Delta^{\lambda\rho}(\nabla_\lambda U^\sigma)(\nabla_\rho U_\sigma) - (U^\lambda \nabla_\lambda U^\sigma)(U^\rho \nabla_\rho U_\sigma), \notag \\
    &= \Delta^{\lambda\rho} \left( \sigma^\sigma_{\ \lambda} + \omega^\sigma_{\ \lambda} + \frac{\theta}{D-2}\Delta^\sigma_\lambda \right) \left( \sigma_{\sigma\rho} + \omega_{\sigma\rho} + \frac{\theta}{D-2}\Delta_{\sigma\rho} \right) - A^\sigma A_\sigma, \notag \\
    &= \sigma^2 + \omega^2 + \frac{\theta^2}{D-2} - A^2.
\end{align}
This leads to
\begin{align}
    U_\sigma \nabla^2 U^\sigma &= - \left( \sigma^2 + \omega^2 + \frac{\theta^2}{D-2} - A^2 \right), \notag \\
    &= A^2 - \sigma^2 - \omega^2 - \frac{\theta^2}{D-2}=\mathcal{O}(1).\label{Exp:Lap:U:Parallel}
\end{align}
At leading order, we get
\begin{equation}
    (U_\sigma \nabla^2 U^\sigma) \sigma_{\mu\nu} = \left( A^2 - \sigma^2 - \omega^2 \right) \sigma_{\mu\nu} +\mathcal{O}(1/D)=\mathcal{O}(1).
\end{equation}
\noindent\textbf{4. Derivation of $\nabla_{\langle\mu}(\nabla^2U_{\nu\rangle})$}
\begin{eqnarray}
\nabla_{\langle\mu}(\nabla^2U_{\nu\rangle}) &=& \left[ \nabla_\mu \left( (\nabla^2 U)^\perp_\nu - U_\nu (U \cdot \nabla^2 U) \right) \right]_{\langle \mu \nu \rangle}, \nonumber \\
&=& \nabla^\perp_{\langle\mu}(\nabla^2 U)^\perp_{\nu\rangle} - (U \cdot \nabla^2 U)\sigma_{\mu\nu}, \nonumber \\
&=& \nabla_{\langle\mu}^{\perp}(\nabla\cdot\sigma)_{\nu\rangle}^{\perp} + \nabla_{\langle\mu}^{\perp}(\nabla\cdot\omega)_{\nu\rangle}^{\perp}+ \mathcal{O}(1). \label{eq:A18}
\end{eqnarray}
We used Eq.~\eqref{Exp:Lap:U:Perp} and \eqref{Exp:Lap:U:Parallel} to obtain the final expression.
  \subsubsection*{5. Derivation of $\nabla_{\langle \mu} K (\nabla^2 U_{\nu \rangle})$}
\begin{align}
    \nabla_{\langle \mu} K (\nabla^2 U_{\nu \rangle}) &= \nabla^\perp_{\langle \mu} K (\nabla^2 U)^\perp_{\nu \rangle}, \notag \\
    &= \nabla^\perp_{\langle \mu} K \left[ (\nabla \cdot \sigma)^\perp + (\nabla \cdot \omega)^\perp \right]_{\nu \rangle} +\mathcal{O}(D),\notag \\
    &= \nabla^\perp_{\langle \mu} K (\nabla \cdot \sigma)^\perp_{\nu \rangle} + \nabla^\perp_{\langle \mu} K (\nabla \cdot \omega)^\perp_{\nu \rangle} + \mathcal{O}(D).
\end{align}
\subsubsection*{6. Derivation of $\nabla^2 U_{\langle \mu} \nabla^2 U_{\nu \rangle}$}
\begin{align}
    \nabla^2 U_{\langle \mu} \nabla^2 U_{\nu \rangle} &= (\nabla^2 U)^\perp_{\langle \mu} (\nabla^2 U)^\perp_{\nu \rangle}, \notag \\
    &= \left[ (\nabla \cdot \sigma)^\perp + (\nabla \cdot \omega)^\perp  \right]_{\langle \mu} \times \left[ (\nabla \cdot \sigma)^\perp+ (\nabla \cdot \omega)^\perp \right]_{\nu \rangle} + \mathcal{O}(D),\notag
\\ &= (\nabla \cdot \sigma)^\perp_{\langle \mu} (\nabla \cdot \sigma)^\perp_{\nu \rangle} + (\nabla \cdot \omega)^\perp_{\langle \mu} (\nabla \cdot \omega)^\perp_{\nu \rangle}+ 2 (\nabla \cdot \sigma)^\perp_{\langle \mu} (\nabla \cdot \omega)^\perp_{\nu \rangle} + \mathcal{O}(D).
\end{align}
\subsubsection*{7. Derivation of  $A_{\langle \mu} \nabla^2 U_{\nu \rangle}$}
\begin{align}
    A_{\langle \mu} (\nabla^2 U_{\nu \rangle}) &= A_{\langle \mu} (\nabla^2 U)^\perp_{\nu \rangle}, \notag \\
    &=A_{\langle \mu} \left[ (\nabla \cdot \sigma)^\perp + (\nabla \cdot \omega)^\perp \right]_{\nu \rangle} +\mathcal{O}(1),\notag \\
    &= A_{\langle \mu}  (\nabla \cdot \sigma)^\perp_{\nu \rangle} + A_{\langle \mu}  (\nabla \cdot \omega)^\perp_{\nu \rangle} + \mathcal{O}(1).
\end{align}
Using the above relation, we get the expression of shear stress tensor as Eq.~\eqref{Exp:pi:mu:nu}.
\subsection{Calculation for bulk part}\label{Bulk_Part}
Taking Eq.~\eqref{pigeo} as a starting point in these section we substitute the landau velocity for the membrane velocity field and perform some necessary transformation, which is a necessary prerequisite for extracting transport coefficient from bulk pressure.
First, we deduce some geometric identities as follows:
\begin{align*}
&[\nabla_\gamma, \nabla_\mu] u^\gamma = R_{\lambda\mu} u^\lambda, \\
\Rightarrow& \nabla_\gamma \nabla_\mu u^\gamma = R_{\mu\lambda} u^\lambda + \nabla_\mu \nabla_\gamma u^\gamma.
\end{align*}
Using Gauss-Codazzi relation $R_{\mu\lambda}=K  K_{\mu\lambda} - K_{\mu\sigma} K^\sigma_{~\lambda}$ we have,
\begin{align*}
\nabla_\gamma \nabla_\mu u^\gamma &= (K K_{\mu\lambda} - K_{\mu\sigma} K^\sigma_{~\lambda}) u^\lambda + \nabla_\mu \nabla_\gamma u^\gamma, \\
&= K K_{\mu\lambda} u^\lambda - K_{\mu\sigma} K^\sigma_{~\lambda} u^\lambda + \nabla_\mu (\nabla \cdot u), \\
\Rightarrow K_{\mu\lambda} u^\lambda &= \frac{\nabla_\gamma \nabla_\mu u^\gamma}{K} + \frac{K_{\mu\sigma} K^\sigma_{~\lambda} u^\lambda}{K} - \frac{\nabla_\mu (\nabla \cdot u)}{K}. 
\end{align*}
Substituting the leading order value of $K_{\mu\lambda} u^\lambda$ at the subleading terms, and using $\nabla \cdot u \sim 1/D$ we get
\begin{align*}
 K_{\mu\lambda} u^\lambda &= \frac{\nabla_\gamma \nabla_\mu u^\gamma}{K} + K_{\mu\sigma} \frac{\nabla_\alpha \nabla^\sigma u^\alpha}{K^2} + \mathcal{O}(1/D^2). \\
u^\mu K_{\mu\lambda} u^\lambda &= \frac{u^\mu}{K} \nabla_\gamma \nabla_\mu u^\gamma + \frac{(\nabla_\alpha \nabla_\sigma u^\alpha)(\nabla_\gamma \nabla^\sigma u^\gamma)}{K^3} + \mathcal{O}(1/D^2), \\
&= \frac{1}{K} \nabla_\gamma (u^\mu \nabla_\mu u^\gamma) - \frac{1}{K} (\nabla_\gamma u^\mu)(\nabla_\mu u^\gamma) + \frac{(\nabla_\alpha \nabla_\sigma u^\alpha)(\nabla_\gamma \nabla^\sigma u^\gamma)}{K^3} + \mathcal{O}(1/D^2). 
\end{align*}
Now, using above expression of $ K_{\mu\lambda} u^\lambda$ and the identity $\nabla_\alpha K^{\alpha\beta}=\nabla^\beta K$, the expression of the following term becomes:
\begin{align*}
\frac{1}{8\pi K} K^{\alpha\beta} \nabla_\alpha u_\beta &= \frac{1}{8\pi K} \left[ \nabla_\alpha (K^{\alpha\beta} u_\beta) - u_\beta \nabla_\alpha K^{\alpha\beta} \right], \\
&= \frac{1}{8\pi K} \left[ \nabla_\alpha (K^{\alpha\beta} u_\beta) - u \cdot \nabla K \right], \\
&= \frac{1}{8\pi K} \left[ \nabla_\alpha \left( \frac{\nabla_\gamma \nabla^\alpha u^\gamma}{K} + K^\alpha_{~\sigma} \frac{\nabla_\delta \nabla^\sigma u^\delta}{K^2} \right) - u \cdot \nabla K \right] + \mathcal{O}(1/D^2), \\
&= \frac{1}{8\pi K} \left[ \nabla_\alpha \left( \frac{\nabla_\gamma \nabla^\alpha u^\gamma}{K} \right) - u \cdot \nabla K \right] + \frac{\nabla_\sigma K}{8\pi K} \frac{\nabla_\delta \nabla^\sigma u^\delta}{K^2} + \mathcal{O}(1/D^2).
\end{align*}
Using the above identities, the expression of bulk pressure can be modified as follows:
\begin{align}\label{Bulk:Geo}
\Pi &= -\frac{1}{16\pi K} \nabla_{\alpha} (u \cdot \nabla u^{\alpha}) + \frac{1}{8\pi K} \Biggl[ \nabla_{\alpha} \left( \frac{\nabla_\gamma \nabla^\alpha u^\gamma}{K} \right) - u \cdot \nabla K+\frac{1}{2} (\nabla_\mu u^\nu)(\nabla_\nu u^\mu)\\
&\quad - \frac{1}{2} \frac{(\nabla_\alpha \nabla_\sigma u^\alpha)(\nabla_\beta \nabla^\sigma u^\beta)}{K^2}  + \frac{\nabla_\sigma K \nabla_\delta \nabla^\sigma u^\delta}{K^2} - \frac{\nabla_\gamma\nabla^\beta u^\gamma}{K} \left( \frac{\nabla_{\beta} K}{K} - \frac{2 \nabla^2 u_{\beta}}{K} \right)  \notag\\
&\quad - \left( \frac{u \cdot \nabla K}{K} - \frac{\nabla_{\mu} (u \cdot \nabla u^{\mu})}{2K} - \frac{K}{2D} \right) \frac{\nabla_{\nu} (u \cdot \nabla u^{\nu})}{K} \Biggr]\notag \\
&\quad - \frac{1}{8\pi (D-2)} \left( \frac{\nabla^2 K}{2 K^2} - \frac{u \cdot \nabla K}{K} + 2 \frac{u_{\alpha} \nabla^2 u^{\alpha}}{K} + \frac{\nabla_{\alpha} (u \cdot \nabla u^{\alpha})}{K} \right) + \widetilde{\pi} + \mathcal{O}(1/D^2).
\end{align}
Expanding the following term, we substitute the membrane velocity field $u^\mu$ in the form of the Landau velocity as, $u^\alpha = U^\alpha - l^\alpha/e$, and putting their leading order value as $l^{\alpha} = -\frac{\nabla^2 U^{\sigma}}{2K} \Delta_{\sigma}^{\alpha}$ and $e = \frac{K}{2}$ yields,
\begin{align}
-\frac{1}{16\pi K} \nabla_{\alpha} (u^{\beta} \nabla_{\beta} u^{\alpha}) &= -\frac{1}{16\pi K} \nabla_{\alpha} \left( \left( U^{\beta} - \frac{l^{\beta}}{e} \right) \nabla_{\beta} \left( U^{\alpha} - \frac{l^{\alpha}}{e} \right) \right), \nonumber \\
&= -\frac{1}{16\pi K} \nabla_{\alpha} \left( U^{\beta} \nabla_{\beta} U^{\alpha} - U^{\beta} \nabla_{\beta} \frac{l^{\alpha}}{e} - \frac{l^{\beta}}{e} \nabla_{\beta} U^{\alpha} \right) + \mathcal{O}(1/D^2), \nonumber \\
&= -\frac{1}{16\pi K} \nabla_{\alpha} (U^{\beta} \nabla_{\beta} U^{\alpha}) - \frac{1}{16\pi K} \nabla_{\alpha} \left( U^{\beta} \nabla_{\beta} \left( \frac{\nabla^2 U^{\sigma}}{K^2} \Delta_{\sigma}^{\alpha} \right) \right), \nonumber \\
&\quad - \frac{1}{16\pi K} \nabla_{\alpha} \left( \frac{\nabla^2 U^{\sigma}}{K^2} \Delta_{\sigma}^{\beta} \nabla_{\beta} U^{\alpha} \right) + \mathcal{O}(1/D^2),\nonumber \\
&\equiv - \frac{1}{16\pi K} \nabla_\alpha A^\alpha + \mathcal{T}_1 + \mathcal{T}_2 + \mathcal{O}(1/D^2) \,,
\end{align}
where $A^\alpha = U^\beta \nabla_\beta U^\alpha$, and we define the subleading corrections $\mathcal{T}_1$ and $\mathcal{T}_2$. To evaluate $\mathcal{T}_1$, we expand the inner derivative acting on the projector. Recognizing that $U_\sigma \nabla^2 U^\sigma \sim \mathcal{O}(1)$ in the large $D$ limit, the tensor structures reduce to:
\begin{align*}
\nabla_\beta \left( \frac{\nabla^2 U^\sigma}{K^2} \Delta^\alpha_\sigma \right) &= \Delta^\alpha_\sigma \nabla_\beta \left( \frac{\nabla^2 U^\sigma}{K^2} \right) + \frac{\nabla^2 U^\sigma}{K^2} (U^\alpha \nabla_\beta U_\sigma + U_\sigma \nabla_\beta U^\alpha), \nonumber \\
&= \Delta^\alpha_\sigma \frac{\nabla_\beta \nabla^2 U^\sigma}{K^2} - \Delta^\alpha_\sigma \frac{2}{K^3} \nabla^2 U^\sigma \nabla_\beta K + \frac{\nabla^2 U^\sigma}{K^2} U^\alpha \nabla_\beta U_\sigma + \mathcal{O}(1/D^2) \,.
\end{align*}
Substituting this expansion back into $\mathcal{T}_1$ and applying the trace scaling $\nabla_\alpha U^\alpha \sim \mathcal{O}(1)$, we bypass subleading product rule evaluations to find:
\begin{align}
\mathcal{T}_1 &= - \frac{1}{16\pi K} \nabla_\alpha \left( \Delta^\alpha_\sigma U^\beta \frac{\nabla_\beta \nabla^2 U^\sigma}{K^2} - \Delta^\alpha_\sigma \frac{2}{K^3} \nabla^2 U^\sigma U^\beta \nabla_\beta K + \frac{\nabla^2 U^\sigma}{K^2} U^\alpha U^\beta \nabla_\beta U_\sigma \right) + \mathcal{O}(1/D^2), \nonumber \\
&= - \frac{1}{16\pi K} \left[ \nabla_\sigma \left( \frac{U^\beta \nabla_\beta \nabla^2 U^\sigma}{K^2} \right) - 2 \nabla_\sigma \left( \frac{\nabla^2 U^\sigma}{K^3} U^\beta \nabla_\beta K \right) \right] + \mathcal{O}(1/D^2), \nonumber \\
&= - \frac{U^\beta}{16\pi K^3} \nabla_\sigma (\nabla_\beta \nabla^2 U^\sigma) + \frac{U^\beta \nabla_\beta K}{8\pi K^4} \nabla_\sigma \nabla^2 U^\sigma + \mathcal{O}(1/D^2) \,.
\end{align}
For the second term $\mathcal{T}_2$, bypassing the scalar gradients which strictly fall to $\mathcal{O}(1/D^2)$ yields:
\begin{align}
\mathcal{T}_2 &= - \frac{1}{16\pi K} \nabla_\alpha \left( \frac{\nabla^2 U^\sigma}{K^2} \Delta^\beta_\sigma \nabla_\beta U^\alpha \right), \nonumber \\
&= - \frac{1}{16\pi K} \Delta^\beta_\sigma \frac{\nabla^2 U^\sigma}{K^2} \nabla_\alpha \nabla_\beta U^\alpha, \nonumber \\
&= - \frac{1}{16\pi K^3} \nabla^2 U^\sigma \nabla_\alpha \nabla_\sigma U^\alpha + \mathcal{O}(1/D^2) \,.
\end{align}
Combining all derived components into the parent equation, the final truncated sum is
\begin{align}\label{Bulk1st}
-\frac{1}{16\pi K} \nabla_{\alpha} (u^{\beta} \nabla_{\beta} u^{\alpha}) &= - \frac{1}{16\pi K} \nabla_\alpha A^\alpha - \frac{U^\beta}{16\pi K^3} \nabla_\sigma (\nabla_\beta \nabla^2 U^\sigma) \nonumber \\
&\quad + \frac{U^\beta \nabla_\beta K}{8\pi K^4} \nabla_\sigma \nabla^2 U^\sigma - \frac{1}{16\pi K^3} \nabla^2 U^\sigma \nabla_\alpha \nabla_\sigma U^\alpha + \mathcal{O}(1/D^2) \,.
\end{align}
Now we decompose the following quantity:
\begin{align}
    \nabla_\alpha \nabla_\gamma \nabla^\alpha \left( \frac{l^\gamma}{K} \right) 
    &= \nabla_\alpha \nabla_\gamma \left( -l^\gamma \frac{\nabla^\alpha K}{K^2} + \frac{\nabla^\alpha l^\gamma}{K} \right) \notag\\
    &= \nabla_\alpha \left( -\nabla_\gamma l^\gamma \frac{\nabla^\alpha K}{K^2} + \frac{\nabla_\gamma \nabla^\alpha l^\gamma}{K} \right) + \mathcal{O}(1) \notag\\
    &= -\frac{\nabla_\gamma l^\gamma}{K^2} \nabla^2 K + \frac{\nabla_\alpha \nabla_\gamma \nabla^\alpha l^\gamma}{K} + \mathcal{O}(1).
    \end{align}
Substituting $l^\gamma = -\frac{\nabla^2 U^\sigma \Delta^\gamma_\sigma}{2K}$, we have
\begin{align}
    \nabla_\alpha \nabla_\gamma \nabla^\alpha \left( \frac{l^\gamma}{K} \right) 
    &= \frac{\nabla_\gamma (\nabla^2 U^\sigma \Delta^\gamma_\sigma)}{2K^3} \nabla^2 K - \frac{1}{K} \nabla_\alpha \nabla_\gamma \nabla^\alpha \left( \frac{\nabla^2 U^\sigma \Delta^\gamma_\sigma}{2K} \right)+ \mathcal{O}(1).
\end{align}
Evaluating the inner tensor components yields
\begin{align}
    \nabla_\gamma (\nabla^2 U^\sigma \Delta^\gamma_\sigma) 
    &= \Delta^\gamma_\sigma \nabla_\gamma \nabla^2 U^\sigma + U_\sigma \nabla^2 U^\sigma \nabla_\gamma U^\gamma \notag\\
    &= \nabla_\sigma \nabla^2 U^\sigma + \mathcal{O}(1), 
\end{align}
and
\begin{align}
    \nabla_\alpha \nabla_\gamma \nabla^\alpha \left( \frac{\nabla^2 U^\sigma}{2K} \Delta^\gamma_\sigma \right) 
    &= \frac{1}{2} \nabla_\alpha \nabla_\gamma \left[ \nabla^\alpha \left( \frac{\nabla^2 U^\sigma}{K} \right) \Delta^\gamma_\sigma + \frac{\nabla^2 U^\sigma}{K} (U_\sigma \nabla^\alpha U^\gamma + U^\gamma \nabla^\alpha U_\sigma) \right], \notag\\
    &= \frac{1}{2} \nabla_\alpha \nabla_\gamma \left[ \left( \frac{1}{K} \nabla^\alpha \nabla^2 U^\sigma - \frac{1}{K^2} \nabla^2 U^\sigma \nabla^\alpha K \right) \Delta^\gamma_\sigma + \frac{\nabla^2 U^\sigma}{K} U^\gamma \nabla^\alpha U_\sigma \right] + \mathcal{O}(D). \notag
\end{align}
As $\nabla \cdot U \sim \mathcal{O}(1),$ at leading order we only get the following combination. Hence, we obtain
\begin{align}
    \nabla_\alpha \nabla_\gamma \nabla^\alpha \left( \frac{\nabla^2 U^\sigma}{2K} \Delta^\gamma_\sigma \right) 
    &= \frac{1}{2} \nabla_\alpha \left[ \frac{1}{K} \nabla_\sigma \nabla^\alpha \nabla^2 U^\sigma - \frac{1}{K^2} (\nabla_\sigma \nabla^2 U^\sigma) \nabla^\alpha K \right] + \mathcal{O}(D), \notag\\
    &= \frac{1}{2K} \nabla_\alpha \nabla_\sigma \nabla^\alpha \nabla^2 U^\sigma - \frac{1}{2K^2} (\nabla_\sigma \nabla^2 U^\sigma )\nabla^2 K + \mathcal{O}(D).
\end{align}
Assembling these derivatives, we obtain
\begin{align}\label{Expression1}
    \nabla_\alpha \nabla_\gamma \nabla^\alpha \left( \frac{l^\gamma}{K} \right) 
    &= \frac{\nabla_\sigma (\nabla^2 U^\sigma)}{K^3} \nabla^2 K - \frac{1}{2K^2} \nabla_\alpha \nabla_\sigma \nabla^\alpha \nabla^2 U^\sigma + \mathcal{O}(1).
\end{align}
Now, we decompose the following term by substituting $u^\gamma = U^\gamma - \frac{l^\gamma}{e}$ and $l^\beta = - \frac{\nabla^2 U^\sigma}{2K} \Delta_\sigma^\beta$, we find
\begin{align}\label{Bulk2nd}
    &\frac{1}{8\pi K} \nabla_\alpha \left( \frac{\nabla_\gamma \nabla^\alpha u^\gamma}{K} \right) - \frac{u \cdot \nabla K}{8\pi K} \notag\\
    &= \frac{1}{8\pi K} \nabla_\alpha \left( \frac{\nabla_\gamma \nabla^\alpha U^\gamma}{K} \right) - \frac{1}{8\pi K} \nabla_\alpha \left( \frac{\nabla_\gamma \nabla^\alpha \left(\frac{l^\gamma}{e}\right)}{K} \right)- \frac{U \cdot \nabla K}{8\pi K} - \frac{(\nabla^2 U^\sigma) \Delta_\sigma^\beta \nabla_\beta K}{8\pi K^3} + \mathcal{O}(1/D^2), \notag\\
    &= \frac{1}{8\pi K} \nabla_\alpha \left( \frac{\nabla_\gamma \nabla^\alpha U^\gamma}{K} \right) - \frac{U \cdot \nabla K}{8\pi K} - \frac{(\nabla^2 U^\sigma) \nabla_\sigma K}{8\pi K^3}  - \frac{U_\sigma \nabla^2 U^\sigma \dot{K}}{8\pi K^3}\notag\\
    &\quad - \frac{\nabla_\alpha \nabla_\gamma \nabla^\alpha \left( \frac{l^\gamma}{K} \right)}{4\pi K^2} + \mathcal{O}(1/D^2). \notag\\
    &\text{Using Eq.~\eqref{Expression1}, we obtain,}\\
    &\frac{1}{8\pi K} \nabla_\alpha \left( \frac{\nabla_\gamma \nabla^\alpha u^\gamma}{K} \right) - \frac{u \cdot \nabla K}{8\pi K}\notag\\
    &= \frac{1}{8\pi K} \nabla_\alpha \left( \frac{\nabla_\gamma \nabla^\alpha U^\gamma}{K} \right) - \frac{U \cdot \nabla K}{8\pi K} - \frac{(\nabla^2 U^\sigma) \nabla_\sigma K}{8\pi K^3}  - \frac{\nabla_\sigma (\nabla^2 U^\sigma) \nabla^2 K}{4\pi K^5} \notag\\
    &\quad + \frac{1}{8\pi K^4} \nabla_\alpha \nabla_\sigma \nabla^\alpha \nabla^2 U^\sigma + \mathcal{O}(1/D^2).
\end{align}
By substituting  $u_\mu$ in terms of $U_\mu$ using Eq~\eqref{LandauVelocity} and the expression of $\tilde{\pi}$ using Eq.~\eqref{Exp:pi:tilde} and substituting Eq.~\eqref{Bulk1st}~and~\eqref{Bulk2nd} in Eq.~\eqref{Bulk:Geo} we get
\begin{align}\label{pi:raw}
    \Pi=&- \frac{1}{16\pi K} \nabla_\alpha A^\alpha - \frac{U^\beta}{16\pi K^3} \nabla_\sigma (\nabla_\beta \nabla^2 U^\sigma)  + \frac{U^\beta \nabla_\beta K}{8\pi K^4} \nabla_\sigma \nabla^2 U^\sigma - \frac{1}{16\pi K^3} \nabla^2 U^\sigma \nabla_\alpha \nabla_\sigma U^\alpha \nonumber \\&\frac{1}{8\pi K} \nabla_\alpha \left( \frac{\nabla_\gamma \nabla^\alpha U^\gamma}{K} \right) - \frac{U \cdot \nabla K}{8\pi K} - \frac{(\nabla^2 U^\sigma) \nabla_\sigma K}{8\pi K^3}- \frac{\nabla_\sigma (\nabla^2 U^\sigma) \nabla^2 K}{4\pi K^5} + \frac{1}{8\pi K^4} \nabla_\alpha \nabla_\sigma \nabla^\alpha \nabla^2 U^\sigma\nonumber \\
&\quad+ \frac{1}{8\pi K} \Biggl[\frac{1}{2} (\nabla_\mu U^\nu)(\nabla_\nu U^\mu)- \frac{1}{2} \frac{(\nabla_\alpha \nabla_\sigma U^\alpha)(\nabla_\beta \nabla^\sigma U^\beta)}{K^2}  + \frac{\nabla_\sigma K \nabla_\delta \nabla^\sigma U^\delta}{K^2}\notag\\
&\quad  - \frac{\nabla_\gamma\nabla^\beta U^\gamma}{K} \left( \frac{\nabla_{\beta} K}{K} - \frac{2 \nabla^2 U_{\beta}}{K} \right) - \left( \frac{U \cdot \nabla K}{K} - \frac{\nabla_{\mu} (U \cdot \nabla U^{\mu})}{2K} - \frac{K}{2D} \right) \frac{\nabla_{\nu} (U \cdot \nabla U^{\nu})}{K} \Biggr]\notag \\
&\quad - \frac{1}{8\pi (D-2)} \left( \frac{\nabla^2 K}{2 K^2} - \frac{U \cdot \nabla K}{K} + 2 \frac{U_{\alpha} \nabla^2 U^{\alpha}}{K} + \frac{\nabla_{\alpha} (U \cdot \nabla U^{\alpha})}{K} \right) \notag\\ &\quad- \frac{\theta}{8\pi(D-2)}+ \frac{\nabla_{\alpha} A^{\alpha}}{8\pi K (D-2)} - \frac{\nabla_\alpha (\nabla^2 U^\alpha)}{8\pi K^2 (D-2)} + \mathcal{O}(1/D^2). 
\end{align}
Now, we evaluate the term 
\begin{align}
\nabla_{\gamma} \nabla^{\alpha} U^{\gamma} &= \nabla_{\gamma} \left( \sigma^{\alpha \gamma} + \omega^{\alpha \gamma} - U^{\alpha} A^{\gamma} + \frac{\theta}{(D-2)} \Delta^{\alpha \gamma} \right), \nonumber \\
&= \nabla_{\gamma} \sigma^{\alpha \gamma} + \nabla_{\gamma} \omega^{\alpha \gamma} - U^{\alpha} \nabla \cdot A - A^{\gamma} (\sigma_{\gamma}^{\;\;\alpha} + \omega_{\gamma}^{\;\;\alpha} - U_{\gamma} A^{\alpha}) + \mathcal{O}(1/D), \nonumber \\
&= \nabla_{\gamma} \sigma^{\alpha \gamma} + \nabla_{\gamma} \omega^{\alpha \gamma} - U^{\alpha} \nabla \cdot A - A^{\gamma} (\sigma_{\gamma}^{\;\;\alpha} + \omega_{\gamma}^{\;\;\alpha}) + \mathcal{O}(1/D).
\end{align}
From Appendix~\ref{Calculation for Shear part}, we have
\begin{align}
U_{\alpha} \nabla^{2} U^{\alpha} &= \left( A^{2} - \sigma^{2} - \omega^{2} - \frac{\theta^{2}}{D-2}\right)  =\mathcal{O}(1),\nonumber \\
\quad \nabla^{2} U^{\alpha} &= \nabla^{\lambda} \sigma_{\lambda}^{\;\;\alpha} + \nabla^{\lambda} \omega_{\lambda}^{\;\;\alpha} + \mathcal{O}(1).
\end{align}
The explicit expressions of the following term at leading order are evaluated as
\begin{align*}\label{KinematicIdentity}
&\quad \frac{1}{2} (\nabla_\mu U^\nu)(\nabla_\nu U^\mu) - \frac{1}{2 K^2} (\nabla_\alpha \nabla_\rho U^\alpha)(\nabla_\beta \nabla^\rho U^\beta) + \frac{1}{K^2} \nabla_\rho K (\nabla_\alpha \nabla^\rho U^\alpha) \nonumber \\
&= \frac{1}{2} \left( \sigma_{\mu\nu}\sigma^{\mu\nu} - \omega_{\mu\nu}\omega^{\mu\nu} + \frac{\theta^2}{D-2} \right) \nonumber \\
&\quad - \frac{1}{2 K^2} \left[ \nabla_\alpha \sigma_\rho^{~\alpha} + \nabla_\alpha \omega_\rho^{~\alpha} - A^\alpha (\sigma_{\alpha\rho} + \omega_{\alpha\rho}) + \frac{1}{D-2} \Delta_\rho^{~\alpha} \nabla_\alpha \theta + U_\rho \left( \frac{\theta^2}{D-2} - \nabla_\alpha A^\alpha \right) \right]^2 \nonumber \\
&\quad + \frac{1}{K^2} \nabla^\rho K \left[ \nabla_\alpha \sigma_\rho^{~\alpha} + \nabla_\alpha \omega_\rho^{~\alpha} - A^\alpha (\sigma_{\alpha\rho} + \omega_{\alpha\rho}) + \frac{1}{D-2} \Delta_\rho^{~\alpha} \nabla_\alpha \theta + U_\rho \left( \frac{\theta^2}{D-2} - \nabla_\alpha A^\alpha \right) \right].
\end{align*}
The term
\begin{align}
\nabla_{\alpha} \left( \frac{\nabla_{\gamma} \nabla^{\alpha} U^{\gamma}}{K} \right) &= - \frac{\nabla_{\alpha} K}{K^{2}} \nabla_{\gamma} \nabla^{\alpha} U^{\gamma} + \frac{1}{K} \nabla_{\alpha} ( \nabla_{\gamma} \nabla^{\alpha} U^{\gamma} ), \nonumber \\
&= - \frac{\nabla_{\alpha} K}{K^{2}} \left( \nabla_{\gamma} \sigma^{\alpha \gamma} + \nabla_{\gamma} \omega^{\alpha \gamma} - U^{\alpha} \nabla \cdot A \right) \nonumber \\
&\quad + \frac{1}{K} \nabla_{\alpha} \left( \nabla_{\gamma} \sigma^{\alpha \gamma} + \nabla_{\gamma} \omega^{\alpha \gamma} - U^{\alpha} \nabla \cdot A - A^{\gamma} (\sigma_{\gamma}^{\;\;\alpha} + \omega_{\gamma}^{\;\;\alpha}) \right) + \mathcal{O}(1/D), \nonumber \\
&= + \frac{1}{K} \nabla_{\alpha} \nabla_{\gamma} \left( \sigma^{\alpha \gamma} + \omega^{\alpha \gamma} \right) - \frac{\nabla_{\alpha} K}{K^{2}} \nabla_{\gamma} \left( \sigma^{\alpha \gamma} + \omega^{\alpha \gamma} \right) + \frac{\dot{K}}{K^{2}} \nabla \cdot A \nonumber \\
&\quad-\frac{\theta}{K} \nabla \cdot A - \frac{1}{K} U^{\alpha} \nabla_{\alpha} (\nabla \cdot A) - \frac{1}{K} A^{\gamma} \nabla_{\alpha} (\sigma_{\gamma}^{\;\;\alpha} + \omega_{\gamma}^{\;\;\alpha}) + \mathcal{O}(1/D).
\end{align}
Substituting all derived expressions into Eq.~\eqref{pi:raw}, and isolating the components, yields
\begin{align}\label{eq:expanded}
\Pi &= - \frac{1}{16\pi K} \nabla \cdot A + \frac{\dot{K}}{8\pi K^4} \nabla_\alpha (\nabla^\lambda \sigma_\lambda^{~\alpha} + \nabla^\lambda \omega_\lambda^{~\alpha}) - \frac{1}{16\pi K^3} (\nabla^\lambda \sigma_\lambda^{~\sigma} + \nabla^\lambda \omega_\lambda^{~\sigma})(\nabla^\gamma \sigma_{\gamma\sigma} - \nabla^\gamma \omega_{\gamma\sigma}) \nonumber \\
&\quad + \frac{1}{8\pi K} \biggl[ \frac{1}{K} \nabla_\alpha \nabla_\gamma (\sigma^{\alpha\gamma} + \omega^{\alpha\gamma}) - \frac{\nabla_\alpha K}{K^2} \nabla_\gamma (\sigma^{\alpha\gamma} + \omega^{\alpha\gamma}) + \frac{\dot{K}}{K^2} \nabla \cdot A - \frac{\theta}{K} \nabla \cdot A \nonumber \\
&\quad - \frac{1}{K} U^\alpha \nabla_\alpha (\nabla \cdot A) - \frac{A^\gamma}{K} \nabla^\lambda (\sigma_{\lambda\gamma} - \omega_{\lambda\gamma}) \biggr] \nonumber \\
&\quad - \frac{\dot{K}}{8\pi K} - \frac{\nabla_\alpha K}{8\pi K^3} \nabla^\lambda (\sigma_\lambda^{~\alpha} + \omega_\lambda^{~\alpha}) - \frac{\nabla^2 K}{4\pi K^5} \nabla_\alpha (\nabla^\lambda \sigma_\lambda^{~\alpha} + \nabla^\lambda \omega_\lambda^{~\alpha}) \nonumber \\
&\quad + \frac{1}{8\pi K} \biggl[ \frac{1}{2} (\sigma_{\mu\nu}\sigma^{\mu\nu} - \omega_{\mu\nu}\omega^{\mu\nu}) - \frac{1}{2K^2} \Big( \nabla^\gamma \sigma_{\gamma}^{~\rho} - \nabla^\gamma \omega_{\gamma}^{~\rho} - U^\rho \nabla \cdot A \Big)^2 \nonumber \\
&\quad + \frac{\nabla_\rho K}{K^2} \Big( \nabla^\gamma \sigma_\gamma^{~\rho} - \nabla^\gamma \omega_\gamma^{~\rho} - U^\rho \nabla \cdot A \Big) \nonumber \\
&\quad - \frac{1}{K} \Big( \nabla^\gamma \sigma_\gamma^{~\beta} - \nabla^\gamma \omega_\gamma^{~\beta} - U^\beta \nabla \cdot A \Big) \left( \frac{\nabla_\beta K}{K} - \frac{2}{K} (\nabla^\lambda \sigma_{\lambda\beta} + \nabla^\lambda \omega_{\lambda\beta}) \right) \nonumber \\
&\quad - \frac{\dot{K}}{K^2} \nabla \cdot A + \frac{1}{2K^2} (\nabla \cdot A)^2 + \frac{1}{2D} \nabla \cdot A \biggr] \nonumber \\
&\quad - \frac{1}{8\pi (D-2)} \left( \frac{\nabla^2 K}{2K^2} - \frac{\dot{K}}{K} + \frac{\nabla \cdot A}{K} \right) - \frac{\theta}{8\pi(D-2)} - \frac{1}{8\pi K^2 (D-2)} \nabla_\alpha (\nabla^\lambda \sigma_\lambda^{~\alpha} + \nabla^\lambda \omega_\lambda^{~\alpha}) \nonumber \\
&\quad - \frac{U^\beta}{16\pi K^3} \nabla_\sigma \nabla_\beta (\nabla^\lambda \sigma_\lambda^{~\sigma} + \nabla^\lambda \omega_\lambda^{~\sigma}) + \frac{1}{8\pi K^4} \nabla_\alpha \nabla_\sigma \nabla^\alpha (\nabla^\lambda \sigma_\lambda^{~\sigma} + \nabla^\lambda \omega_\lambda^{~\sigma}) + \mathcal{O}(1/D^2).
\end{align}
To explicitly reduce the above expression, we use the orthogonality relations as $ U^{\alpha}A_{\alpha} = 0, \quad U^{\alpha}\sigma_{\lambda\alpha} = 0, \quad U^{\alpha}\omega_{\lambda\alpha} = 0,$ and define the spatial divergence vectors $V^\mu \equiv \nabla_\lambda \sigma^{\lambda\mu}$ and $W^\mu \equiv \nabla_\lambda \omega^{\lambda\mu}$, and expanding the following term yields
\begin{align}
- \frac{1}{16\pi K^3} \Big( V^\rho - W^\rho - U^\rho \nabla \cdot A \Big)^2 &= - \frac{1}{16\pi K^3} \Big( (V - W)^2 + U^2 (\nabla \cdot A)^2 \Big), \nonumber \\
&= - \frac{1}{16\pi K^3} (V^2 - 2V \cdot W + W^2) + \frac{1}{16\pi K^3} (\nabla \cdot A)^2.
\end{align}
Now, the term
\begin{align}
+ \frac{1}{8\pi K^3} \nabla_\rho K \Big( V^\rho - W^\rho - U^\rho \nabla \cdot A \Big) &= \frac{1}{8\pi K^3} \nabla_\rho K (V^\rho - W^\rho) - \frac{\dot{K}}{8\pi K^3} \nabla \cdot A.
\end{align}
Also expanding the following results in
\begin{align}
&- \frac{1}{8\pi K^3} \Big( V^\beta - W^\beta - U^\beta \nabla \cdot A \Big) \Big( \nabla_\beta K - 2(V_\beta + W_\beta) \Big) \nonumber \\
&= - \frac{1}{8\pi K^3} \nabla_\beta K (V^\beta - W^\beta) + \frac{\dot{K}}{8\pi K^3} \nabla \cdot A + \frac{2}{8\pi K^3} (V^\beta - W^\beta)(V_\beta + W_\beta), \nonumber \\
&= - \frac{1}{8\pi K^3} \nabla_\beta K (V^\beta - W^\beta) + \frac{\dot{K}}{8\pi K^3} \nabla \cdot A + \frac{2}{8\pi K^3} (V^2 - W^2).
\end{align}
Substituting these distributed components back into Eq.~\eqref{eq:expanded} permits direct algebraic term counting. Summing the $\dot{K} \nabla \cdot A$ coefficients yields exact cancellation
\begin{equation}
\left( \frac{1}{8\pi K^3} \right) - \left( \frac{1}{8\pi K^3} \right) - \left( \frac{1}{8\pi K^3} \right) + \left( \frac{1}{8\pi K^3} \right) = 0.
\end{equation}
Summing the coefficients of $\nabla_\mu K$ contracted with spatial divergences gives
\begin{align}
&- \frac{\nabla_\alpha K}{8\pi K^3} (V^\alpha - W^\alpha) - \frac{\nabla_\alpha K}{8\pi K^3} (V^\alpha + W^\alpha) + \frac{\nabla_\rho K}{8\pi K^3} (V^\rho - W^\rho) - \frac{\nabla_\beta K}{8\pi K^3} (V^\beta - W^\beta) \nonumber \\
&= - \frac{\nabla_\alpha K}{8\pi K^3} (2V^\alpha) = - \frac{1}{4\pi K^3} V^\alpha \nabla_\alpha K.
\end{align}
Summing the quadratic spatial divergences $(V, W)^2$ yields
\begin{align}
&- \frac{1}{16\pi K^3} (V^\delta + W^\delta)(V_\delta - W_\delta) - \frac{1}{16\pi K^3} (V^\rho - W^\rho)^2 + \frac{2}{8\pi K^3} (V^\beta - W^\beta)(V_\beta + W_\beta) \nonumber \\
&= \frac{1}{16\pi K^3} \Bigl[ - (V^2 - W^2) - (V^2 - 2V \cdot W + W^2) + 4(V^2 - W^2) \Bigr], \nonumber \\
&= \frac{1}{8\pi K^3} \Bigl[ V^2 + V \cdot W - 2W^2 \Bigr].
\end{align}
The quadratic and linear $\nabla \cdot A$ components combine as follows:
\begin{align}
\left( \frac{1}{16\pi K^3} + \frac{1}{16\pi K^3} \right) (\nabla \cdot A)^2 &= \frac{1}{8\pi K^3} (\nabla \cdot A)^2, \\
- \left( \frac{1}{16\pi K} - \frac{1}{16\pi D K} + \frac{1}{8\pi K(D-2)} \right) \nabla \cdot A &= - \frac{1}{16\pi K} \nabla \cdot A+\mathcal{O}(1/D^2).
\end{align}
Combining all the $\dot{K}$ terms gives:
\begin{equation}
- \frac{\dot{K}}{8\pi K} + \frac{\dot{K}}{8\pi K(D-2)} = - \frac{D-3}{8\pi (D-2)} \frac{\dot{K}}{K}.
\end{equation}
Furthermore, the expression of double divergence of the antisymmetric vorticity tensor $\omega^{\alpha\gamma}$ vanishes. As it reduces to the commutator of the covariant derivatives, evaluated using the Ricci identity
\begin{equation}
    \nabla_{\alpha}\nabla_{\gamma}\omega^{\alpha\gamma} = \frac{1}{2} [\nabla_{\alpha}, \nabla_{\gamma}] \omega^{\alpha\gamma} = \frac{1}{2} \left( R^{\alpha}_{\;\;\mu\alpha\gamma} \omega^{\mu\gamma} + R^{\gamma}_{\;\;\mu\alpha\gamma} \omega^{\alpha\mu} \right) = R_{\mu\nu} \omega^{\mu\nu} = 0.
\end{equation}
This contraction vanishes because the Ricci tensor $R_{\mu\nu}$ is symmetric. Substituting all simplified algebraic terms directly into the following expression:
\begin{align}\label{pi:final}
\Pi &= - \frac{1}{16\pi K} \nabla \cdot A + \frac{1}{16\pi K} (\sigma_{\mu\nu}\sigma^{\mu\nu} - \omega_{\mu\nu}\omega^{\mu\nu})  - \frac{D-3}{8\pi (D-2)} \frac{\dot{K}}{K} + \frac{(\nabla \cdot A)^2}{8\pi K^3} \nonumber \\
&\quad + \frac{\dot{K}}{8\pi K^4} \nabla_{\alpha} \nabla_{\lambda} \sigma^{\lambda\alpha} - \frac{\theta}{8\pi K^2} \nabla \cdot A - \frac{1}{8\pi K^2} U^{\alpha} \nabla_{\alpha} (\nabla \cdot A) - \frac{1}{16\pi(D-2)} \frac{\nabla^2 K}{K^2}\nonumber \\
&\quad  - \frac{\nabla^2 K}{4\pi K^5} \nabla_{\alpha} \nabla_{\lambda} \sigma^{\lambda\alpha}- \frac{\theta}{8\pi(D-2)} - \frac{1}{8\pi K^2 (D-2)} \nabla_{\alpha} \nabla_{\lambda} \sigma^{\lambda\alpha} \nonumber \\
&\quad + \frac{1}{8\pi K^3} \Bigl[ (\nabla^{\lambda} \sigma_{\lambda}^{\;\;\alpha})(\nabla^{\gamma} \sigma_{\gamma\alpha}) + (\nabla^{\lambda} \sigma_{\lambda}^{\;\;\alpha})(\nabla^{\gamma} \omega_{\gamma\alpha}) - 2 (\nabla^{\lambda} \omega_{\lambda}^{\;\;\alpha})(\nabla^{\gamma} \omega_{\gamma\alpha}) \Bigr] \nonumber \\
&\quad - \frac{1}{4\pi K^3} (\nabla^{\lambda} \sigma_{\lambda}^{\;\;\alpha}) \nabla_{\alpha} K + \frac{1}{8\pi K^2} \nabla_{\alpha} \nabla_{\gamma} \sigma^{\alpha \gamma} - \frac{1}{8\pi K^2} A^{\gamma} (\nabla^{\lambda} \sigma_{\lambda\gamma} - \nabla^{\lambda} \omega_{\lambda\gamma}) \nonumber \\
&\quad - \frac{U^\beta}{16\pi K^3} \nabla_\sigma \nabla_\beta (\nabla^\lambda \sigma_\lambda^{~\sigma} + \nabla^\lambda \omega_\lambda^{~\sigma}) + \frac{1}{8\pi K^4} \nabla_\alpha \nabla_\sigma \nabla^\alpha (\nabla^\lambda \sigma_\lambda^{~\sigma} + \nabla^\lambda \omega_\lambda^{~\sigma}) + \mathcal{O}(1/D^2).
\end{align}
Now, the following higher-order derivative terms can be further simplified:
\subsection*{Term 1}
\begin{align}
U^\beta \nabla_\alpha \nabla_\beta (\nabla^\lambda \sigma_\lambda^{~\alpha} + \nabla^\lambda \omega_\lambda^{~\alpha}) &= U^\beta \nabla_\beta \nabla_\alpha (\nabla^\lambda \sigma_\lambda^{~\alpha} + \nabla^\lambda \omega_\lambda^{~\alpha}) \nonumber \\
&\quad + U^\beta R_{\alpha\beta\rho}^{~~~~\alpha} (\nabla_\lambda \sigma^{\lambda\rho} + \nabla_\lambda \omega^{\lambda\rho}), \nonumber \\
&= U^\beta \nabla_\beta (\nabla_\alpha \nabla_\lambda \sigma^{\lambda\alpha}) - U^\beta R_{\beta\rho} (\nabla^\lambda \sigma_\lambda^{~\rho} + \nabla^\lambda \omega_\lambda^{~\rho}). \nonumber \\
\intertext{Using the Gauss-Codazzi relation to substitute the Ricci tensor:}
U^\beta \nabla_\alpha \nabla_\beta (\nabla^\lambda \sigma_\lambda^{~\alpha} + \nabla^\lambda \omega_\lambda^{~\alpha}) &= U^\beta \nabla_\beta (\nabla_\alpha \nabla_\lambda \sigma^{\lambda\alpha}) \nonumber \\
&\quad - U^\beta (K K_{\beta\rho} - K_{\beta\delta} K^\delta_\rho) (\nabla^\lambda \sigma_\lambda^{~\rho} + \nabla^\lambda \omega_\lambda^{~\rho}), \nonumber \\
&= U^\beta \nabla_\beta (\nabla_\alpha \nabla_\lambda \sigma^{\lambda\alpha}) \nonumber \\
&\quad - K \frac{\nabla_\lambda \nabla_\rho U^\lambda}{K} (\nabla^\lambda \sigma_\lambda^{~\rho} + \nabla^\lambda \omega_\lambda^{~\rho}) + \mathcal{O}(D).
\end{align}
To further simplify this expression, we substitute the standard kinematic decomposition of the velocity gradient, 
\begin{align}
\nabla_\lambda \nabla_\rho U^\lambda &= \nabla_\lambda \left( \sigma_\rho^{~\lambda} + \omega_\rho^{~\lambda} - U_\rho A^\lambda + \frac{\theta}{D-2} \Delta_\rho^{~\lambda} \right), \nonumber \\
&= \nabla^\lambda \sigma_{\lambda\rho} - \nabla^\lambda \omega_{\lambda\rho} - U_\rho \nabla \cdot A + \mathcal{O}(1), \nonumber \\
&= V_\rho - W_\rho - U_\rho \nabla \cdot A + \mathcal{O}(1).
\end{align}
Using this in the following algebraic expression yields the following:
\begin{align}
- (\nabla_\lambda \nabla_\rho U^\lambda) (\nabla^\gamma \sigma_\gamma^{~\rho} + \nabla^\gamma \omega_\gamma^{~\rho}) &= - (V_\rho - W_\rho - U_\rho \nabla \cdot A)(V^\rho + W^\rho), \nonumber \\
&= - (V_\rho - W_\rho)(V^\rho + W^\rho) + \mathcal{O}(D), \nonumber \\
&= - (V_\rho V^\rho + V_\rho W^\rho - W_\rho V^\rho - W_\rho W^\rho) + \mathcal{O}(D), \nonumber \\
&= - (V^2 - W^2) + \mathcal{O}(D).
\end{align}
So, the complete simplified expression for the total derivative block is given as follows:
\begin{align}
U^\beta \nabla_\alpha \nabla_\beta (\nabla^\lambda \sigma_\lambda^{~\alpha} + \nabla^\lambda \omega_\lambda^{~\alpha}) &= U^\beta \nabla_\beta (\nabla_\alpha \nabla_\lambda \sigma^{\lambda\alpha}) \nonumber \\
&\quad - \left[ (\nabla^\lambda \sigma_\lambda^{~\rho})(\nabla^\gamma \sigma_{\gamma\rho}) - (\nabla^\lambda \omega_\lambda^{~\rho})(\nabla^\gamma \omega_{\gamma\rho}) \right] + \mathcal{O}(D).
\end{align}
\subsection*{Term 2}
Using the Gauss-Codazzi relations $R_{\alpha\rho} = K K_{\alpha\rho} - K_{\alpha\delta} K^\delta_\rho,\quad \nabla_\alpha K^{\alpha \beta}=\nabla^\beta K$ and large $D$ scaling, the leading order expression of the following term is extracted as:
\begin{align}
\nabla_\alpha \nabla_\delta \nabla^\alpha (\nabla^\lambda \sigma_\lambda^{~\delta} + \nabla^\lambda \omega_\lambda^{~\delta}) &= \nabla_\alpha \nabla^\alpha \nabla_\delta (\nabla^\lambda \sigma_\lambda^{~\delta} + \nabla^\lambda \omega_\lambda^{~\delta})  + \nabla_\alpha \left( R_{\delta~~~\rho}^{~\alpha\delta} (\nabla^\lambda \sigma_\lambda^{~\rho} + \nabla^\lambda \omega_\lambda^{~\rho}) \right), \nonumber \\
&= \nabla^2 (\nabla_\delta \nabla_\lambda \sigma^{\lambda\delta})  + \nabla^\alpha \Big( R_{\alpha\rho} (\nabla^\lambda \sigma_\lambda^{~\rho} + \nabla^\lambda \omega_\lambda^{~\rho}) \Big), \nonumber \\
&= \nabla^2 (\nabla_\delta \nabla_\lambda \sigma^{\lambda\delta})  + (\nabla^\lambda \sigma_\lambda^{~\rho} + \nabla^\lambda \omega_\lambda^{~\rho}) \nabla^\alpha (K K_{\alpha\rho} - K_{\alpha\delta} K^\delta_\rho), \nonumber \\
&= \nabla^2 (\nabla_\delta \nabla_\lambda \sigma^{\lambda\delta})  + \nabla^\lambda (\sigma_\lambda^{~\rho} + \omega_\lambda^{~\rho}) (K \nabla_\rho K) + \mathcal{O}(D^2).
\end{align}
Incorporating those changes, the final expression of bulk pressure becomes
\begin{align}\label{pi:final_simplified}
\Pi &= - \frac{1}{16\pi K} \nabla \cdot A + \frac{1}{16\pi K} (\sigma_{\mu\nu}\sigma^{\mu\nu} - \omega_{\mu\nu}\omega^{\mu\nu}) - \frac{D-3}{8\pi (D-2)} \frac{\dot{K}}{K} + \frac{(\nabla \cdot A)^2}{8\pi K^3} \nonumber \\
&\quad + \frac{\dot{K}}{8\pi K^4} \nabla_{\alpha} \nabla_{\lambda} \sigma^{\lambda\alpha} - \frac{\theta}{8\pi K^2} \nabla \cdot A - \frac{1}{8\pi K^2} U^{\alpha} \nabla_{\alpha} (\nabla \cdot A) \nonumber \\
&\quad - \frac{1}{16\pi(D-2)} \frac{\nabla^2 K}{K^2} - \frac{\nabla^2 K}{4\pi K^5} \nabla_{\alpha} \nabla_{\lambda} \sigma^{\lambda\alpha} - \frac{\theta}{8\pi(D-2)} - \frac{1}{8\pi K^2 (D-2)} \nabla_{\alpha} \nabla_{\lambda} \sigma^{\lambda\alpha} \nonumber \\
&\quad + \frac{1}{16\pi K^3} \Bigl[ 3(\nabla^{\lambda} \sigma_{\lambda}^{\;\;\alpha})(\nabla^{\gamma} \sigma_{\gamma\alpha}) + 2(\nabla^{\lambda} \sigma_{\lambda}^{\;\;\alpha})(\nabla^{\gamma} \omega_{\gamma\alpha}) - 5(\nabla^{\lambda} \omega_{\lambda}^{\;\;\alpha})(\nabla^{\gamma} \omega_{\gamma\alpha}) \Bigr] \nonumber \\
&\quad - \frac{1}{8\pi K^3} \nabla_{\alpha} K (\nabla^{\lambda} \sigma_{\lambda}^{\;\;\alpha} - \nabla^{\lambda} \omega_{\lambda}^{\;\;\alpha})  + \frac{1}{8\pi K^2} \nabla_{\alpha} \nabla_{\gamma} \sigma^{\alpha \gamma} - \frac{1}{8\pi K^2} A^{\gamma} (\nabla^{\lambda} \sigma_{\lambda\gamma} - \nabla^{\lambda} \omega_{\lambda\gamma}) \nonumber \\
&\quad - \frac{U^\beta}{16\pi K^3} \nabla_\beta (\nabla_{\alpha} \nabla_{\lambda} \sigma^{\lambda\alpha}) \nonumber + \frac{1}{8\pi K^4} \nabla^2 (\nabla_{\alpha} \nabla_{\lambda} \sigma^{\lambda\alpha}) + \mathcal{O}(1/D^2).
\end{align}

\bibliographystyle{JHEP}
\bibliography{biblio}

\end{document}